\documentclass{jfm}
\usepackage{amsmath}
\usepackage{mathrsfs}
\usepackage{hyperref}
\hypersetup{colorlinks=true, allcolors=blue}
\usepackage{multirow}

\usepackage{color}
\usepackage[authormarkup=superscript]{changes}\definechangesauthor[color=purple]{Jie}

\newcommand{\ie}{\textit{i.e.~}}
\newcommand{\PrEc}{\mbox{\textit{PrEc}}} 
\newcommand{\Ec}{\mbox{\textit{Ec}}} 
\newcommand{\Ma}{\mbox{\textit{Ma}}} 
\newcommand\pp{\partial}
\newcommand\dd{\mathrm{d}}

\title{Linear instability of Poiseuille flows with highly non-ideal fluids}
\shorttitle{Linear instability of Poiseuille flows with highly non-ideal fluids}
\author{Jie Ren\aff{1}\corresp{\email{j.ren-1@tudelft.nl}}, Song Fu\aff{2}
		\and Rene Pecnik\aff{1}\corresp{\email{r.pecnik@tudelft.nl}}}
\shortauthor{J. Ren, S. Fu and R. Pecnik}
\affiliation{\aff{1}Process and Energy Department, Delft University of Technology, Leeghwaterstraat 39, 2628 CB Delft, The Netherlands \aff{2}School of aerospace engineering, Tsinghua University, Beijing 100084, China}

\newcommand{\red}[1]{{\color{black}{#1}}}

\begin{document}
\graphicspath{{Figures/}}
\maketitle

\begin{abstract}
The objective of this work is to investigate linear modal and algebraic instability in Poiseuille flows with fluids close to their vapour-liquid critical point. Close to this critical point, the ideal gas assumption does not hold and large non-ideal fluid behaviours occur. As a representative non-ideal fluid, we consider supercritical carbon dioxide (CO$_2$) at pressure of 80 bar, which is above its critical pressure of 73.9 bar. 
The Poiseuille flow is characterized by the Reynolds number ($\Rey=\rho_{w}^{*}u_{r}^{*}h^{*}/\mu_{w}^{*}$), the product of Prandtl ($\Pran=\mu_{w}^{*}C_{pw}^{*}/\kappa_{w}^{*}$) and Eckert number ($\Ec=u_{r}^{*2}/C_{pw}^{*}T_{w}^{*}$), and the wall temperature that in addition to pressure determines the thermodynamic reference condition. For low Eckert numbers, the flow is essentially isothermal and no difference with the well-known stability behaviour of incompressible flows is observed. However, if the Eckert number increases, the viscous heating causes gradients of thermodynamic and transport properties, and non-ideal gas effects become significant. Three regimes of the laminar base flow can be considered, subcritical (temperature in the channel is entirely below its pseudo-critical value),  transcritical, and supercritical temperature regime. If compared to the linear stability of an ideal gas Poiseuille flow, we show  that the base flow is more unstable in the subcritical regime, inviscid unstable in the transcritical regime, while significantly more stable in the supercritical regime. Following the corresponding states principle, we expect that qualitatively similar results will be obtained for other fluids at equivalent thermodynamic states. 
\end{abstract}

\begin{keywords}
non-ideal gas, absolute/convective instability, compressible flows
\end{keywords}

\section{Introduction}\label{Sec1}

% 1 - aplications
Many processes in industrial applications constitute of flows with fluids that do not follow the ideal gas law. For example, flows in vapour power systems, re-entry of spacecrafts, supercritical dyeing, refrigeration and heat pump systems \citep[examples in supercritical fluids can be found in][]{Brunner2010}. The non-ideality of fluids is especially significant in the thermodynamic region close to the vapour critical point. As such, it is of great importance to understand the fundamental physics that are related to flows with these fluids.

% 2 - turbulence 
Recently, researchers have studied how non-ideal gas effects influence turbulence and heat transfer. For example, \citet{Kawai2015,Kawai2016} studied turbulent boundary layers with supercritical pressures and transcritical temperatures. They found that the mean velocity profiles (with density weighted Van Driest transformation) coincide with the same log-law as seen in an ideal gas. \citet{Sciacovelli2017,Sciacovelli2017b,Sciacovelli2016} comprehensively studied turbulence dynamics and near wall turbulence of flows with molecularly complex fluids in the dense gas regime using direct numerical simulations. They found that dense-gas flows with a heavy fluorocarbon exhibit almost negligible friction heating (in channel flows), weakening of compressive (and enhancement of expanding) structures (in homogeneous isotropic turbulence). \citet{Patel2016} studied the influence of variable properties on fully developed turbulent channel flows and derived a velocity transformation that allows to collapse velocity profiles for heated or cooled non-ideal fluids. Moreover, \citet{Rinaldi2017} provided an explanation of near wall turbulence modulation, especially the intercomponent energy transfer that has been observed by, \eg \citet{Morinishi2004}, \citet{Pirozzoli2008}, \citet{Duan2010}. \citet{Nemati2016,Peeters2016} studied turbulent heat transfer to supercritical CO$_2$, indicating that both the mean and instantaneous property variations have significant effects on turbulent structures and their self-regeneration processes in near-wall turbulence. \citet{Alferez2017} have studied the refraction properties of compression shock waves in non-ideal gases. One of the new regimes found is that, due to the non-ideality of the fluid it is possible that acoustic modes can be completely damped by a compression shock, leading to so-called `quiet shocks'.

\begin{figure}
\begin{center}
\includegraphics[width=0.49\linewidth]{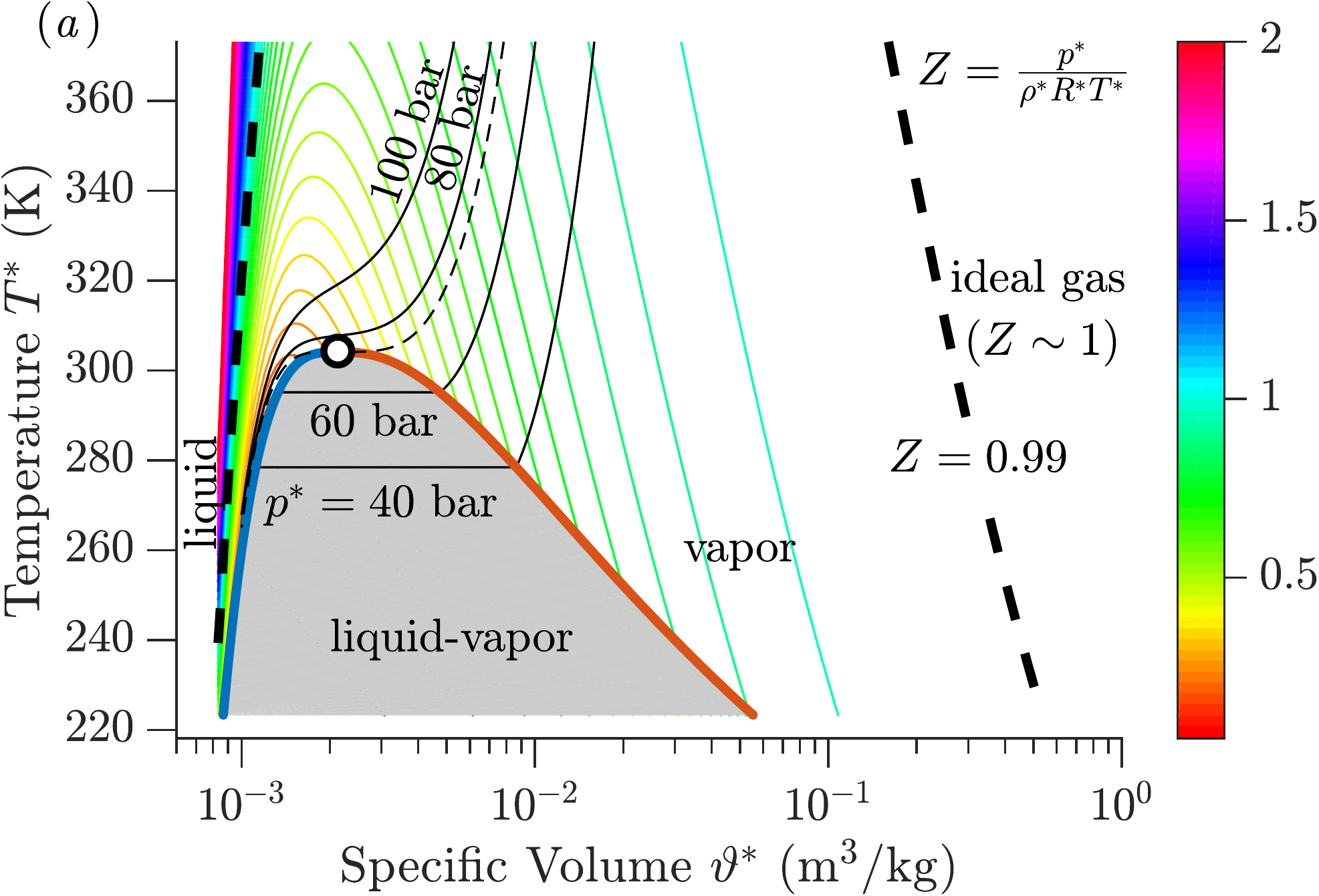}
\includegraphics[width=0.49\linewidth]{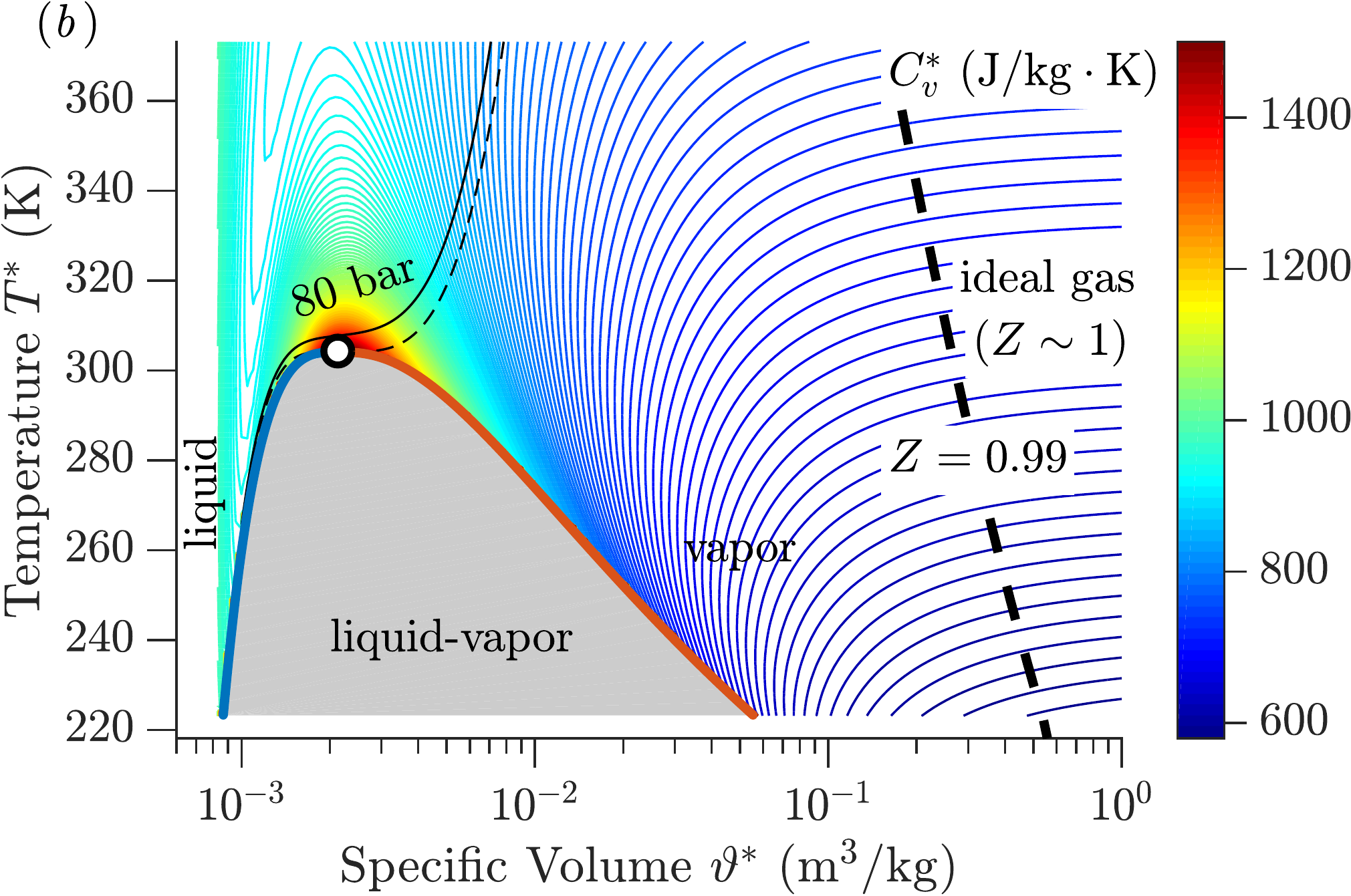}\\
\includegraphics[width=0.49\linewidth]{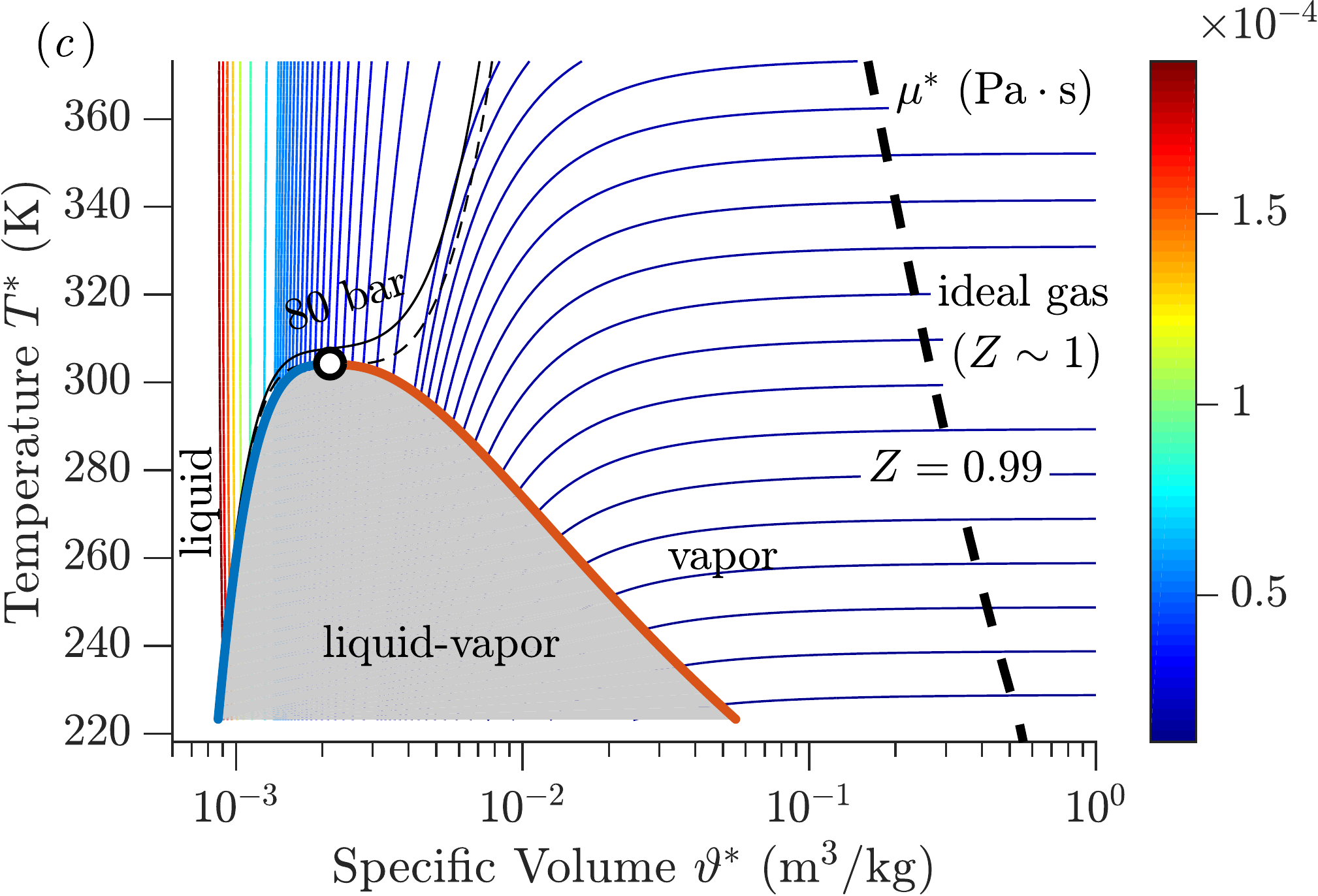}
\includegraphics[width=0.49\linewidth]{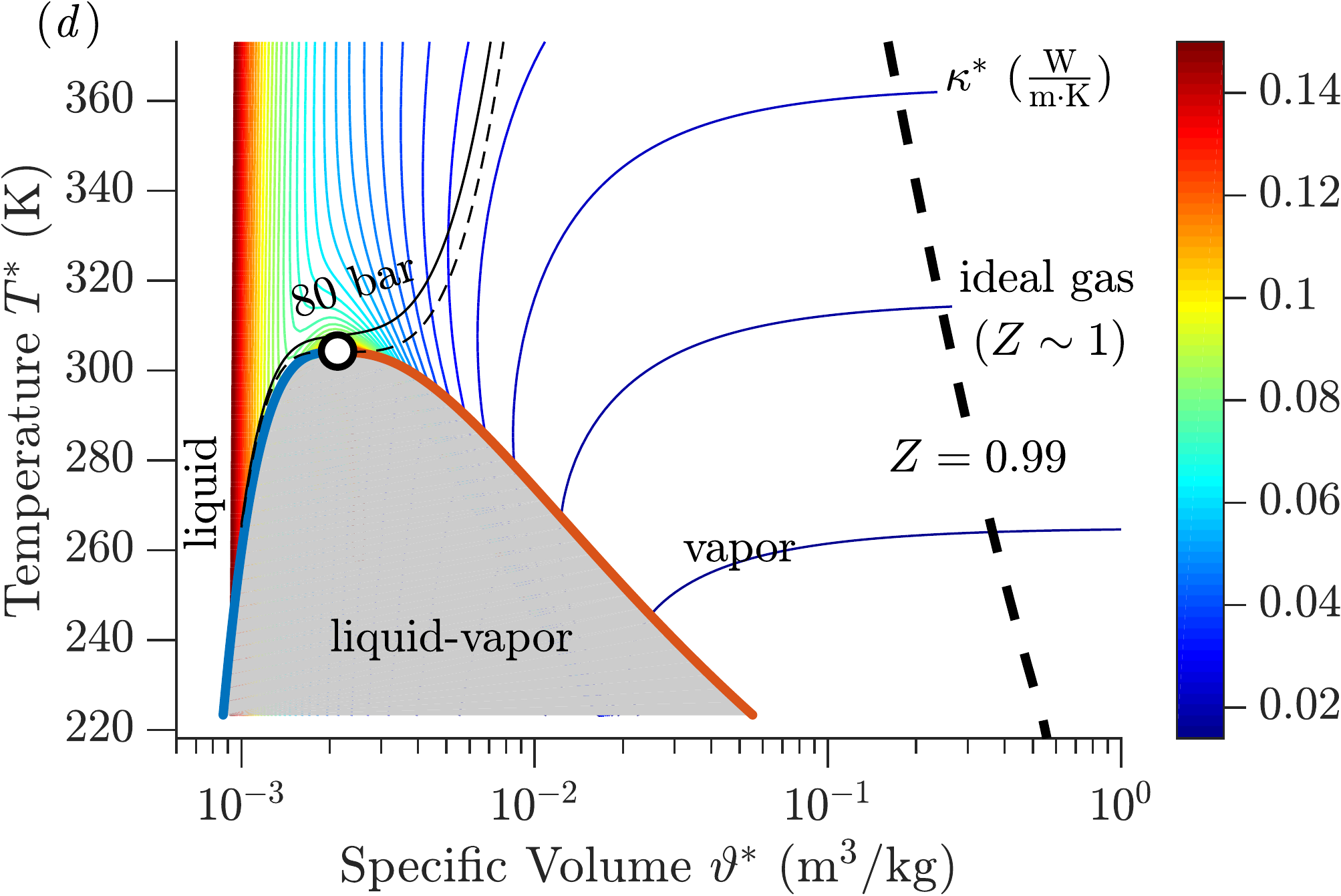}
\end{center}
\caption{$T-\vartheta$ diagram of CO$_2$ along with the critical point (white circle), the saturation curves (blue and red solid lines), the liquid-vapour region (grey area), the critical isobar (black thin dashed line), isobars of 40, 60, 80 and 100 bar (black thin lines), compressibility factor $Z=0.99$ (black thick dashed line) as well as colored contour lines of (a) $Z=p^*/\rho^* R^*T^*$, (b) the specific heat capacity at constant volume $C_v^*$, (c) the dynamic viscosity $\mu^*$, and (d) the thermal conductivity $\kappa^*$.}
\label{Fig1}
\end{figure}

% 3 - Ideal / non-ideal gas 
For ideal gases, the thermodynamic properties are associated with a simple equation of state (EOS). Additionally, the transport properties (namely, the viscosity and thermal conductivity) can be estimated as unary functions of the temperature (\eg the widely used power law or Sutherland's law). To assess to which degree the ideal gas law holds, it is possible to evaluate the compressibility factor, defined as $Z=p^*/\rho^* R^*T^*$. Figure~\ref{Fig1} shows the $T-\vartheta$ diagram (temperature - specific volume diagram, $\vartheta=1/\rho$) of carbon dioxide CO$_2$. The white circle in each subplot indicates the critical point, which for CO$_2$ is at a pressure and temperature of $p^*_c=73.9$ bar and $T^*_c=304.25$ K. In this paper, we denote dimensional and critical quantities with superscript `$*$' and subscript `$c$', respectively. Figure \ref{Fig1}(a) shows the critical isobar (black thin dashed line), four isobars of 40 to 100 bar (black thin lines), and the compressibility factor $Z$ (colored contour lines). Close to the critical point, the non-ideality is clearly indicated by low values of $Z$, while the boundary between ideal and non-ideal gas behavior is roughly indicated by the thick dashed line of $Z=0.99$. The distribution of the thermodynamic and transport properties (specific heat capacity at constant volume $C_v^*$, dynamic viscosity $\mu^*$ and thermal conductivity $\kappa^*$) are shown in figure \ref{Fig1}(b,c,d). In the ideal gas region, these contour lines become quasi-parallel to the $x$-axis, indicating that they can be regarded as functions of temperature only. On the other hand, near the critical point, the gradients of these properties with respect to temperature and specific volume become significant. 

% 4 - Stability
In view of its great simplicity, most of the present knowledge on stability and laminar-turbulent transition is limited to the ideal gas \citep{Fedorov2011} or incompressible flows, where thermodynamic properties are constant. On the other hand, numerical simulations of real gas effects (high-enthalpy effects) in hypersonic flows has just gone through an initial stage \citep{Zhong2012,Marxen2013,Marxen2014}. In fact, the well-known Orr-Sommerfeld equation \citep[][often termed the O-S equation]{Orr1907, Sommerfeld1908} was derived by applying the linear stability theory (LST) to the incompressible parallel plane shear flow. Solved as an eigenvalue problem (in the time/space-asymptotic limit), the growth rate and profiles of the perturbation are obtained from the most unstable mode as its eigenvalue and eigenvector. This is known as the modal stability problem. The critical Reynolds number $\Rey_c$, below which the flow is stable,  regardless of the wavenumber and frequency of the perturbation, is often determined and emphasized in such modal stability analysis. For example, in plane Poiseuille flow, $\Rey_c$  is numerically determined to be 5772.22 \citep{Thomas1953, Orszag1971}. Here the Reynolds number is based on the half-channel height and the centerline flow velocity. Due to the non-normality of most practical linear systems, the modal stability analysis cannot cover the full behavior of the linear instability \citep{Schmid2001, Schmid2014}. Instead of solving the eigenvalue problem, the stability equation can be formulated as an initial value problem under the framework of constrained optimization. Maximizing the energy growth in a finite domain of time or space, leads to the optimal perturbation, which grows transiently even below the critical Reynolds number $\Rey_c$. This is termed the transient growth or algebraic growth. Accordingly, a ``critical'' Reynolds number can be defined for the algebraic growth as well. Also for plane Poiseuille flow, this number is 49.6 \citep{Joseph1969, Busse1969}. 

Studies of viscosity stratified flows, where the viscosity depends on temperature, has recently received attention, readers may refer to \citet{Govindarajan2014} for a review. Based on the modified O-S equations, early studies show that including a linear temperature profile destabilizes the Poiseuille flow \citep{Potter1972} and stabilizes/destabilizes the water boundary layer flow (depend on wall heating/cooling) \citep{Wazzan1972}. However, viscosity and temperature perturbations were ignored in both studies and were later examined by \citet{Pinarbasi1995}. \citet{Wall1996,Wall1997} investigated the effects of different viscosity models, indicating that the flow can either be more stable or unstable. The study on wall heating and viscosity-stratification has also been extended to transient growth, secondary instability \red{as well as instabilites in other types of flows} \citep{Chikkadi2005,SAMEEN2007,SAMEEN2011,sahu2011,sahu2014}. For compressible plane Couette flow, \citet{Malik2008} showed that the flow is more stable with viscosity stratification, while recently, a further study on this flow is given by \citet{Saikia2017}, in which the effects of individual/combined viscosity-thermal conductivity stratification are elucidated. The influence of viscosity gradients on the edge state is recently studied by \citet{Rinaldi2018}, showing that in minimal channel flows, the kinetic energy level and the driving force of self-sustained cycle of the edge state depends on viscosity distribution. The above studies are based on the incompressible flow assumption or the ideal gas equation-of-state (EoS), at the same time, transport properties are estimated as functions of temperature only. 

% 5 - Motivation
Since there is very limited knowledge on flow stability with highly non-ideal fluids, we investigate Poiseuille flows with fluids close to the thermodynamic vapour-liquid critical point. In \S \ref{Sec2}, the gas model, the formulation of the stability analysis and the related numerical methods are outlined in detail. The results and discussions on the base flow are provided in \S \ref{Sec3}, followed by the modal growth and algebraic instability in \S \ref{Sec4} and \ref{Sec5}, respectively. The paper is concluded in \S \ref{Sec6}.

\section{Governing equations}\label{Sec2}

\subsection{Flow conservation equations}\label{Sec2-1}
The laws of conservation of mass, momentum and energy (the Navier-Stokes (N-S) equations), in dimensionless form, are given by 
\begin{equation}\label{NS1}
\frac{\partial\rho}{\partial t}+\frac{\partial\left(\rho u_{j}\right)}{\partial x_{j}}=0,
\end{equation}
\begin{equation}\label{NS2}
\frac{\partial\left(\rho u_{i}\right)}{\partial t}+\frac{\partial\left(\rho u_{i}u_{j}+p\delta_{ij}-\tau_{ij}\right)}{\partial x_{j}}=F_i,
\end{equation}
\begin{equation}\label{NS3}
\frac{\partial\left(\rho E\right)}{\partial t}+\frac{\partial\left(\rho Eu_{j}+pu_{j}+q_{j}-u_{i}\tau_{ij}\right)}{\partial x_{j}}=u_jF_j,
\end{equation}
where $x_i=(x,y,z)$ are the coordinates in the streamwise, wall-normal and spanwise directions, $u_i=(u,v,w)$ are the velocity components, $t$ the time, $\rho$ the fluid density, $E=e+u_iu_i/2$ the total energy, $e$ the internal energy, $F_i$ the body force and $p$ is the pressure. The viscous stress tensor, $\tau_{ij}$, and the heat flux, $q_{j}$, are given by 
\begin{equation}\label{stresstensor}
\tau_{ij}=\frac{\mu}{\Rey}\left(\frac{\partial u_{i}}{\partial x_{j}}+\frac{\partial u_{j}}{\partial x_{i}}\right)+\frac{\lambda}{\Rey}\delta_{ij}\frac{\partial u_{k}}{\partial x_{k}}, \quad
%\end{equation}
%\begin{equation}
q_{j}=-\frac{\kappa}{\Rey\PrEc}\frac{\partial T}{\partial x_{j}}.
\end{equation}
Here $\mu$ is the dynamic viscosity, $\lambda=\mu_b -2/3 \mu$ the second viscosity, $\mu_b$ the bulk viscosity, and $\kappa$ is the thermal conductivity. Results presented in the following sections are subject to $\mu_b=0$. However, we will discuss the influence of the bulk viscosity on the linear stability in Appendix \ref{appC}. 

The above equations have been non-dimensionalized by reference values, as follows
\begin{gather}
u=\frac{u^{*}}{u_{r}^{*}},~
x_{i}=\frac{x_{i}^{*}}{h^{*}},~
t=\frac{t^{*}u_{r}^{*}}{h^{*}},~
p=\frac{p^{*}}{\rho_{w}^{*}u_{r}^{*2}},~
\rho=\frac{\rho^{*}}{\rho_{w}^{*}},~\nonumber \\
T=\frac{T^{*}}{T_{w}^{*}},~
E=\frac{E^{*}}{u_{r}^{*2}},
\mu=\frac{\mu^{*}}{\mu_{w}^{*}},~
\kappa=\frac{\kappa^{*}}{\kappa_{w}^{*}},
\end{gather}
which leads to the definition of the Reynolds number, $\Rey$, Prandtl number, $\Pran$, Eckert number, $\Ec$ and the Mach number, $\Ma$, which are given as 
\begin{equation}
\Rey=\frac{\rho_{w}^{*}u_{r}^{*}h^{*}}{\mu_{w}^{*}},~
\Pran=\frac{\mu_{w}^{*}C_{pw}^{*}}{\kappa_{w}^{*}},~
\Ec=\frac{u_{r}^{*2}}{C_{pw}^{*}T_{w}^{*}},~
\Ma=\frac{u_{r}^{*}}{c^*_w}.
\end{equation}
The subscript $w$ denotes wall values, $h^*$ is the half channel height, $c^*_w$ is the speed of sound at the wall, $u^*_r$ is the reference velocity. Note that for an ideal gas $\Ec=(\gamma-1)\Ma^2$, where $\gamma$ is the heat capacity ratio. In this study, both walls are at the same temperature. \red{Discussions on the choice of different reference scalings are provided in Appendix \ref{appD}. }

\subsection{Fluid equations of state}
In order to find a closed form of the conservation equations, an equation of state for the fluid has to be specified. As a representative example of non-ideal fluids, the study is performed with CO$_2$ at a pressure of $p^*= 80$ bar, which is above the critical pressure, within the highly non-ideal thermodynamic region (see the isobar in figure \ref{Fig1}). To account for the non-ideal gas effects, the NIST REFPROP library \citep{Lemmon_2002} has been used to obtain the most accurate thermodynamic and transport properties along with their gradients.
\red{The multi-parameter EoS (in functional forms) used in REFPROP are developed with an optimization algorithm. The EoS are suitable for a broad variety of fluids while high accuracy can be maintained. Readers shall refer to \citet{Span2003a} for the derivation of the EoS. To build the linear stability equations (see Appendix \ref{appA}), the temperature $T_0$ and density $\rho_0$ are provided as input, while the required properties and their derivatives are obtained as output from REFPROP.} 
Moreover, as a direct method to determine the thermodynamic properties, several cubic EoS (see Appendix \ref{appB}), i.e. van der Waals \citep{VdW_1873}, Redlich-Kwong \citep{RK_1949} and Peng-Robinson \citep{PR_1976}, are used for the stability analysis as comparison. All results with the non-ideal EOS are also compared with an ideal gas model (IG). \red{A constant specific heat ratio $\gamma=1.289$ is used for the IG model.} All the fluid models are summarized in table \ref{Table1}. 
\begin{table}
\begin{center}
\begin{tabular}{lll}
Fluid model 	& EoS 			& Transport properties\vspace{5pt}\\
RP    		& REFPROP   		& REFPROP\\
PR    		& Peng-Robinson 	& REFPROP\\
RK    		& Redlich-Kwong 	& REFPROP\\
VW    		& van der Waals 	& REFPROP\\
IG    		& ideal gas  		& Power/Sutherland law
\end{tabular}
\caption{Fluid models studied in this paper. Gradients of the properties (with respect to temperature and density) are calculated analytically (see Appendix \ref{appB}) or numerically with a finite-difference algorithm (REFPROP). As shown in figure~\ref{Fig2}(c,d), there is no  discernible difference using the power or Sutherland law for the IG model, therefore results presented in this study for IG are based on the power law.}
\label{Table1}
\end{center}
\end{table}

\begin{figure}
\begin{center}
\vspace{10pt}
\includegraphics[width=0.95\linewidth]{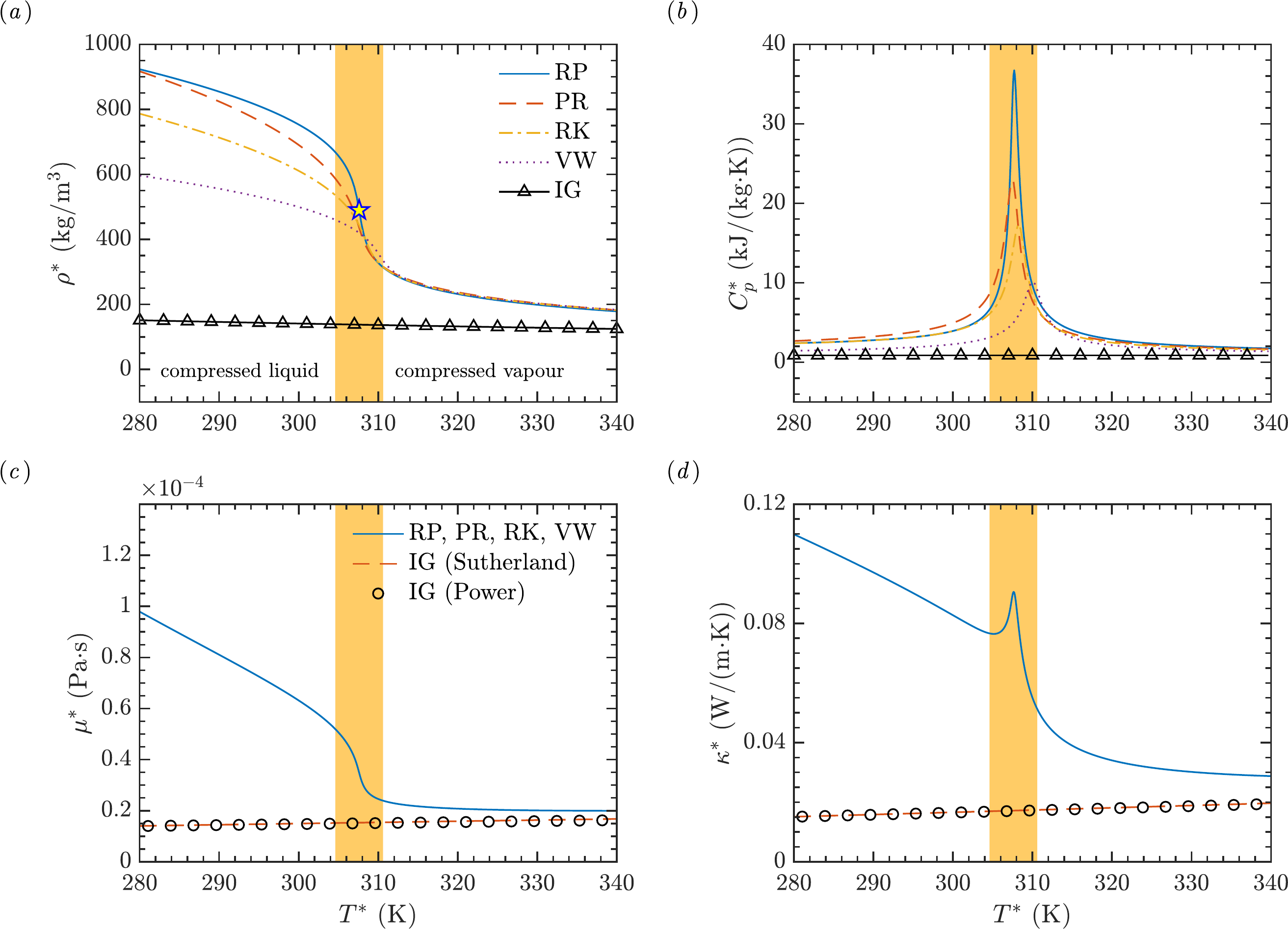}
\end{center}
\caption{Thermodynamic and transport properties of CO$_2$ at $p^*=80$ bar for the fluid models in table~\ref{Table1}. Sub-figures show the distribution of (a) density $\rho^*$, (b) heat capacity at constant pressure $C_p^*$, (c) viscosity $\mu^*$ and, (d) thermal conductivity $\kappa^*$ versus temperature $T^*$. The pentagram shows the pseudo-critical temperature $T^*_{pc}$ (RP model). The shaded area indicates the pseudo-critical transition.}
\label{Fig2}
\end{figure}

Figure \ref{Fig2} shows the thermodynamic and transport properties of CO$_2$ at a pressure of 80 bar. The pentagram in subplot (a) shows the pseudo-critical temperature ($T_{pc}^*=307.7$~K, RP model), which is defined as the point on a supercritical isobar where $C_p^*$ reaches a maximum. Near $T_{pc}^*$, all properties show large gradients, which do not exist in an ideal gas. As shown in figure \ref{Fig2}(a,b), the Peng-Robinson (PR) EoS is closest to the highly accurate multiparameter EoS of CO$_2$ as implemented in REFPROP (RP). % for $\rho^*$ and $C_p^*$ near $T_{pc}^*$. 
In general, the cubic EoSs do capture key features of the thermodynamic property variations. 
%However, as we will show later, they are not capable to quantitatively predict the linear instability.
In figure~\ref{Fig2}(c,d), the power law~\eqref{power} and Sutherland law~\eqref{suther}, which fall on top of each other, are compared to the distributions from RP. The power and Sutherland laws for dynamic viscosity and thermal conductivity are given as 
\begin{equation}\label{power}
\frac{\mu^*}{\mu^*_{\textrm{ref}}}=\left(\frac{T^*}{T^*_\textrm{ref}}\right)^{n_{1}},~
\frac{\kappa^*}{\kappa^*_\textrm{ref}}=\left(\frac{T^*}{T^*_\textrm{ref}}\right)^{n_{2}},~n_1=0.79,~
n_2=1.30,~
{T_{\textrm{ref}}^{*}}=273~\textrm{K},
\end{equation}
\begin{equation}\label{suther}
\frac{\mu^{*}}{\mu_{\textrm{ref}}^{*}}=\left(\frac{T^{*}}{T_{\textrm{ref}}^{*}}\right)^{\frac{3}{2}}\frac{T_{\textrm{ref}}^{*}+S_{1}^{*}}{T^{*}+S_{1}^{*}},~
\frac{\kappa}{\kappa_\textrm{ref}}=\left(\frac{T^{*}}{T_{\textrm{ref}}^{*}}\right)^{\frac{3}{2}}\frac{T_{\textrm{ref}}^{*}+S_{2}^{*}}{T^{*}+S_{2}^{*}},
\end{equation}
where
\begin{equation}
\left.\begin{array}{c}
{T_{\textrm{ref}}^{*}}=273~\textrm{K},~
\mu^*_{\textrm{ref}}=1.37\times10^{-5}~\textrm{Pa}\cdot\textrm{s},~
\kappa^*_{\textrm{ref}}=0.0146~\textrm{W}/(\textrm{m}\cdot\textrm{K}),~\\
n_1=0.79,~
n_2=1.30,~
S_1^*=222~\textrm{K},~
S_2^*=1800~\textrm{K}.
\end{array}\right\} 
\end{equation}
In general, as temperature increases from subcritical to supercritical values, the fluid continuously transitions from \red{compressed liquid} to \red{compressed vapour} and finally reaches values that can be described by an ideal gas.

\subsection{The linearized stability equations}\label{Sec2-2}
Following the common procedure, the flow field is decomposed into the base flow and a perturbation, as 
\begin{equation}\label{decompose}
\left.\begin{array}{c}
\rho=\rho_{0}+\rho^{\prime}\\
u_{i}=u_{i0}+u_{i}^{\prime}\\
T=T_{0}+T^{\prime}\\
p=p_{0}+p^{\prime}\\
E=E_{0}+E^{\prime}\\
\mu=\mu_{0}+\mu^{\prime}\\
\kappa=\kappa_{0}+\kappa^{\prime}
\end{array}\right\} 
\end{equation}
It is known that for simple compressible systems (\eg pure substances, uniform mixture of nonreacting gases), the thermodynamic state is defined by two independent thermodynamic properties. In this study, we keep $\rho$ and $T$ as the two basic thermodynamic variables, while the other thermodynamic and transport properties (\eg $E$, $p$, $\mu$, $\kappa$) are determined as functions of $\rho$ and $T$. For example, the pressure perturbation $p^{\prime}$ is expanded by a Taylor-series with respect to $\rho_0$ and $T_0$ in the following way
\begin{multline}
p^{\prime}=\left.\frac{\partial p_{0}}{\partial\rho_{0}}\right|_{T_{0}}\rho^{\prime}+\left.\frac{\partial p_{0}}{\partial T_{0}}\right|_{\rho_{0}}T^{\prime}\\
+\frac{1}{2}\left(\left.\frac{\partial^{2}p_{0}}{\partial\rho_{0}^{2}}\right|_{T_{0}}\rho^{\prime}\rho^{\prime}+2\left(\left.\frac{\partial}{\partial\rho_{0}}\right|_{T_{0}}\left.\frac{\partial}{\partial T_{0}}\right|_{\rho_{0}}p_{0}\right)\rho^{\prime}T^{\prime}+\left.\frac{\partial^{2}p_{0}}{\partial T_{0}^{2}}\right|_{\rho_{0}}T^{\prime}T^{\prime}\right)+\cdots
\end{multline}

For the sake of brevity, the partial derivative of a quantity with respect to $T$ at constant $\rho$, will be written as $\left.\partial/\partial T\right|_{\rho_{0}} \equiv \partial/\partial T$, and accordingly  $\left.\partial/\partial \rho\right|_{T_{0}} \equiv \partial/\partial \rho$. The stability equation is derived by substituting \eqref{decompose} into the N-S equations \eqref{NS1}, \eqref{NS2} and \eqref{NS3}, and  then subtracting the governing equations of the base flow. With the nonlinear terms neglected, the linear stability equations are formulated as
\begin{multline}\label{Stability}
\mathscr{L}_{t}\frac{\partial\boldsymbol{q}}{\partial t}+\mathscr{L}_{x}\frac{\partial\boldsymbol{q}}{\partial x}+\mathscr{L}_{y}\frac{\partial\boldsymbol{q}}{\partial y}+\mathscr{L}_{z}\frac{\partial\boldsymbol{q}}{\partial z}+\mathscr{L}_{q}\boldsymbol{q} \\
+\mathscr{V}_{xx}\frac{\partial^{2}\boldsymbol{q}}{\partial x^{2}}+\mathscr{V}_{xy}\frac{\partial^{2}\boldsymbol{q}}{\partial x\partial y}+\mathscr{V}_{xz}\frac{\partial^{2}\boldsymbol{q}}{\partial x\partial z}+\mathscr{V}_{yy}\frac{\partial^{2}\boldsymbol{q}}{\partial y^{2}}+\mathscr{V}_{yz}\frac{\partial^{2}\boldsymbol{q}}{\partial y\partial z}+\mathscr{V}_{zz}\frac{\partial^{2}\boldsymbol{q}}{\partial z^{2}}=0.
\end{multline}
Here $\boldsymbol{q}=\left(\rho^{\prime},u^{\prime},v^{\prime},w^{\prime},T^{\prime}\right)^{T}$ is the perturbation vector and $\mathscr{L}_t$, $\mathscr{L}_x$, $\mathscr{L}_y$, $\mathscr{L}_z$, $\mathscr{L}_q$, $\mathscr{V}_{xx}$, $\mathscr{V}_{yy}$, $\mathscr{V}_{zz}$, $\mathscr{V}_{xy}$, $\mathscr{V}_{yz}$, and $\mathscr{V}_{xz}$ are matrices of size $5\times5$. The detailed expressions for these matrices are provided in Appendix~\ref{appA}. As can be seen, they are functions of the base flow, the thermodynamic and transport properties, $\Rey$ and $\PrEc$. The gradients of the properties are either calculated analytically using cubic EoS (see Appendix~\ref{appB}) or numerically employing finite-difference method within the REFPROP library. 

\subsection{Modal and algebraic stability}\label{Sec2-3}
In modal stability, the perturbation is assumed to have the form
\begin{equation}\label{Wave}
\boldsymbol{q}\left(x,y,z,t\right)=\hat{\boldsymbol{q}}\left(y\right)\exp\left(i\alpha x+i\beta z-i\omega t\right)+c.c.
\end{equation}
where $c.c.$ stands for the complex conjugate. Substituting \eqref{Wave} into \eqref{Stability} results in 
\begin{multline}\label{Stability2}
(-i\omega\mathscr{L}_{t}+i\alpha\mathscr{L}_{x}+\mathscr{L}_{y}D+i\beta\mathscr{L}_{z}+\mathscr{L}_{q}\\
-\alpha^{2}\mathscr{V}_{xx}+i\alpha\mathscr{V}_{xy}D-\alpha\beta\mathscr{V}_{xz}+\mathscr{V}_{yy}D^{2}+i\beta \mathscr{V}_{yz}D-\beta^{2}\mathscr{V}_{zz})\hat{\boldsymbol{q}}=0,
\end{multline}
where $D={\rm d}/{\rm d}y$. The equation \eqref{Stability2} is solved as an eigenvalue problem, which describes the development of the perturbations in temporal or spatial domain, \ie
\begin{equation}\label{Eigen}
\begin{cases}
L_{T}\hat{\boldsymbol{q}}=\omega\mathscr{L}_{t}\hat{\boldsymbol{q}} & \mathrm{(temporal)},\\
L_{S}\hat{\boldsymbol{q}}=\alpha\left(\beta\mathscr{V}_{xz}-i\mathscr{V}_{xy}D-i\mathscr{L}_{x}\right)\hat{\boldsymbol{q}}+\alpha^{2}\mathscr{V}_{xx}\hat{\boldsymbol{q}} & (\mathrm{spatial}),
\end{cases}
\end{equation}
where
\begin{eqnarray}
L_{T}&=&\alpha\mathscr{L}_{x}-i\mathscr{L}_{y}D+\beta\mathscr{L}_{z}-i\mathscr{L}_{q} \nonumber\\ 
&+&i\alpha^{2}\mathscr{V}_{xx}+\alpha\mathscr{V}_{xy}D+i\alpha\beta\mathscr{V}_{xz}-iD^{2}\mathscr{V}_{yy}+\beta D\mathscr{V}_{yz}+i\beta^{2}\mathscr{V}_{zz}, \\
L_{S}&=&-i\omega\mathscr{L}_{t}+\mathscr{L}_{y}D+i\beta\mathscr{L}_{z}+\mathscr{L}_{q}+D^{2}\mathscr{V}_{yy}+i\beta D\mathscr{V}_{yz}-\beta^{2}\mathscr{V}_{zz}. 
\end{eqnarray}
Here we consider the temporal problem only, therefore $\alpha$ and $\beta$ are the prescribed streamwise and spanwise wave numbers. $\omega=\omega_r+i\omega_i$ is solved as the eigenvalue, where $\omega_r$ and $\omega_i$ give the angular frequency and growth rate of the perturbation. \red{The domain $0\leq y \leq 2$ is discretized with Chebyshev collocation points, defined by 
\begin{equation}
y_j=1-\cos\dfrac{\pi j}{N}~~~j=0,1,...,N-1,N.
\end{equation}
The differentiation of \eqref{Eigen} is accomplished with the matrix form of Chebyshev collocation derivatives. Numerical tests indicate that, typically $N=200$ (used here) are sufficient to give a grid-independent solution of the physical modes.}

With regard to the algebraic stability, following \citet{Schmid2001}, the optimal energy amplification is defined as:
\begin{equation}
G\left(t\right)=\underset{\boldsymbol{q}_{0}}{\max}\frac{E\left(\boldsymbol{q}\left(t\right)\right)}{E\left(\boldsymbol{q}_{0}\right)},~
G\left(x\right)=\underset{\boldsymbol{q}_{0}}{\max}\frac{E\left(\boldsymbol{q}\left(x\right)\right)}{E\left(\boldsymbol{q}_{0}\right)},~
\end{equation}
Here $E\left(\boldsymbol{q}\right)$ is the disturbance energy with the definition as given in \eqref{Norm}. \red{The perturbation $\boldsymbol{q}$ is expanded by the eigenvector obtained from the modal stability. The calculation of the optimal energy amplification $G$ and the corresponding optimal perturbation (the input), as well as the resulting perturbation (the output), lead to a singular value problem, which is solved with the same Chebyshev differentiation method as in the modal growth \citep{Schmid2001,Schmid2014}. }

The (modal and algebraic) perturbations are solved subjected to the boundary condition: $u^{\prime}=v^{\prime}=w^{\prime}=T^{\prime}=0$ at the lower ($y=0$) and upper wall ($y=2$).

\section{The laminar base flow}\label{Sec3}
The base flow \red{is driven by a constant body force in the streamwise direction} and is obtained by solving \eqref{NS1}, \eqref{NS2} and \eqref{NS3} with the assumption that the flow is fully developed, spanwise and streamwise independent, steady and parallel, \ie $
\partial\left(\right)/\partial x=0,~
\partial\left(\right)/\partial z=0,~
\partial\left(\right)/\partial t=0,~
v=w=0$. 
The N-S equations are thus simplified as 
\begin{equation}\label{EqBaseflow1}
\frac{\pp}{\pp y}\left(\mu\frac{\pp u}{\pp y}\right)=-\Rey F=-\hat{F},
\end{equation}
\begin{equation}\label{EqBaseflow2}
\frac{\pp p}{\pp y}=0,
\end{equation}
\begin{equation}\label{EqBaseflow3}
\frac{\pp}{\pp y}\left(\frac{\kappa}{\PrEc}\frac{\pp T}{\pp y}+\mu u\frac{\pp u}{\pp y}\right)=-\Rey F\cdot u=-\hat{F}\cdot u.
\end{equation}
It is worth noting that the above equations are independent of density, therefore, $\rho_0$ can be separately determined by the EoS. We assume the body force, $\hat{F}$, which drives the flow, to be uniform. \red{To obtain a solution of the base flow, an initial temperature field is assumed, \eg $T=T_w=$ constant, $\mu$ and $\kappa$ are determined from REFPROP according to the temperature and pressure. First, the velocity is solved using equation \eqref{EqBaseflow1}, followed by an update of temperature by solving \eqref{EqBaseflow3}. $\mu$ and $\kappa$ are then updated using the obtained temperature. This procedure is repeated until the solution is converged.}

\subsection{The isothermal limit}\label{Sec3-1}
\begin{figure}
\begin{center}
\vspace{20pt}
\includegraphics[width=0.65\linewidth]{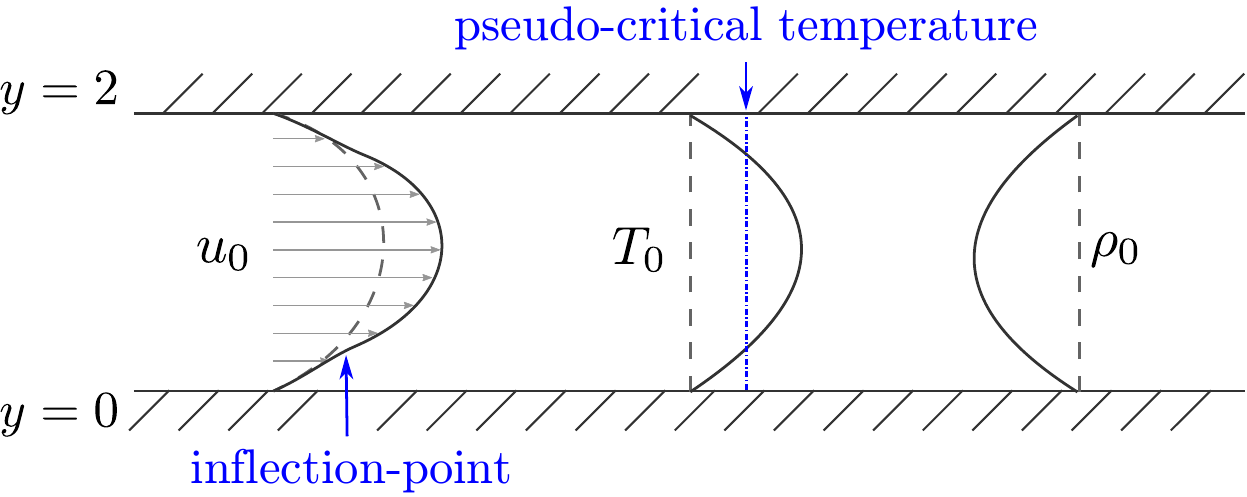}
\end{center}
\caption{Sketch of the laminar base flow. Dashed lines show the isothermal limit with $\PrEc\rightarrow0$, such that $u_0=y(2-y)$, $T_0=1$, $\rho_0=1$. Solid lines represent a transcritical case, in which $T_0$ crosses $T_{pc}$ and $u_0$ is inflectional. }
\label{Fig3}
\end{figure}

When $\PrEc\rightarrow0$, the viscous heating is negligible if compared to the thermal conduction. Therefore, the temperature, as well as the other thermodynamic properties, remain constant, namely $T_0=1,~\rho_0=1,~\mu_0=1,~\kappa_0=1$. 
The flow is thus simply governed by ${\pp^{2}u}/{\pp y^{2}}=-\hat{F}$. Choosing $u^*_r$ as the centerline velocity, leads to setting $\hat{F}=2$. As a result, the dimensionless base flow is independent of any parameters (\eg $T_w$, $\hat{F}$ and $PrEc$) and is given by $u_0=y(2-y)$. A sketch of this base flow, which is free from any non-ideal gas effects, is shown in figure~\ref{Fig3} (dashed lines). 

\subsection{The compressible base flow}\label{Sec3-2}
Equations \eqref{EqBaseflow1}, \eqref{EqBaseflow2} and \eqref{EqBaseflow3} show that the compressible base flow is determined by $\PrEc$, $\hat{F}$, and $T^*_w$. Either by increasing $\PrEc$ or $\hat{F}$, the compressibility effects become more significant. Without loss of the generality, a constant body force $\hat{F}=2$ is specified in this work, while $\PrEc$ is varied from the isothermal limit (we assume $\PrEc=10^{-5}$) to a typical compressible state with $\PrEc=0.1$. For example, setting $\PrEc=0.1$ and $T_w^*=290,~300,$ or $310$ K, the Mach number is $\Ma=0.40,~0.58,$ or $1.35$, respectively. In this work, the wall temperature $T^*_w$ is considered in a range from 265 to 320 K. Note, given our non-dimensionalization, the base flow is free from the choice of the Reynolds number. 

\begin{figure}
\begin{center}
\includegraphics[width=0.48\linewidth,clip]{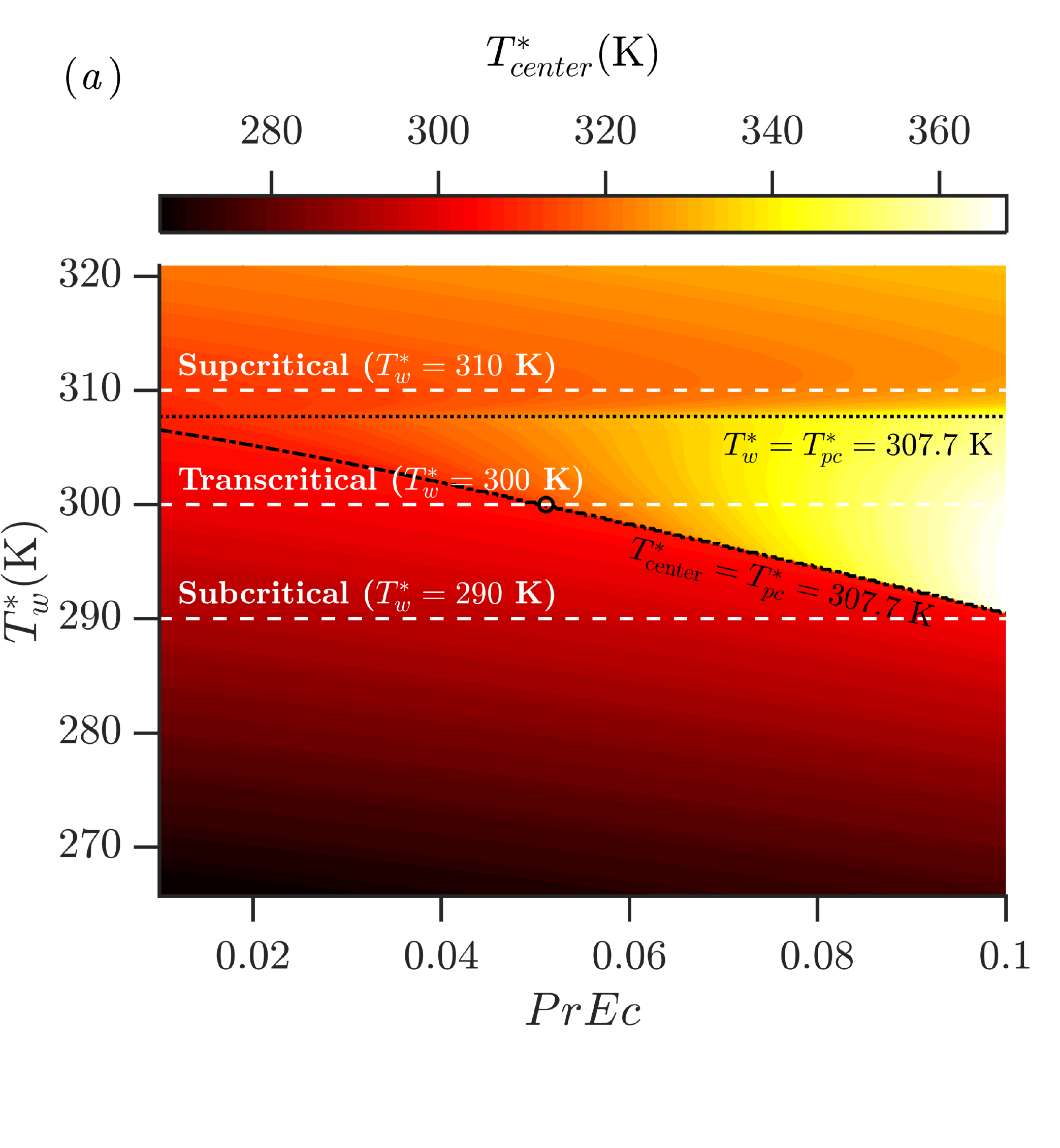}
\includegraphics[width=0.48\linewidth,clip]{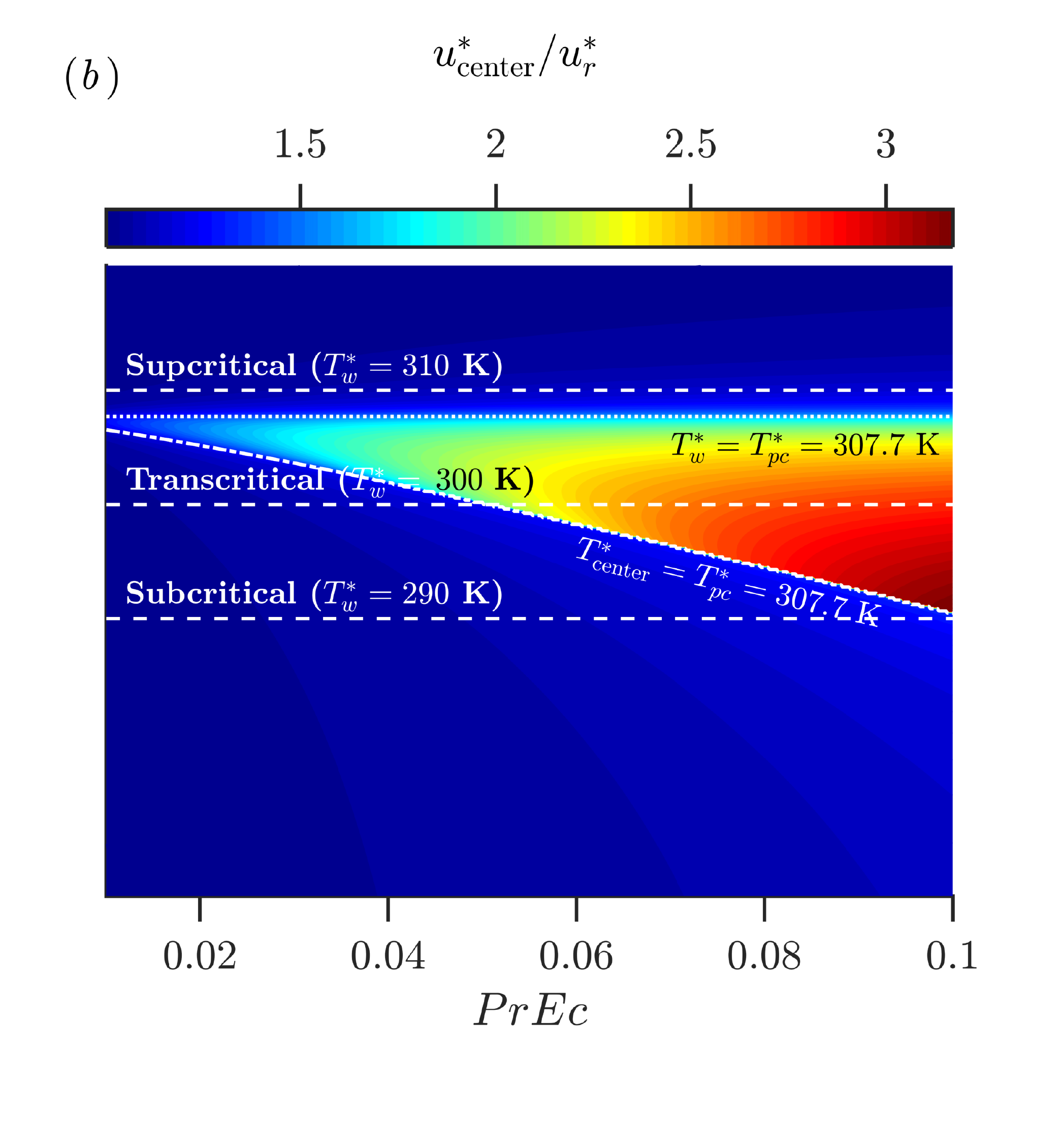}
\end{center}
\caption{Centerline (a) temperature $T^*_{\rm center}$ and (b) velocity $u_{\rm center}$ as functions of wall temperature $T_w^*$ and $\PrEc$. Model RP, $p^*=80$ bar, $\hat{F}=2$.}
\label{Fig4}
\end{figure}

Figure~\ref{Fig4} shows the contours of the centerline temperature $T_{\rm center}^*$ and velocity $u_{\rm center}=u_{\rm center}^*/u_r^*$ as a function of wall temperature $T_w^*$ and $\PrEc$. 
Regardless of the wall temperature, an increase of $\PrEc$ is accompanied with an increase of $T_{\rm center}^*$ and $u_{\rm center}$  as compressible effects become more prominent. Interestingly, a distinct right-angled triangular area emerges in each subplot of figure~\ref{Fig4}. At the hypotenuse of this triangle, the centerline temperature, $T_{\rm center}^*$, and velocity, $u_{\rm center}$, suddenly increase, forming a discontinuity in the $PrEc$--$T_w^*$ plane. 

It is also interesting to note that the hypotenuse of the triangle almost coincides with the line where $T^*_{\rm center}$ reaches the pseudo critical temperature $T_{pc}^*=307.7$ K (shown with the dot-dashed line). Likewise, the upper boundary of the triangle coincides with the dotted line where $T^*_w=T_{pc}^*$. 

\begin{figure}
\begin{center}
\includegraphics[width=0.8\linewidth,clip]{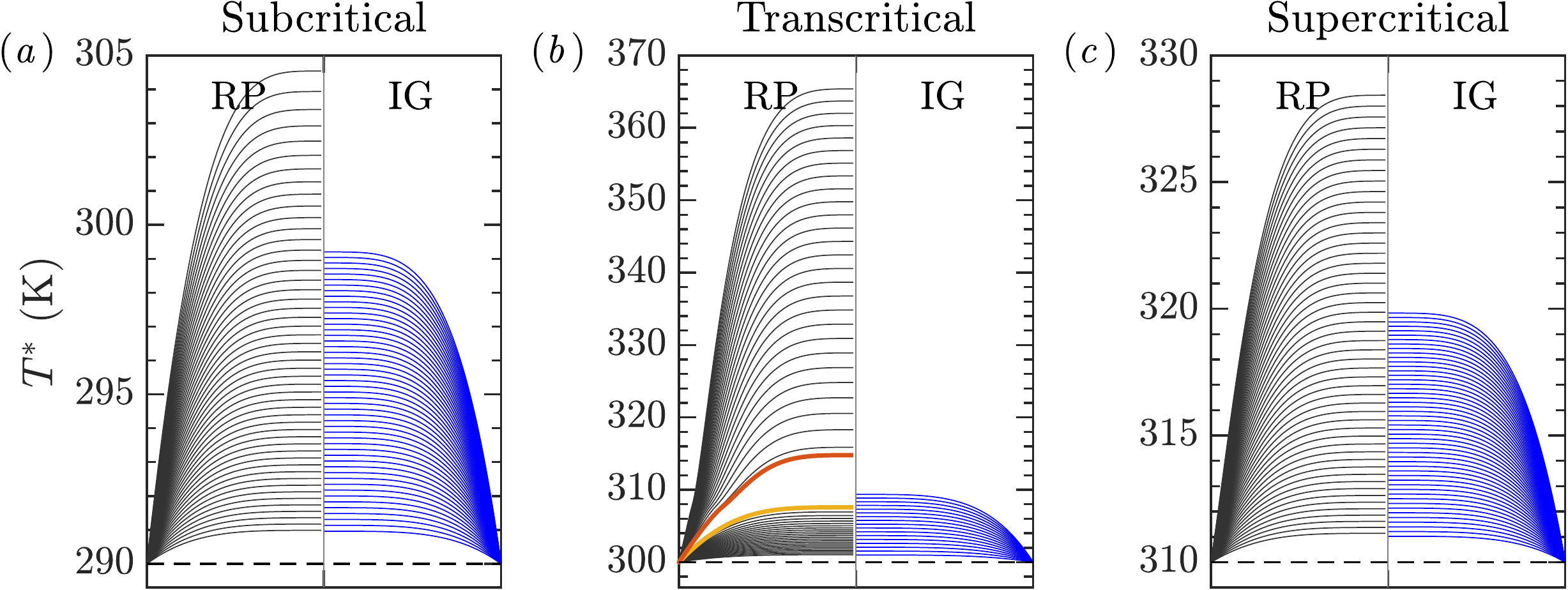}\\
\includegraphics[width=0.8\linewidth,clip]{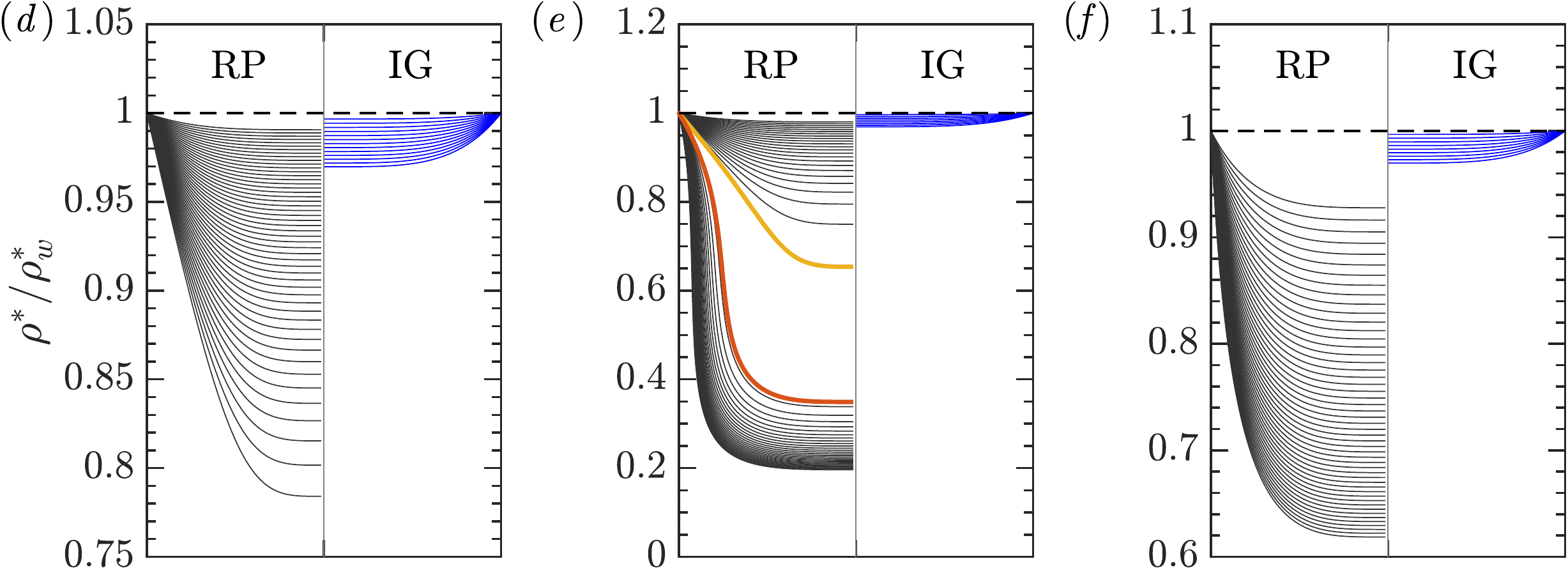}\\
\includegraphics[width=0.8\linewidth,clip]{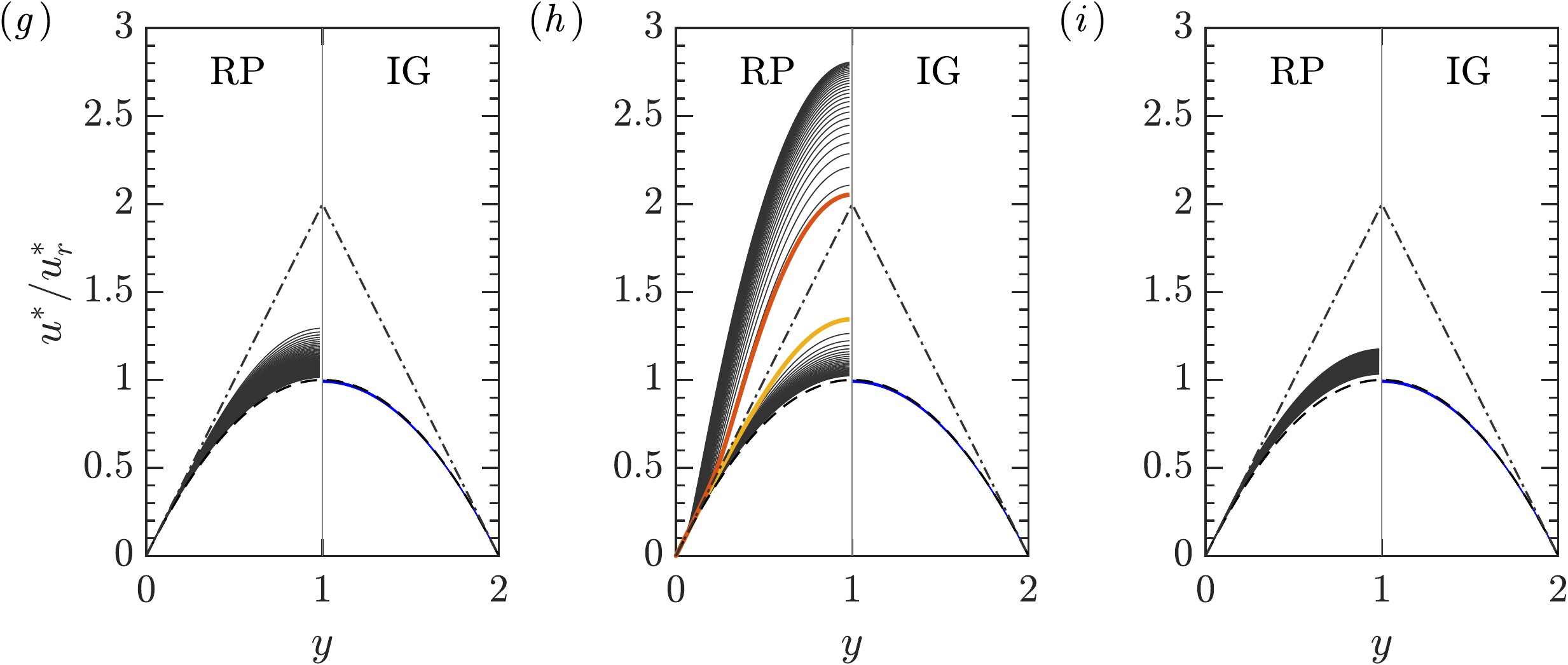}
\end{center}
\caption{Temperature (a-c), density (d-f) and velocity (g-i) profiles of the base flow. The wall temperature is (a,d,g) $T^*_w=290$ K,
(b,e,h) $T^*_w=300$ K,
(c,f,i) $T^*_w=310$ K respectively. $\PrEc$ increases uniformly from 0.01 to 0.1. The black and blue lines on the left and right half denote the non-ideal (RP) and ideal (IG) gases respectively. The dashed lines in each subplot shows the isothermal limit. The REFPROP library is used for the transport and thermodynamic properties of the non-ideal gas. The orange and red lines in subplot (b,e,h) show the profiles at $\PrEc=$ 0.05115 and 0.05116 respectively. The dash-dotted lines (the triangle) in subplot (g,h,i) show the lines of constant gradient $|{\partial u}/{\partial y}|=\hat{F}$.} 
\label{Fig5}
\end{figure}

For a more detailed discussion, we will now define three cases with different wall temperatures that are summarized in table~\ref{Table2} and highlighted by dashed lines in figure~\ref{Fig4}. These cases will also be used in the subsequent sections regarding the linear modal and algebraic instability analysis. The wall temperature for these cases has been set to 290, 300 and 310 K, such that their temperature profile in the considered range of $PrEc$ is either subcritical, transcritical or supercritical, respectively. 
%Three cases are defined, subcritical, transcritical and supercritical, in which the wall temperature is set to 290, 300 and 310 K, respectively. 
Their base flow profiles are plotted in figure~\ref{Fig5}, together with the incompressible limit, indicated by the dashed line in each subplot. The profiles on the left half (black lines) and right half (blue lines) represent the base flow of the non-ideal (RP) and ideal (IG) gases, respectively. As $\PrEc$ uniformly increases from 0.01 to 0.1 it can be seen that the temperature and velocity increase, while the density decreases. For the transcritical case, however, a sudden jump of the base flow profiles can be observed. This jump occurs between $\PrEc=$ 0.05115 and 0.05116, as highlighted by the orange and red lines in figure~\ref{Fig5}(b,e,h). Note, the jump is caused by an inflectional velocity profile as highlighted by the red line in figure~\ref{Fig5}(h). 
\begin{table}
\begin{center}\def~{\hphantom{0}}
\begin{tabular}{rcccc}
Case 	 	& $T_w^*$ & $\PrEc$  & $\Ma$		& Temperature range \vspace{5pt} \\
Subcritical   & 290 K &$\PrEc\leq0.1$ & $\Ma\leq0.40$	& 290 K (wall) - 304.9 K (center)\\
Transcritical & 300 K &$\PrEc\leq0.1$ & $\Ma\leq0.58$	& 300 K (wall) - 366.2 K (center)\\
Supercritical & 310 K &$\PrEc\leq0.1$ & $\Ma\leq1.35$	& 310 K (wall) - 328.6 K (center)\\
\end{tabular}
\caption{Cases investigated in this study.}
\label{Table2}
\end{center}
\end{table}

The discontinuous behaviour with respect to $PrEc$ can be explained as follows. Integrating \eqref{EqBaseflow1} gives $\mu{\partial u}/{\partial y}=-\hat{F}y+C.$ Applying the symmetry condition at the channel center ($y=1$), it follows that $C=\hat{F}$. Therefore, \eqref{EqBaseflow1} can be written as 
\begin{equation}\label{EqBaseflow5}
\mu\frac{\partial^{2}u}{\partial y^{2}}=-\hat{F}-\frac{\partial\mu}{\partial y}\frac{\partial u}{\partial y}=
-\hat{F}\left(1+\frac{1}{\mu}\frac{\partial\mu}{\partial y}(1-y)\right). 
\end{equation}
Based on \eqref{EqBaseflow5}, it can be seen that an inflectional velocity profile occurs if the viscosity gradient is large enough to change the sign within the parenthesis in (\ref{EqBaseflow5}), namely if 
\begin{equation}\label{EqBaseflow6}
\frac{1}{\mu}\left|\frac{\partial\mu}{\partial y}\right|>1. 
\end{equation}
In the cases considered herein, it appears that the viscosity gradient at the wall is large enough to cause an inflectional profile to occur when the temperature in the channel center reaches $T_{pc}^*$. Recall figure \ref{Fig2}(c), a sharp gradient of the viscosity (${\partial\mu}/{\partial T}\ll0$) is seen close to the pseudo-critical point. As $\PrEc$ increases, ${\partial T}/{\partial y}$ increases at the wall, such that ${\partial\mu}/{\partial y}\cong({\partial\mu}/{\partial T})({\partial T}/{\partial y})$ can drop below -1 at the wall, leading to inflectional velocity profiles. The jump of the base flow solution can thus be explained by referring to figure \ref{Fig5}(h). Since, $\mu |{\partial u}/{\partial y}|$ at the wall is equal to the constant forcing $\hat{F}$, regardless of $\PrEc$ and wall temperature, the velocity profiles with/without inflectional points are isolated by the line of constant gradient $\hat{F}$ (the dash-dotted lines that form a triangle in figure \ref{Fig5}(g-i)). Therefore, a velocity profile without an inflection point cannot reach the apex of the triangle ($|{\partial u}/{\partial y}|$ decreases towards channel center) and the sudden increase of the centerline velocity appears once an inflection point is formed. 

In general, the base flow solutions can be summarized as follows:
\begin{itemize}
\item In the {\bf subcritical} case, the wall temperature is much lower than $T_{pc}^*$, and in the range of $\PrEc$ considered, $T_{\rm center}^*$ is always less than $T_{pc}^*$. Hence, the velocity profile is not inflectional.
\item In the {\bf transcritical} case, the wall temperature is close to $T_{pc}^*$, such that for large enough $\PrEc$, $T_{\rm center}^*$ reaches $T_{pc}^*$. Consequently, a jump of the solution with respect to $\PrEc$ occurs and the velocity profile becomes inflectional. From figure~\ref{Fig4}(b), it can be inferred that the lower the wall temperature $T_w^*$, the larger the discontinuity will be.
\item In the {\bf supercritical} case, the wall temperature is higher than $T_{pc}^*$. The properties of the fluid are gas-like \red{(compressed vapour)} and the velocity is not inflectional. 
\end{itemize}

\section{Linear modal instability}\label{Sec4}

Depending on the cases discussed below, we will use the definition of dynamic and thermodynamic modes, as 
\begin{equation}
\begin{cases}
\rho^{\prime}=0,\:\mathrm{and}\:T^{\prime}=0 & \textrm{(dynamic modes)} \\ 
\rho^{\prime}\neq0,\:\mathrm{or}\:T^{\prime}\neq0 & \textrm{(thermodynamic modes). } 
\end{cases}
\end{equation}

\subsection{The isothermal limit}\label{Sec4-1} 
\begin{figure}
\begin{center}
\includegraphics[width=0.465\linewidth,clip]{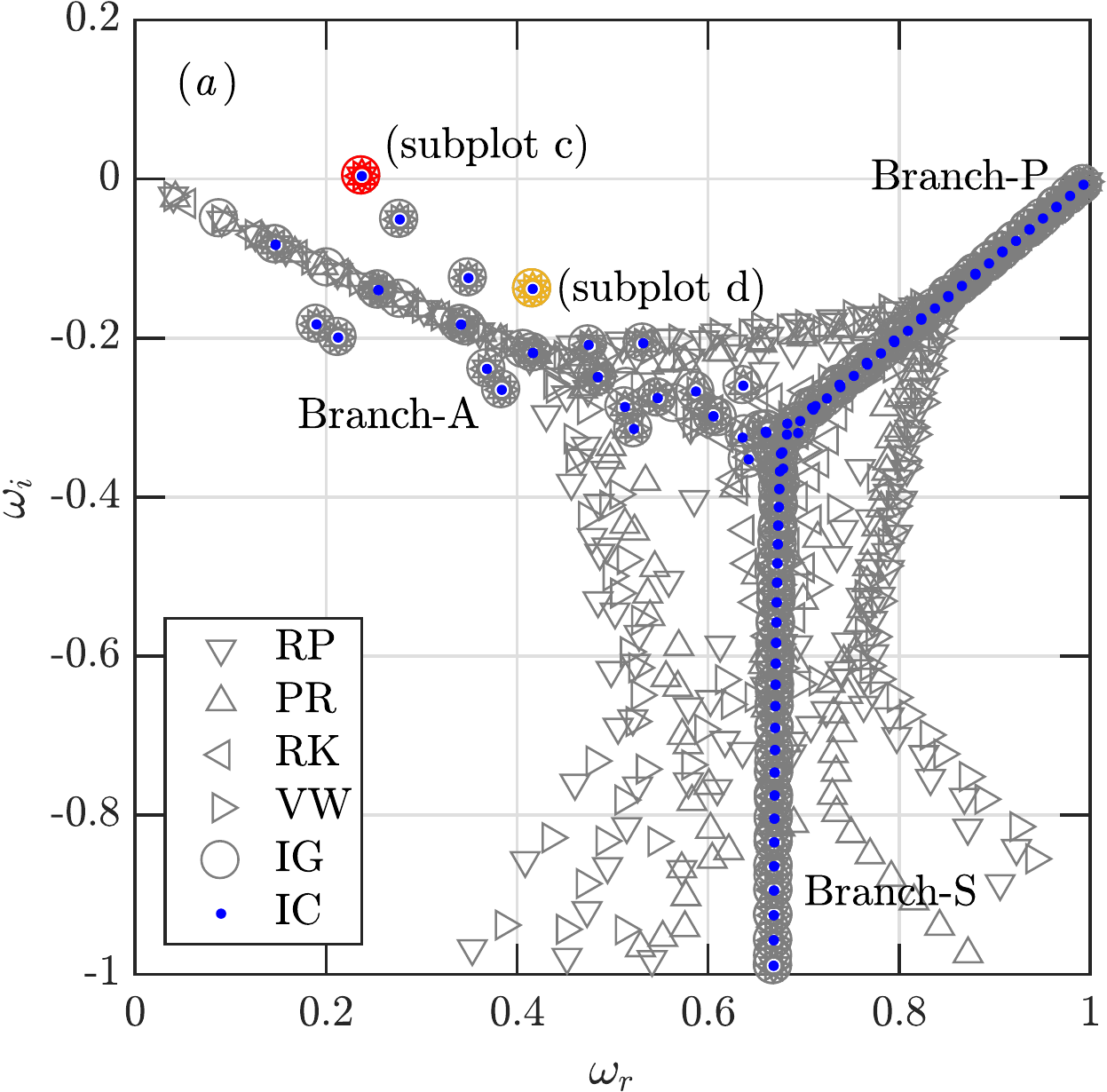} 
\includegraphics[width=0.48\linewidth,clip]{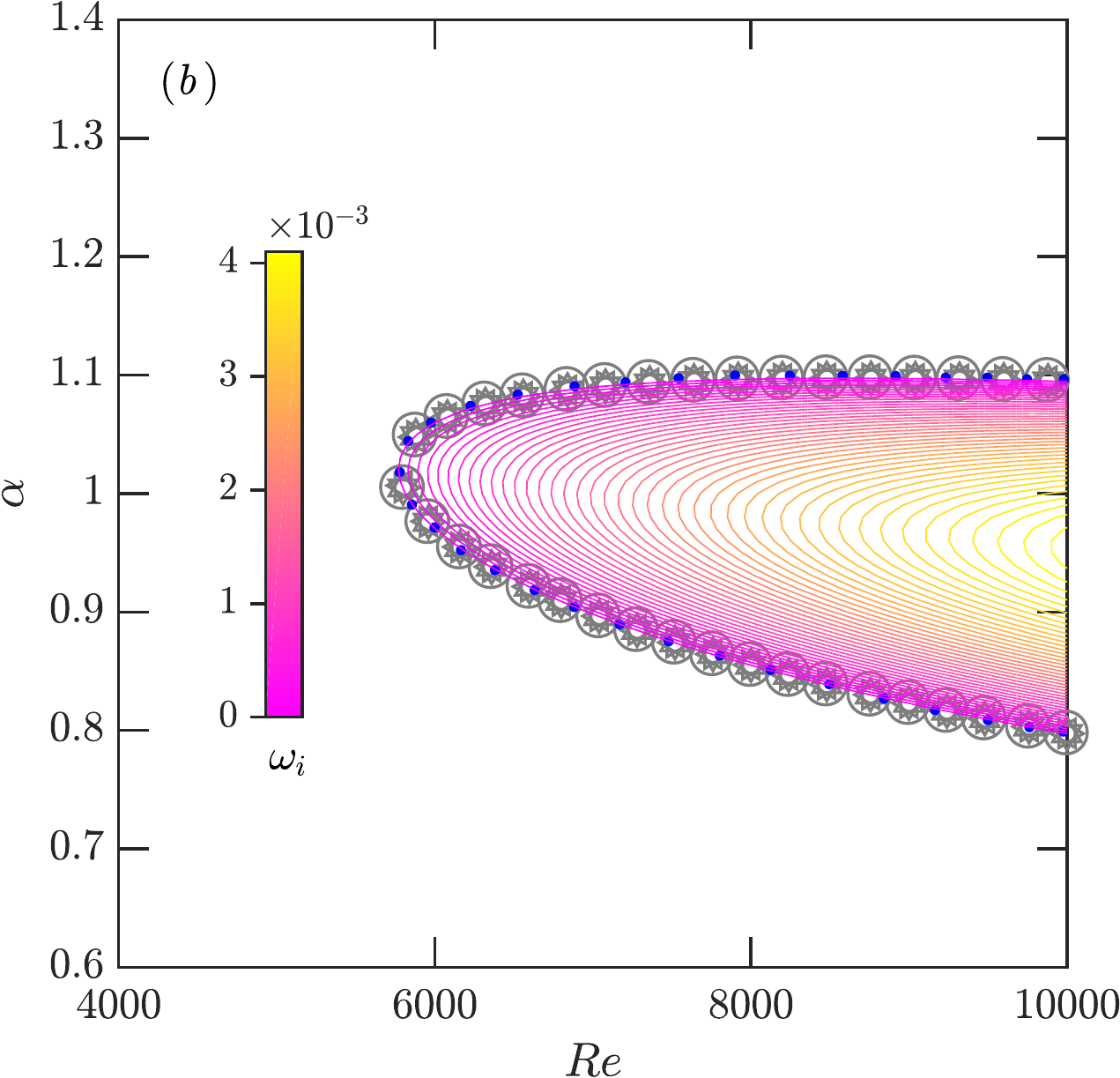} \\ 
\includegraphics[width=0.45\linewidth,clip]{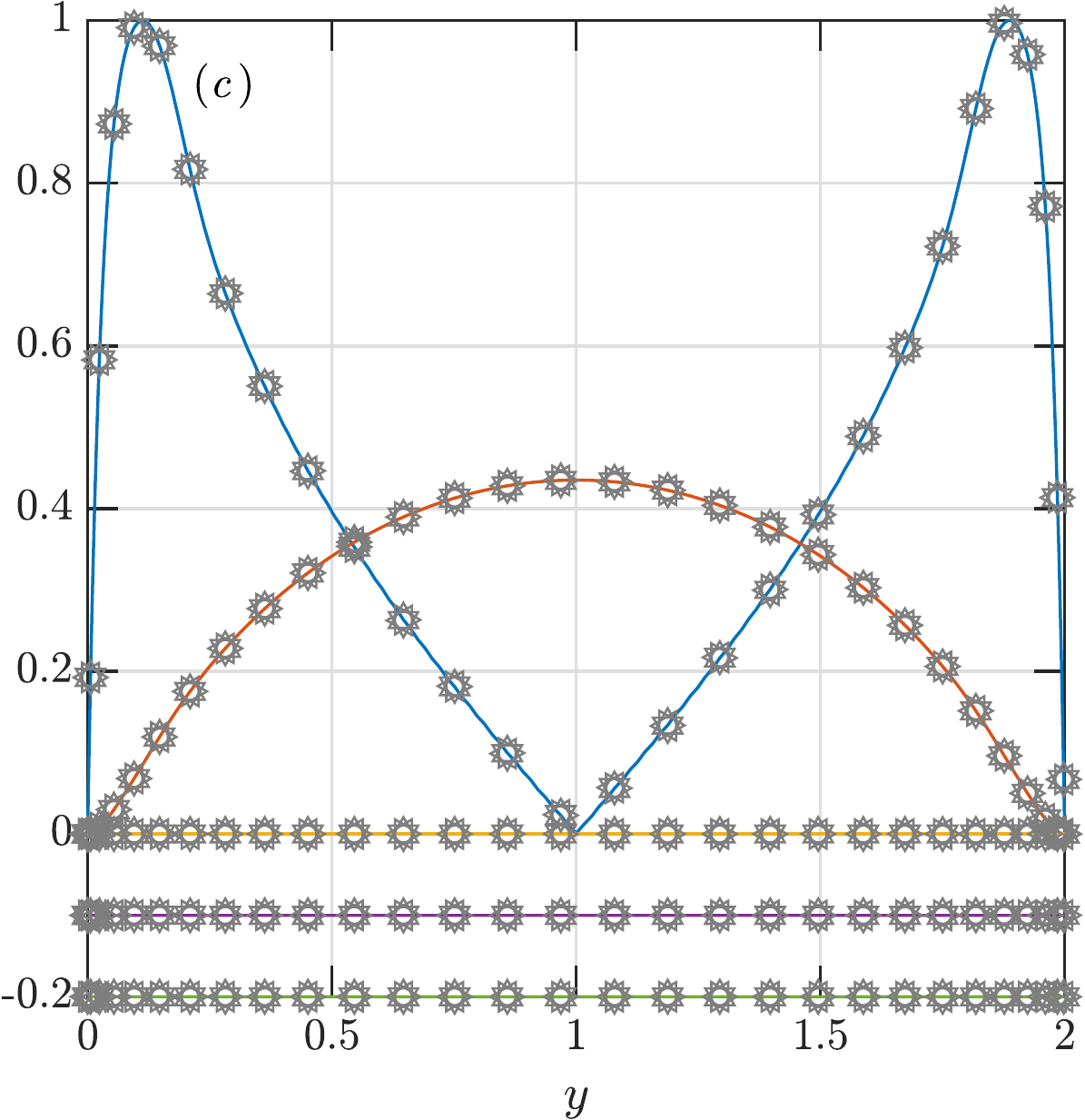} \hspace{4pt} 
\includegraphics[width=0.45\linewidth,clip]{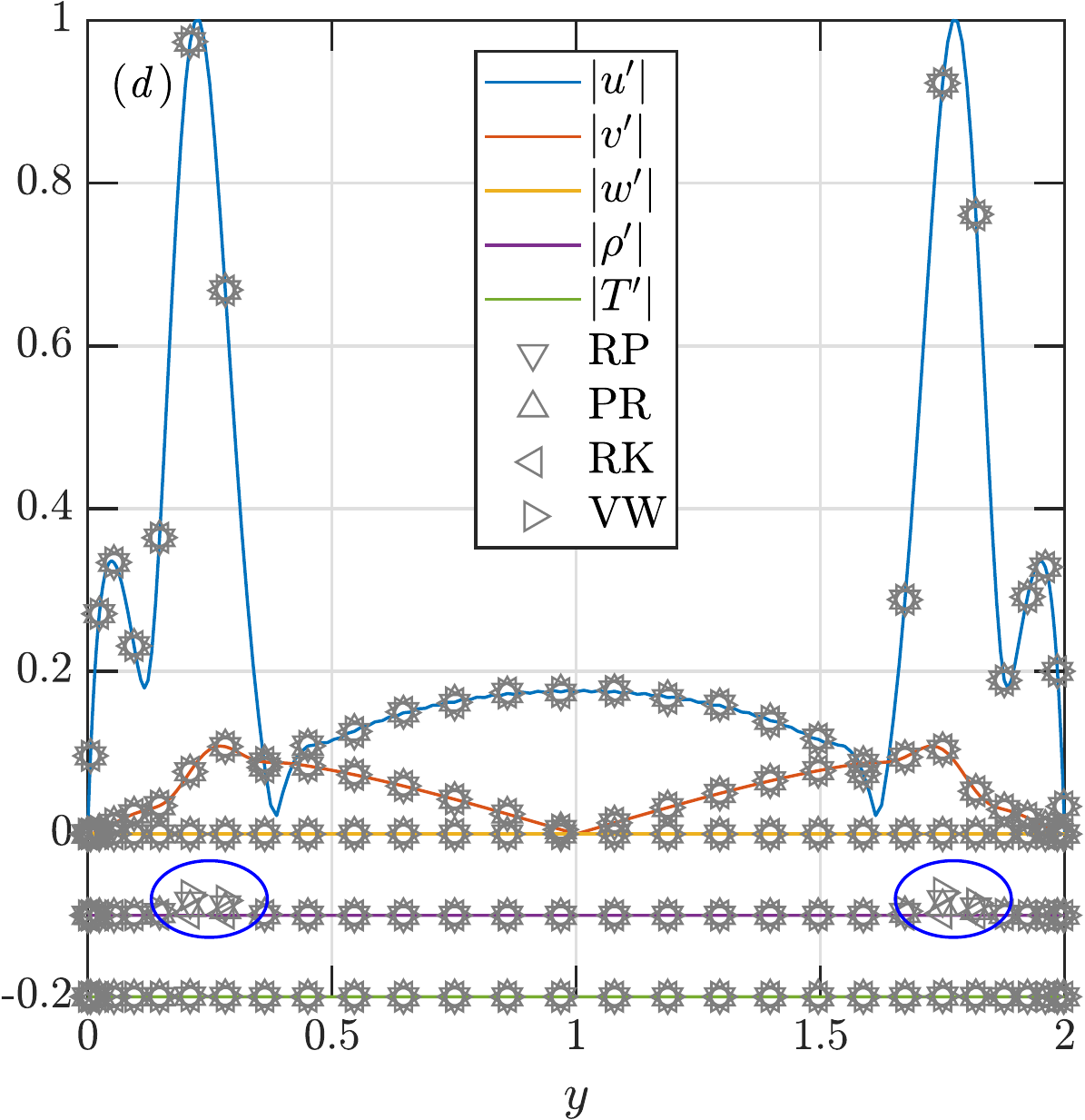} 
\end{center}
\caption{Eigen spectrum (a) and neutral curve (b) for the isothermal limit. The eigen spectrum is subject to $\alpha=1$, $\beta=0$, and $\Rey=10000$. The neutral curve is solved for 2-D perturbations ($\beta=0$). Symbols show results using different fluid models (RP, PR, RK, VW, IG) and incompressible equations (IC). IC in subplot (b) shows the results given by \citet[][pp. 71]{Schmid2001}. In subplot (c) and (d), profiles of the unstable mode ($\omega=0.2375+0.0037i$) and one of the stable modes ($\omega=0.4164-0.1382i$, highlighted in orange in the spectrum) are shown. The perturbations are normalized by $|u^\prime|$. An offset of -0.1 and -0.2 is applied to $|\rho^\prime|$ and $|T^\prime|$. The solid lines are results with fluid model IG.}
\label{Fig6}
\end{figure}

With the base flow obtained in section~\ref{Sec3-1}, we solve the stability equations \eqref{Stability} for the isothermal limit with different fluid models (RP, PR, RK, VW, IG), as well as for the incompressible equations (IC). As shown in figure~\ref{Fig6}(a), at $\Rey=10000$, $\alpha=1$ and $\beta=0$, the A-, P- and S- branches \citep[originally named by][]{Mack_1976} are reproduced by incompressible equations. Comparing the results using different equations of state, the eigenvalues fall on top of the incompressible counterparts, verifying the correct behaviour of the compressible models at low Eckert (Mach) numbers. One of the modes (highlighted in red) is exclusively unstable. Despite being solved with different thermodynamic models, this mode is shown to be a dynamic mode, which leads to identical neutral curve and eigenfunctions as shown in figure~\ref{Fig6}(b,c). The contour lines in figure \ref{Fig6}(b) show the growth rate $\omega_i$ (RP model).  In fact, inspecting the stability equations \eqref{Stability} (see Appendix~\ref{appA}), it can be shown that the thermodynamic and transport properties do not influence the dynamic modes in the isothermal limit. For instance, gradients of properties, which vary among different models, are multiplied with thermodynamic components of the perturbations. 

On the other hand, more stable modes emerge when the compressible equations are solved. By looking into the corresponding eigenfunctions (not shown), thermodynamic components become important in these modes, and as such dependent on the non-ideal gas properties. We plot one of the stable modes in figure \ref{Fig6}(d), where density perturbations are captured by compressible equations (indicated by blue ellipses).

\subsection{Compressible flows}\label{Sec4-2}
\begin{figure}
\begin{center}
\includegraphics[width=0.85\linewidth]{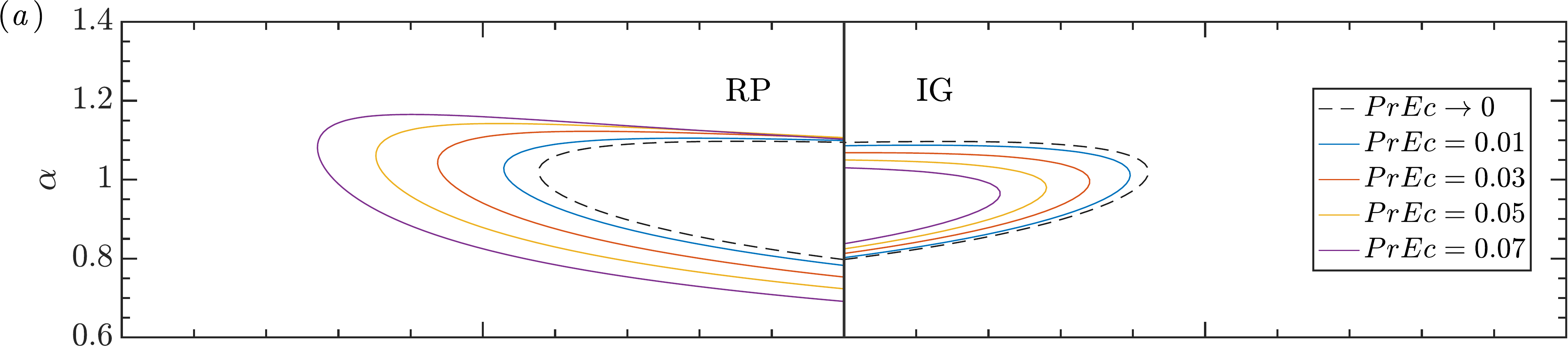}\\
\includegraphics[width=0.85\linewidth]{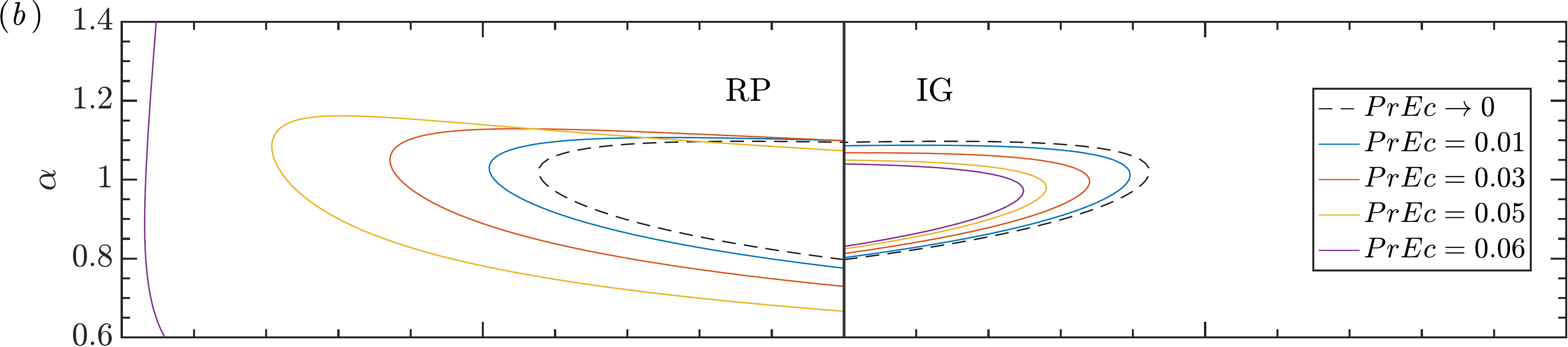}\\
\includegraphics[width=0.85\linewidth]{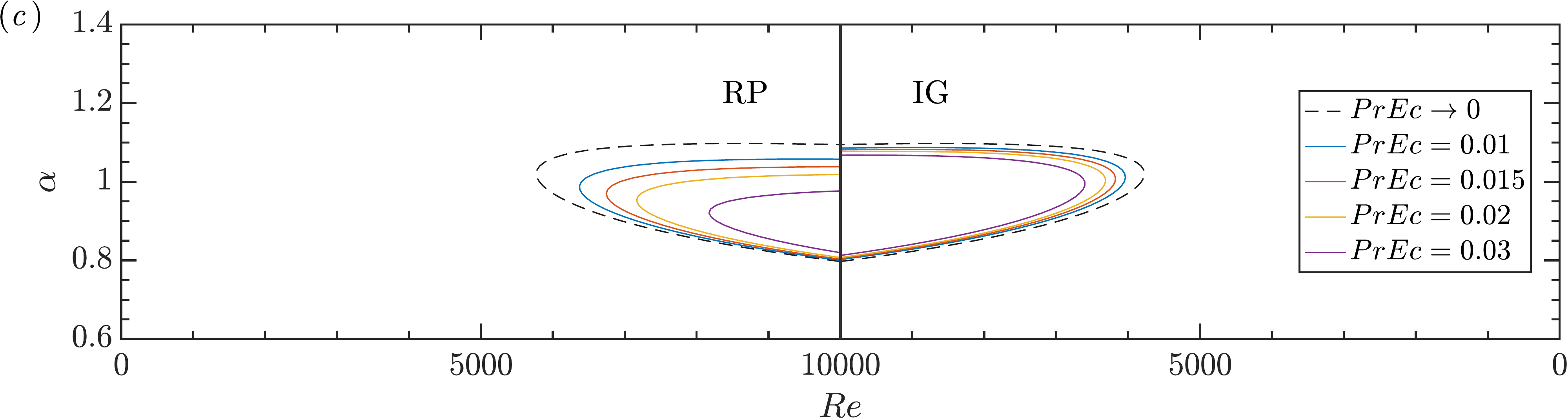}\\
\vspace{0.2cm}
\includegraphics[width=0.85\linewidth]{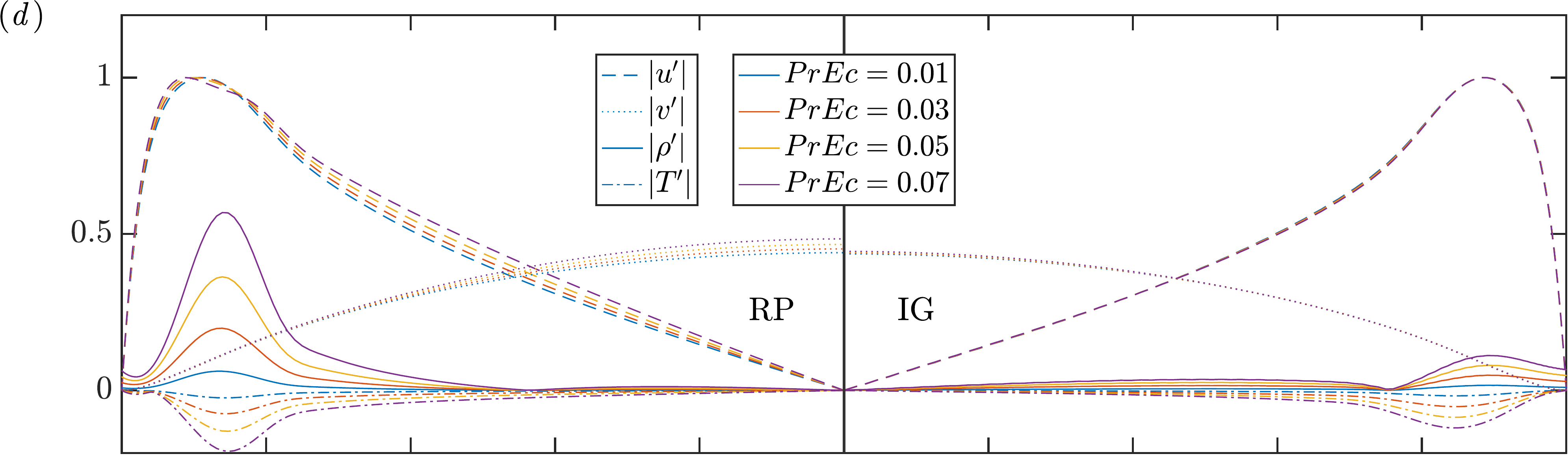}\\
\includegraphics[width=0.85\linewidth]{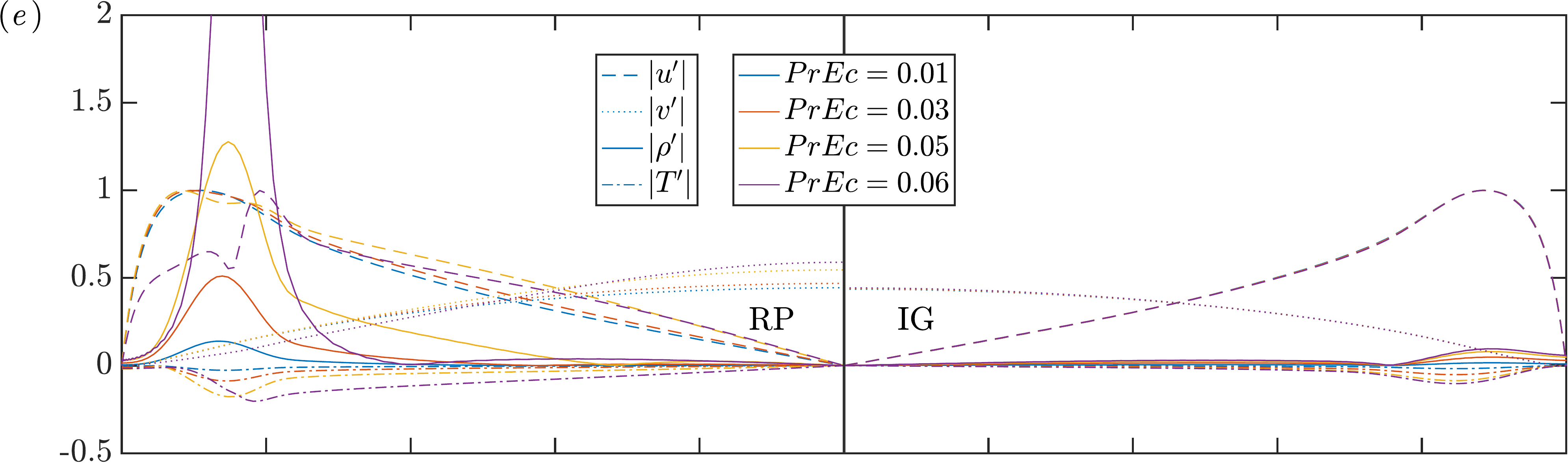}\\
\includegraphics[width=0.85\linewidth]{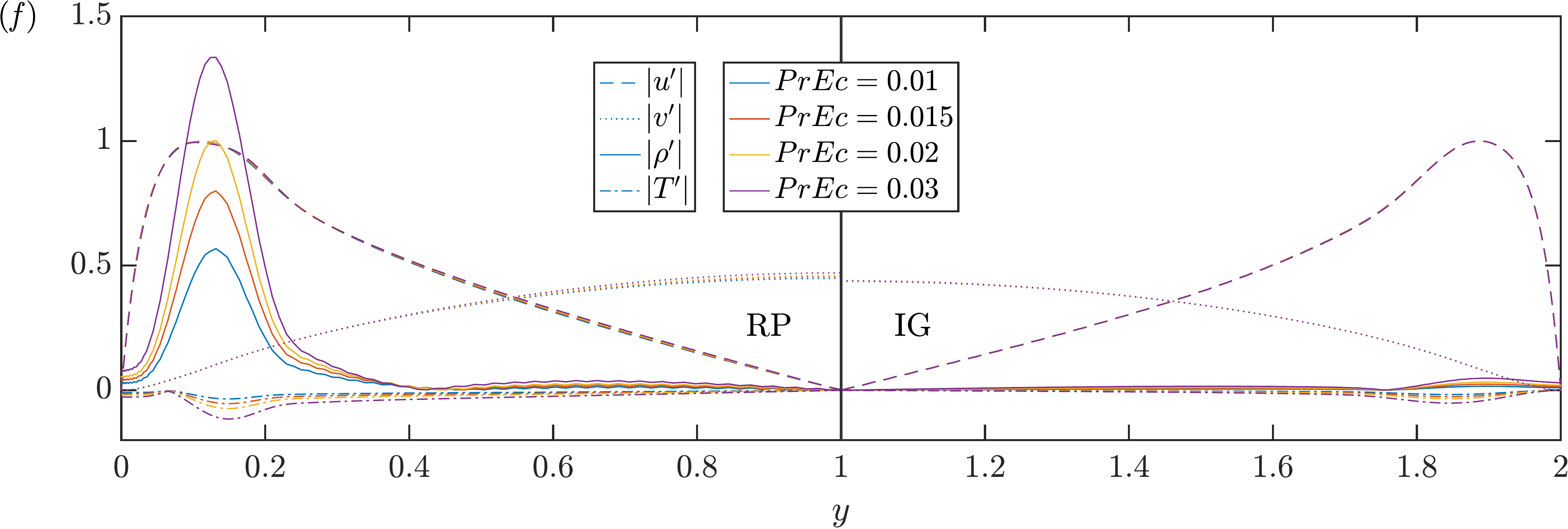}
\end{center}
\caption{Neutral curves and profiles of perturbations for the non-ideal gas (RP model) and ideal gas (IG model).  (a,d) $T_w^*=290$ K, (b,e) $T_w^*=300$ K, (c,f) $T_w^*=310$ K. The neutral curves are obtained for 2-D perturbations ($\beta=0$). The profiles shown are subject to $\alpha=1$, $\beta=0$ and $\Rey=10000$, and they are normalized with $|u^\prime|_{\max}$. The left and right half shows the non-ideal and ideal gas respectively.}
\label{Fig7}
\end{figure}

To achieve a first impression of the non-ideal gas effects, the problem is first studied with the RP model, where thermodynamic and transport properties are taken from the REFPROP library. Figure~\ref{Fig7} shows the neutral curves (a-c) as well as eigenfunctions (d-f) at representative parameters. As discussed in \S~\ref{Sec3-2}, the temperature is subcritical, transcritical and supercritical with $T^*_w=$ 290K, 300K and 310K respectively. The results are compared with ideal gas (IG).

By increasing $\PrEc$, the base flow of the ideal gas becomes more stable as the critical Reynolds number increases, regardless of $T_w$ specified. In fact, despite the difference in wall temperature, the dimensionless thermodynamic and transport properties (scaled with wall values) remain much the same. On the other hand, the behaviour of the non-ideal cases is different for the three cases investigated. In the subcritical case, the flow becomes more unstable when $\PrEc$ is increased. This is manifested by the enlargement of the neutral curve. Similarly, the transcritical case becomes more unstable as $\PrEc$ increases. However, once $\PrEc$ reaches the critical value (in this case $\PrEc=0.05115$), the base flow becomes inflectional. The flow is thus inviscid unstable and the critical Reynolds number is substantially reduced. For instance, the flow is unstable for $\Rey<1000$ and $\PrEc=0.06$. %However, by performing a DNS, it is possible to validate this result. For example, an inflectional velocity downstream of a laminar separation bubble has been studied using DNS by \citet{Marxen2010,Marxen2013}, where the inviscid Kelvin-Helmholtz mechanism plays an essential role. 
In the supercritical case, the increase of $\PrEc$ stabilizes the base flow and the non-ideal gas is even more stable than the ideal gas. In this case, when $\PrEc$ reaches 0.03, the modal instability is found after $\Rey>8000$. Interestingly, a weak influence of $\PrEc$ on the velocity perturbations is observed (see figure~\ref{Fig7} (d-f)), while the amplitudes of density and temperature perturbations are considerably larger if $\PrEc$ increases. For the transcritical case,  when the flow enters the triangular zone, the density perturbation are the most dominant.

\begin{figure}
\includegraphics[scale=0.46,clip]{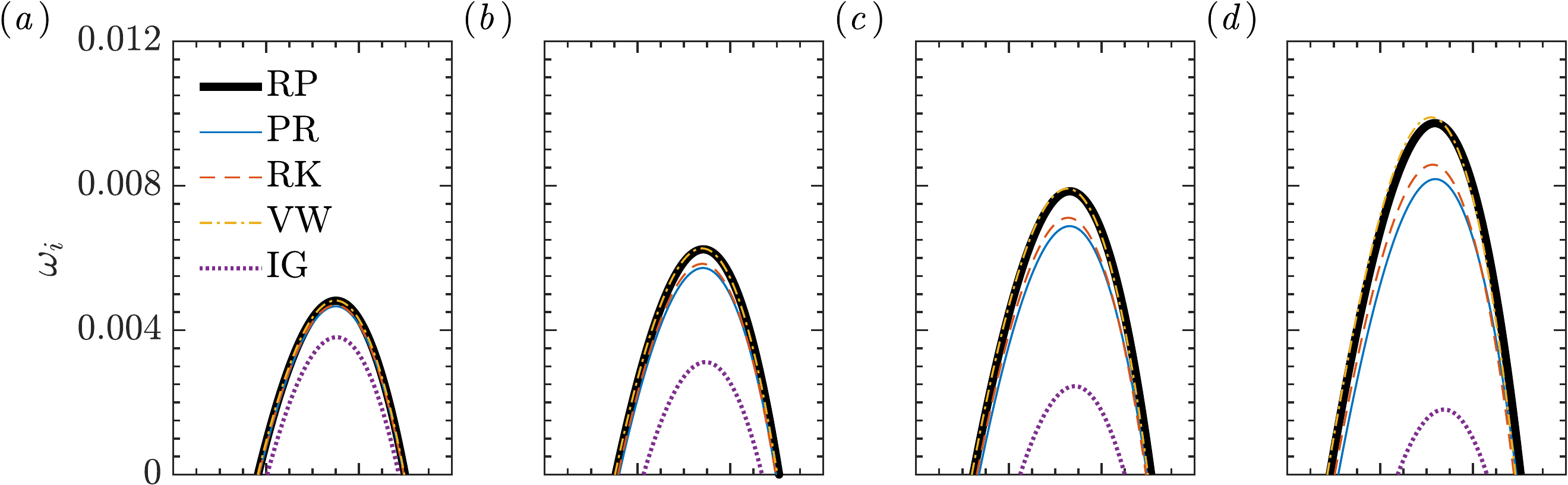}
\includegraphics[scale=0.46,clip]{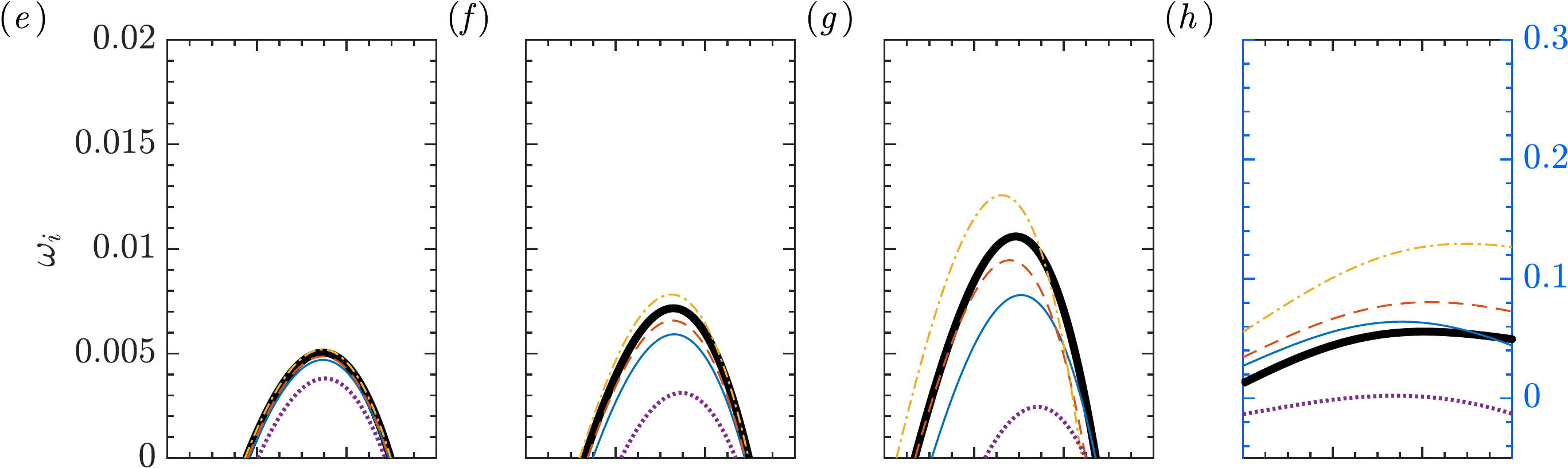}
\includegraphics[scale=0.46,clip]{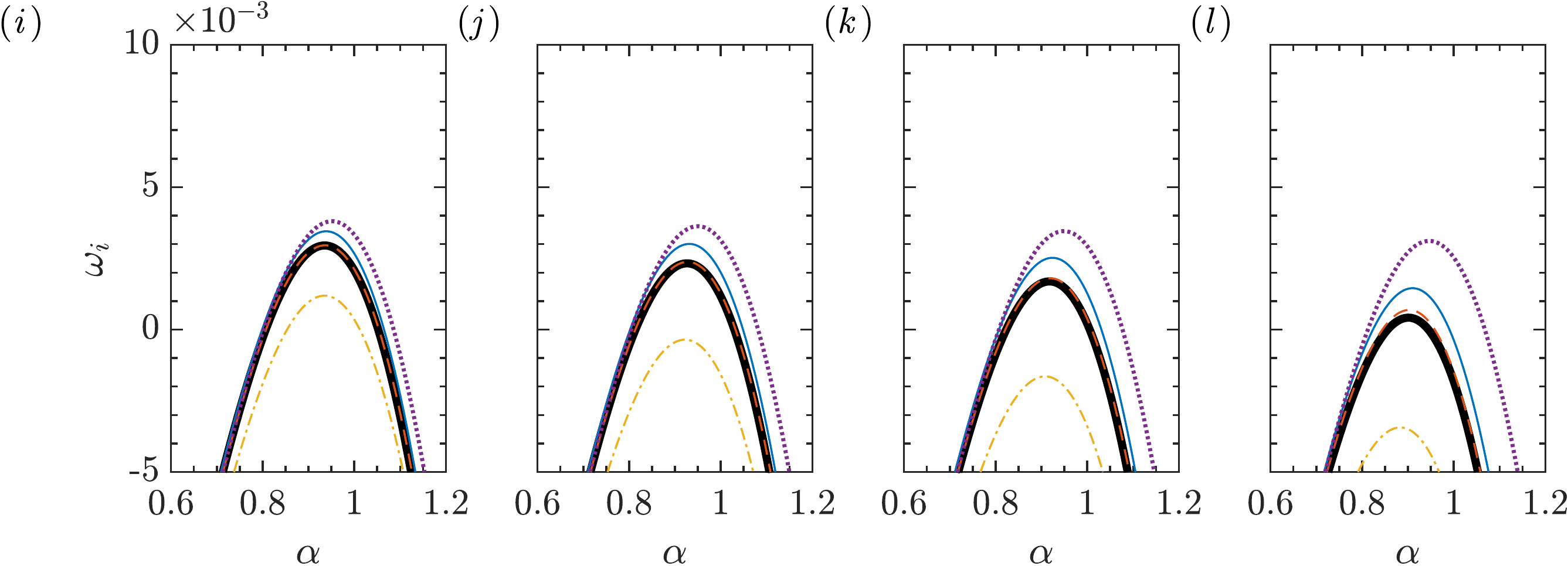}
\caption{Growth rates of the perturbation for different gas models.  Results shown are at $\Rey=10000$, $\beta=0$ for the subcritical case (a-d, $\PrEc=$ 0.01, 0.03, 0.05 and 0.07), transcritical case  (e-h, $\PrEc=$ 0.01, 0.03, 0.05 and 0.06) and supercritical case (i-l, $\PrEc=$ 0.01, 0.015, 0.02 and 0.03). Note that the y-coordinate of subplot (h) is different from the others.}
\label{Fig8}
\end{figure}

Below, we compare the fidelity of the cubic EoS models with the EoS model from REFPROP. The solutions for the ideal EoS are also shown to highlight the difference with respect to the results obtained with the non-ideal EoS models. Figure~\ref{Fig8} shows the growth rate of the unstable modes for all EoS models. Recall the discussion in \S \ref{Sec4-1}, all these curves collapse under the isothermal limit. As can be inferred from each row of figure~\ref{Fig8}, the differences between these models magnify when $\PrEc$ is increased. In all three cases, the cubic EoS models predict the correct trend that the flow becomes more unstable in sub-/transcritical cases, and more stable in supercritical cases as $\PrEc$ increases. %\sout{The cubic  EoS models are capable of giving a correct order of magnitude of growth rates for the inflectional profile.} 
Specifically, the van der Waals EoS shows a good agreement with the RP EoS model in the subcritical case, while both Peng-Robinson and Redlich-Kwong EoS predict a lower growth rate (shown in figure~\ref{Fig8}). In the transcritical case, the van der Waals and Redlich-Kwong give acceptable growth rates if compared to RP. When the base flow becomes inflectional ($\PrEc=0.06$), the Peng-Robinson EoS shows the best approximation. In the supercritical case, Redlich-Kwong produces the best results, while the van der Waals EoS gives a much lower growth rate. Given these observations, it can be concluded that all non-ideal EoS models give the same trends. However, it is not possible to state the fidelity of the cubic EoS models in terms of the growth rate. 
%the qualitatively the   it is difficult to make a meaningful conclusion on the fidelity to the cubic EoS models. 
%However, we shall state that cubic EoS cannot give a quantitatively accurate growth rate if compared with the EoS provided by RefProp. 
%\red{No practically meaningful conclusion can be made among different EoS (on the stability calculation) shall be made. }
%\sout{As a matter of fact, the density component of the base flow is dependent on the EoS in use. This further caused different foundations, based on which gradients of thermodynamic and transport properties are calculated. The deviation on the inputs of the stability analysis (base flow, properties) may thus accumulate or even cancel out. }

\subsection{\red{The kinetic energy budget}}
\red{To further understand the instability mechanism of the non-ideal fluids, we perform a kinetic energy budget analysis for the 2D perturbation. The energy balance equation is the sum of the x-momentum perturbation equation, multiplied with $\hat{u}^\dagger$, and the y- equation, multiplied with $\hat{v}^\dagger$. Here, dagger stands for the complex conjugate. The continuity equation is used to substitute the temporal growth of density, which appears in the x-momentum equation. This gives the following kinetic energy balance equation: 
\begin{equation}\label{budget}
K=\Theta+P+T+V,
\end{equation}
where
\begin{equation}
K=-i\omega\int\rho_{0}\left(\hat{u}\hat{u}^{\dagger}+\hat{v}\hat{v}^{\dagger}\right)\dd y,
\end{equation}
\begin{equation}
\Theta=-i\alpha\int\rho_{0}u_{0}\left(\hat{u}\hat{u}^{\dagger}+\hat{v}\hat{v}^{\dagger}\right)\dd y,
\end{equation}
\begin{equation}
P=-\int\rho_{0}\frac{\partial u_{0}}{\partial y}\hat{v}\hat{u}^{\dagger}\dd y,
\end{equation}
\begin{equation}
\begin{alignedat}{1}T=-\int\left[i\alpha\frac{\partial p_{0}}{\partial\rho_{0}}\hat{\rho}\hat{u}^{\dagger}+i\alpha\frac{\partial p_{0}}{\partial T_{0}}\hat{T}\hat{u}^{\dagger}+\frac{\partial p_{0}}{\partial\rho_{0}}\frac{\partial\hat{\rho}}{\partial y}\hat{v}^{\dagger}+\frac{\partial p_{0}}{\partial T_{0}}\frac{\partial\hat{T}}{\partial y}\hat{v}^{\dagger}+\right.\\
\left.\left(\frac{\partial^{2}p_{0}}{\partial\rho_{0}^{2}}\frac{\partial\rho_{0}}{\partial y}+\frac{\partial^{2}p_{0}}{\partial\rho_{0}\partial T_{0}}\frac{\partial T_{0}}{\partial y}\right)\hat{\rho}\hat{v}^{\dagger}+\left(\frac{\partial^{2}p_{0}}{\partial T_{0}^{2}}\frac{\partial T_{0}}{\partial y}+\frac{\partial^{2}p_{0}}{\partial\rho_{0}\partial T_{0}}\frac{\partial\rho_{0}}{\partial y}\right)\hat{T}\hat{v}^{\dagger}\right]\dd y,
\end{alignedat}
\end{equation}
\begin{equation}
\begin{alignedat}{1}V=\frac{1}{Re}\int\left[-\alpha^{2}\left(2\mu_{0}+\lambda_{0}\right)\hat{u}\hat{u}^{\dagger}+\mu_{0}\frac{\partial^{2}\hat{u}}{\partial y^{2}}\hat{u}^{\dagger}+i\alpha\left(\mu_{0}+\lambda_{0}\right)\frac{\partial\hat{v}}{\partial y}\hat{u}^{\dagger}\right.\\
+i\alpha\frac{\partial\mu_{0}}{\partial y}\hat{v}\hat{u}^{\dagger}+\frac{\partial\mu_{0}}{\partial\rho_{0}}\frac{\partial u_{0}}{\partial y}\frac{\partial\hat{\rho}}{\partial y}\hat{u}^{\dagger}+\frac{\partial\mu_{0}}{\partial y}\frac{\partial\hat{u}}{\partial y}\hat{u}^{\dagger}+\frac{\partial\mu_{0}}{\partial T_{0}}\frac{\partial u_{0}}{\partial y}\frac{\partial\hat{T}}{\partial y}\hat{u}^{\dagger}\\
+\frac{\partial\mu_{0}}{\partial\rho_{0}}\frac{\partial^{2}u_{0}}{\partial y^{2}}\hat{\rho}\hat{u}^{\dagger}+\frac{\partial u_{0}}{\partial y}\left(\frac{\partial^{2}\mu_{0}}{\partial\rho_{0}^{2}}\frac{\partial\rho_{0}}{\partial y}+\frac{\partial^{2}\mu_{0}}{\partial\rho_{0}\partial T_{0}}\frac{\partial T_{0}}{\partial y}\right)\hat{\rho}\hat{u}^{\dagger}\\
+\frac{\partial\mu_{0}}{\partial T_{0}}\frac{\partial^{2}u_{0}}{\partial y^{2}}\hat{T}\hat{u}^{\dagger}+\frac{\partial u_{0}}{\partial y}\left(\frac{\partial^{2}\mu_{0}}{\partial T_{0}^{2}}\frac{\partial T_{0}}{\partial y}+\frac{\partial^{2}\mu_{0}}{\partial T_{0}\partial\rho_{0}}\frac{\partial\rho_{0}}{\partial y}\right)\hat{T}\hat{u}^{\dagger}\\
-\alpha^{2}\mu_{0}\hat{v}\hat{v}^{\dagger}+\left(2\mu_{0}+\lambda_{0}\right)\frac{\partial^{2}\hat{v}}{\partial y^{2}}\hat{v}^{\dagger}+i\alpha\left(\mu_{0}+\lambda_{0}\right)\frac{\partial\hat{u}}{\partial y}\hat{v}^{\dagger}\\
\left.+i\alpha\frac{\partial\mu_{0}}{\partial\rho_{0}}\frac{\partial u_{0}}{\partial y}\hat{\rho}\hat{v}^{\dagger}+i\alpha\frac{\partial\lambda_{0}}{\partial y}\hat{u}\hat{v}^{\dagger}+i\alpha\frac{\partial\mu_{0}}{\partial T_{0}}\frac{\partial u_{0}}{\partial y}\hat{T}\hat{v}^{\dagger}+\left(2\frac{\partial\mu_{0}}{\partial y}+\frac{\partial\lambda_{0}}{\partial y}\right)\frac{\partial\hat{v}}{\partial y}\hat{v}^{\dagger}\right]\dd y.
\end{alignedat}
\end{equation}
The real part of the equation \eqref{budget} describes the balance of the kinetic energy growth. In particular, $K_r$ is the temporal growth of the kinetic energy, $\Theta$ is purely imaginary and does therefore not contribute to the temporal growth, $P_r$ is the production term, $T_r$ is the thermodynamic term, and $V_r$ is the viscous dissipation.}

\begin{figure}
\includegraphics[width=0.45\linewidth]{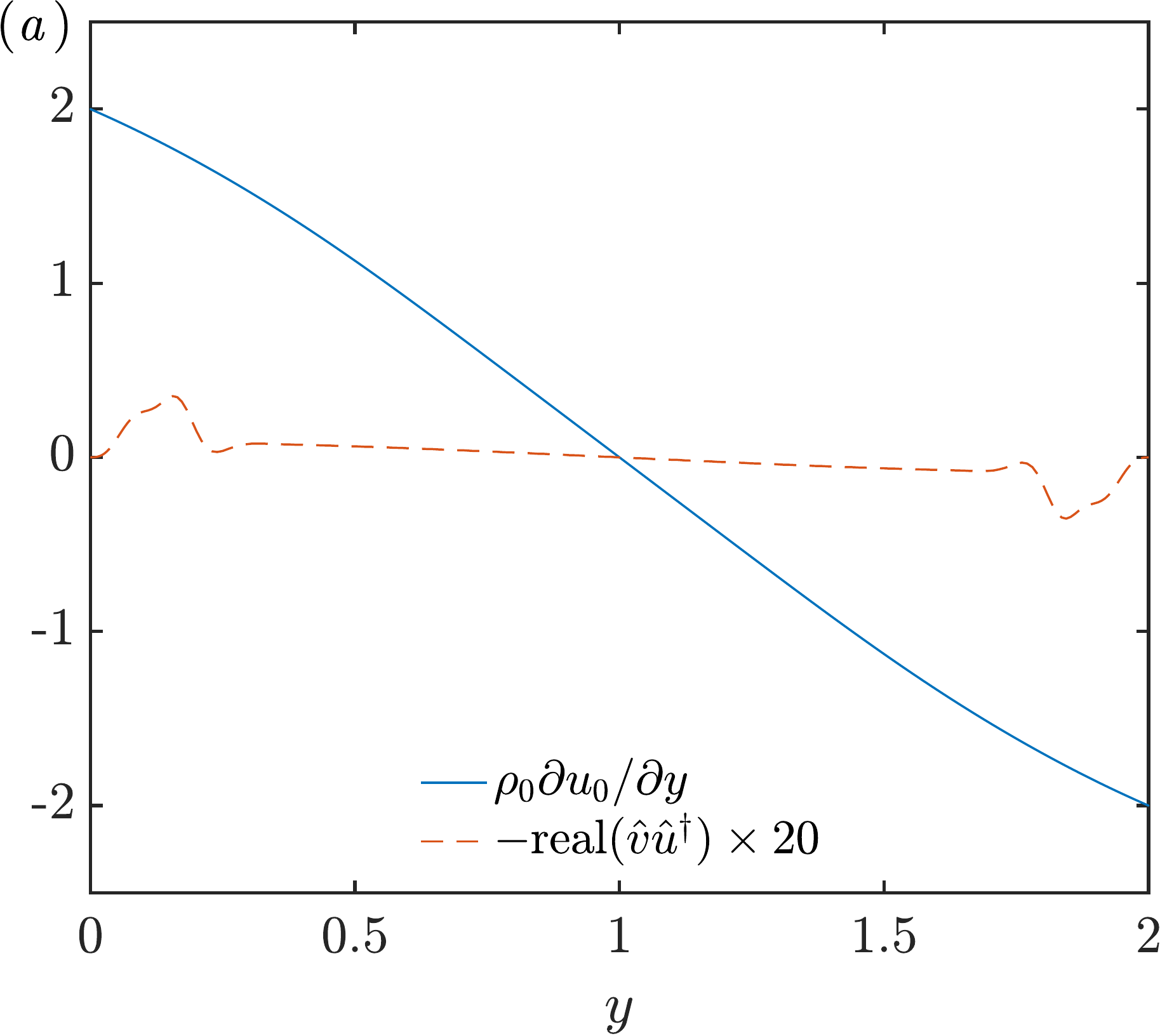}
\includegraphics[width=0.45\linewidth]{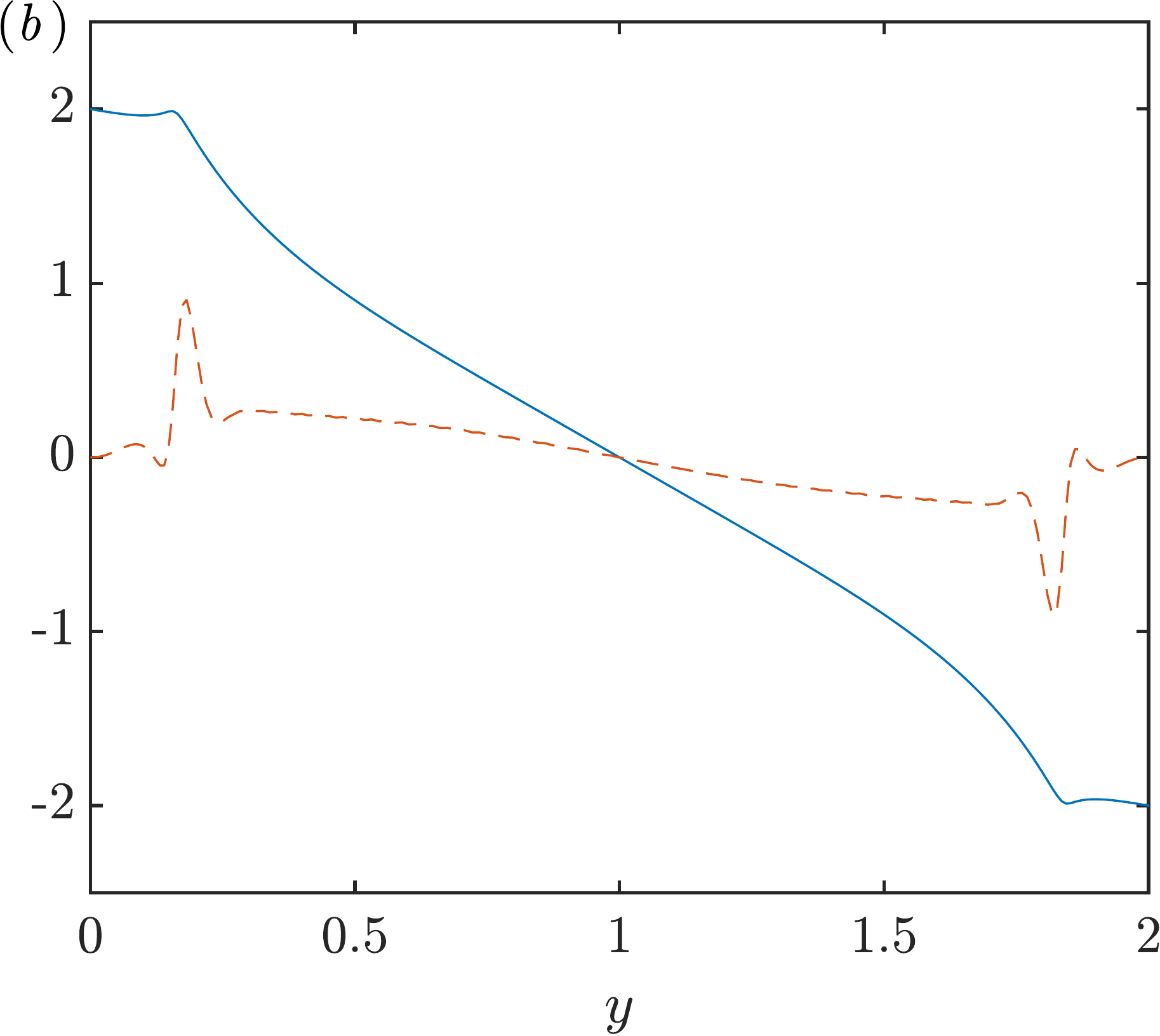}
\caption{Production of the kinetic perturbation energy with $T_w^*=300$ K, $\alpha=1$, $\Rey=10000$. (a) $\PrEc=0.05$, (b) $\PrEc=0.06$. }
\label{Fig9}
\end{figure}

\red{The results of the kinetic energy budget analysis are summarized in table~\ref{Table3}. The analysis is performed for all three cases at $\alpha=1$ and $\Rey=10000$. It clearly shows that for all the cases, the energy growth $K_r$ originates from the production term $P_r$. The thermodynamic term $T_r$ slightly reduces the growth. The viscous dissipation $V_r$ is not sensitive to the parameters and remains almost constant, except in the transcritical case ($T_w^*=300$ K, $\PrEc=0.06$), which has a considerably larger growth rate (as also shown in figure~\ref{Fig7} and \ref{Fig8}). The reason for this lies in a much larger production and a smaller viscous dissipation. Figure~\ref{Fig9} compares the production of the two cases with $\PrEc=0.05$ and 0.06 at $T_w^*=300$ K. It can be inferred that the inflectional velocity profile ($\PrEc=0.06$) has caused a larger $\rho_0\pp u_0/\pp y$ near both walls, the amplitude of the velocity perturbation $\hat{v}\hat{u}^\dagger$ is larger as well. Therefore, a large production term $-\int\rho_{0}\frac{\partial u_{0}}{\partial y}\hat{v}\hat{u}^{\dagger}\dd y$ and accordingly the large growth rate can be explained. }
\begin{table} 
\begin{center}
\begin{tabular}{llrrrr}
\multicolumn{2}{l}{Cases} & \multicolumn{4}{l}{Budgets ($\times10^{-3}$)} \vspace{5pt} \\  \cline{1-2} \cline{3-6}
\\
			  $T_w^*$  &\PrEc  & $K_r$ 	& $P_r$  & $T_r$   &  $V_r$ \vspace{8pt}\\
\multirow{4}{*}{290 K} & 0.01  & 3.4  & 8.2  &  0.0 &  -4.8 \\
  					   & 0.03  & 4.5  & 9.4  &  0.0 &  -4.9 \\
  					   & 0.05  & 5.6  & 10.7 &  0.0 &  -5.1 \\
  					   & 0.07  & 6.8  & 12.1 & -0.1 &  -5.2\vspace{5pt} \\ 
\multirow{4}{*}{300 K} & 0.01  & 3.5  & 8.4  & -0.1 &  -4.8 \\
					   & 0.03  & 5.0  & 10.1 & -0.1 &  -5.0 \\
					   & 0.05  & 6.2  & 11.6 & -0.2 &  -5.2 \\
					   & 0.06  & 17.5 & 21.1 & -2.4 &  -1.2\vspace{5pt} \\
\multirow{4}{*}{310 K} & 0.01  & 1.8  & 6.6  &  0.0 &  -4.8 \\
					   & 0.015 & 1.2  & 6.1  & -0.2 &  -4.7 \\
					   & 0.02  & 0.6  & 5.5  & -0.2 &  -4.7 \\
					   & 0.03  &-0.7  & 4.3  & -0.3 &  -4.7 \\
\end{tabular}
\caption{\red{Kinetic energy budget analysis for two-dimensional perturbations. $\alpha=1$, $\Rey=10000$.}}
\label{Table3}
\end{center}
\end{table}

\section{Algebraic growth}\label{Sec5}
\subsection{Choice of the energy norm}
\begin{figure}
\begin{center}
\includegraphics[width=0.55\linewidth,clip]{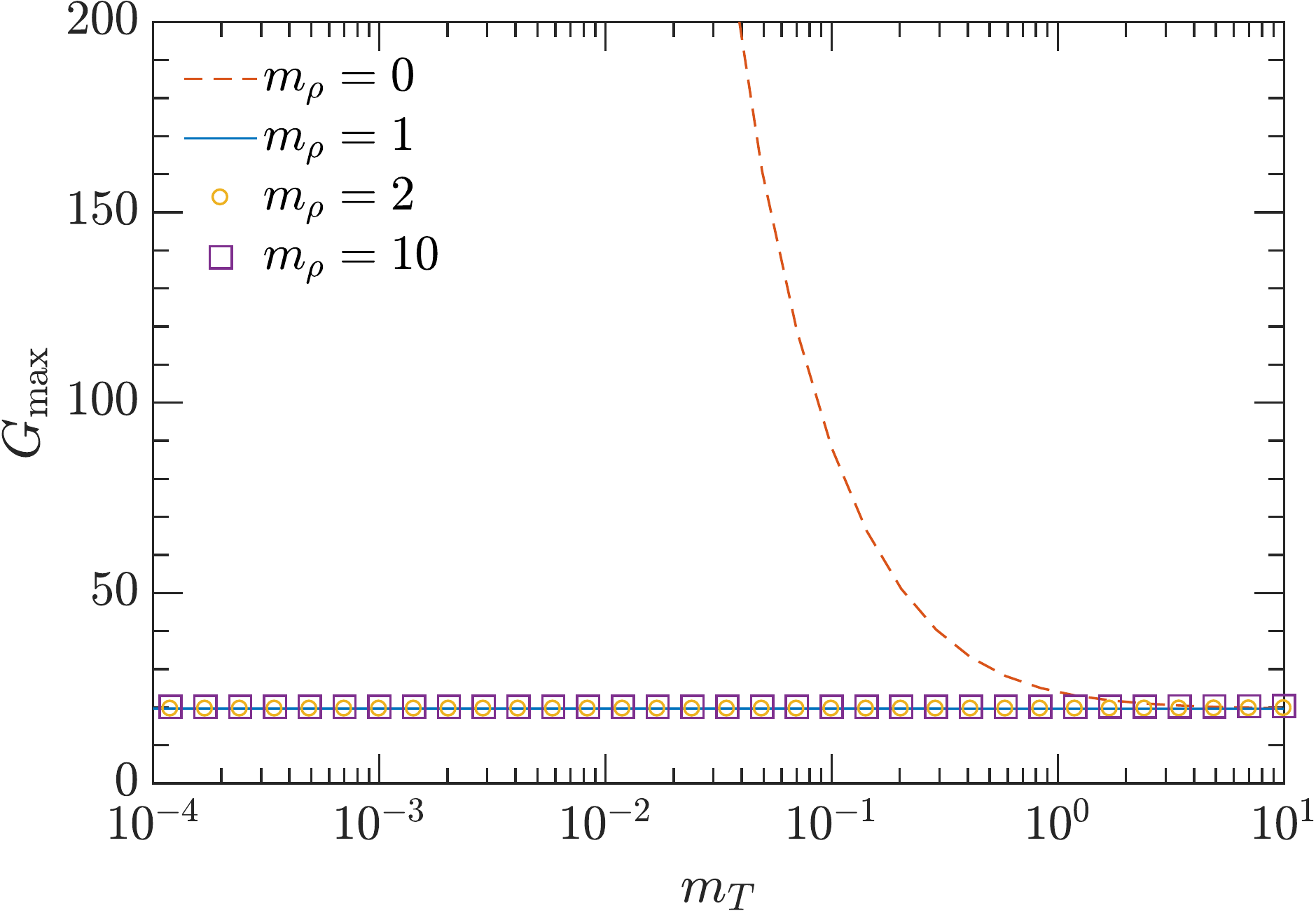}
\end{center}
\caption{Maximum energy growth $G_{\max}$ versus $m_{T}$ using the energy norm \eqref{Norm}. $m_{\rho}=0$, 1, 2 and 10. The fluid (with RP model) is at $\PrEc=0.05$, $\Rey=2000$, $\alpha=1.0$, $\beta=0.25$ and $T^*_w=290$ K.}
\label{Fig10}
\end{figure}

The Mack's energy norm \citep{Mack1969, Hanifi1996} has been extensively used in compressible flows. The norm is designed under the ideal gas assumption, therefore the pressure-related energy transfer terms can be eliminated by choosing suitable coefficients for each components. \red{In fact, Mack's norm is equivalent to Chu's norm \citep{Chu1965,George2011}}. In the current non-ideal gas flows, the equation of states can be different (PR, RK, VW, IG), or even implicit (look-up table) as in the case of the RP EoS model. \red{Therefore, we choose a general form of the norm:}
\begin{equation}\label{Norm}
E\left(\boldsymbol{q}\right)=\int \left(u^{\prime\dagger}u^{\prime}+v^{\prime\dagger}v^{\prime}+w^{\prime\dagger}w^{\prime}\right)+m_{\rho}\rho^{\prime\dagger}\rho^{\prime}+m_{T}T^{\prime\dagger}T^{\prime}~{\rm d}V,
\end{equation}
where $\dagger$ denotes the complex conjugate. This norm has been tested for the compressible ideal/non-ideal gas flows at various conditions. Figure \ref{Fig10} shows the optimal energy growth $G_{\max}$ (the maximum of $G(t)$ over time $t$) as a function of $m_{T}$ and $m_{\rho}$, for $\PrEc=0.05$ (thermodynamic components become important) and a wall temperature of $T^*_w=290$~K. When $m_{\rho}$ is set to 0, $G_{\max}$ converges to a constant value when $m_T$ is large enough. On the other hand, the energy norm is shown to be rather robust when the density component is properly accounted for, \eg $m_{\rho}=1$. Therefore, the results presented in this section are mainly obtained for $m_{\rho}=m_{T}=1$. \red{A comparison with Mack's energy norm ($m_{\rho}=T_0/(\rho_0^2\gamma\Ma^2)$, $m_T=1/(\gamma(\gamma-1)T_0\Ma^2)$) is proivded at the end of this section}. 

\subsection{The isothermal limit}
\begin{figure}
\begin{center}
\includegraphics[width=0.55\linewidth,clip]{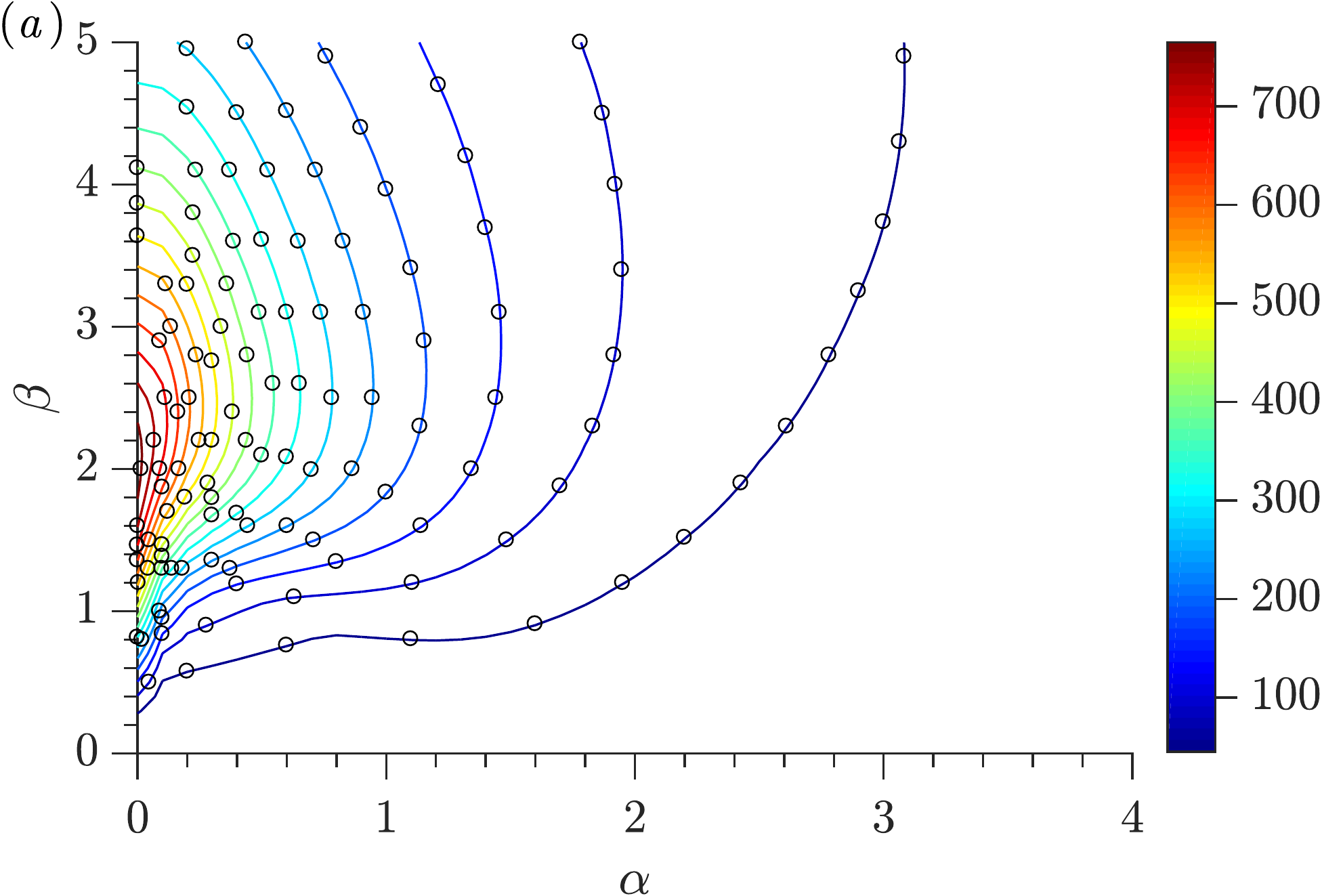}\vspace{4pt}
\includegraphics[width=0.75\linewidth,clip]{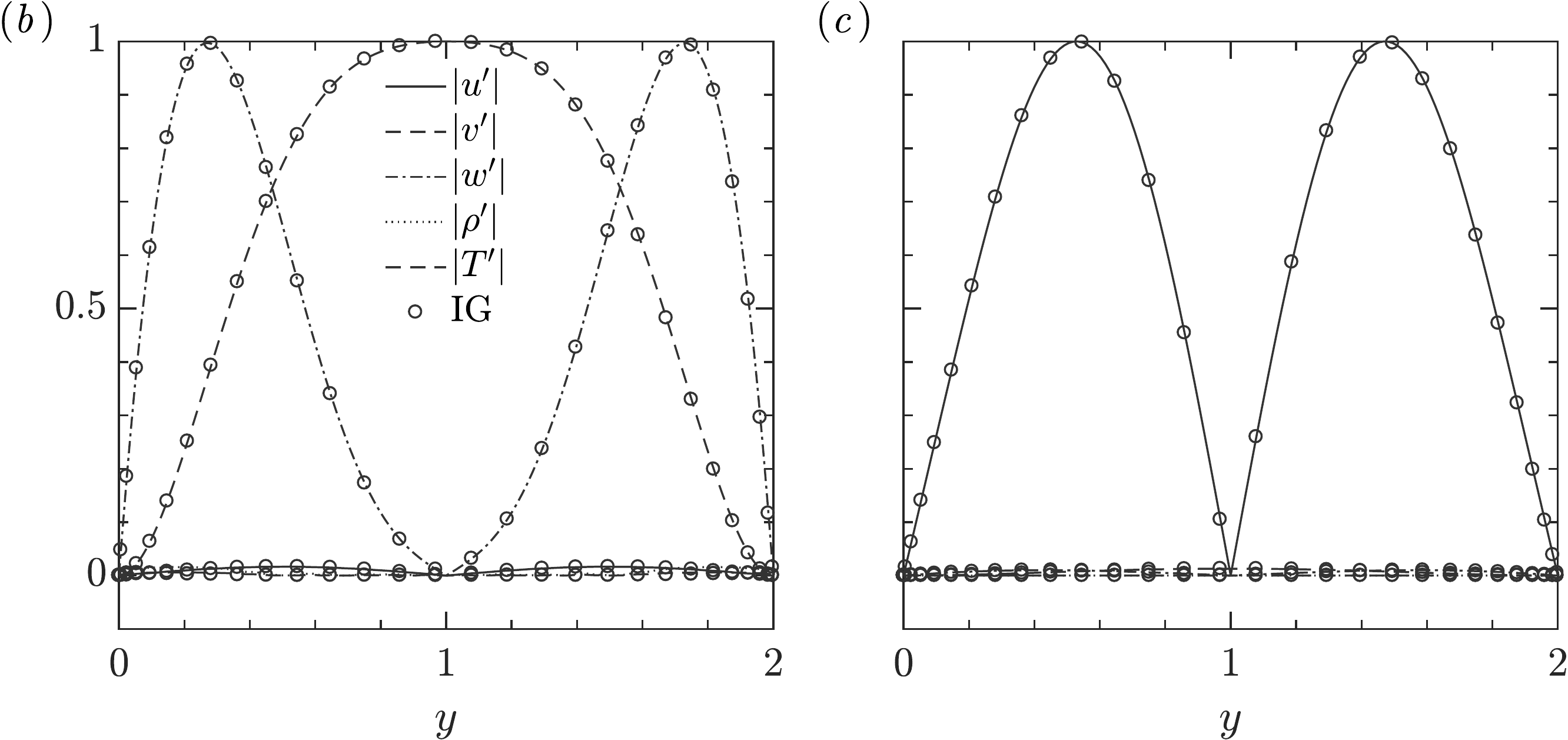}
\end{center}
\caption{Transient growth in the isothermal limit. $\PrEc\rightarrow0$, $\Rey=2000$, $T_w^*=290$ K. (a) Contour plot of $G_{\max}$. (b) The optimal perturbation (input) for $\alpha=0$, $\beta=2$. (c) The corresponding output. Lines and circle symbols show results of non-ideal (RP) and ideal gas (IG) respectively.}
\label{Fig11}
\end{figure}

Although all EoS considered in this work give the same most unstable mode in the isothermal limit (discussed in \S \ref{Sec4-1}), their eigenvalue spectrum can be rather different (see figure \ref{Fig6}(a)). Their corresponding eigenfunctions form the basis of the optimal perturbation and the algebraic growth. 
%In the isothermal limit, as discussed in \S \ref{Sec4-1}, though the most unstable mode is the same, the spectrum of the eigenvalues can be rather different among distinct fluid models (see figure \ref{Fig6}(a)), their corresponding eigenfunctions form the basis of the optimal perturbation and the algebraic growth. 
We show the contour plot of $G_{\max}$ in $\alpha-\beta$ diagram in figure~\ref{Fig11}(a). Lines and circles show results of RP and IG models, respectively. It is evident that they fall on top of each other. In fact, all five models (RP, PR, RK, VW, IG) show the same results, and correspond to the results using incompressible equations. The largest transient growth occurs at $\alpha=0$ and $2\leq\beta\leq2.1$, which is well-known for ideal gas. The optimal perturbation and the corresponding output are shown in figure \ref{Fig11}(b,c) for $\alpha=0$, $\beta=2$. The classic streamwise vortices (the optimal perturbation) and streaks (the corresponding output) are recovered. There is no discernible difference between the non-ideal and ideal gases under the isothermal limit.

\subsection{Compressible flows}

\begin{figure}
\begin{center}
\includegraphics[width=0.8\linewidth,clip]{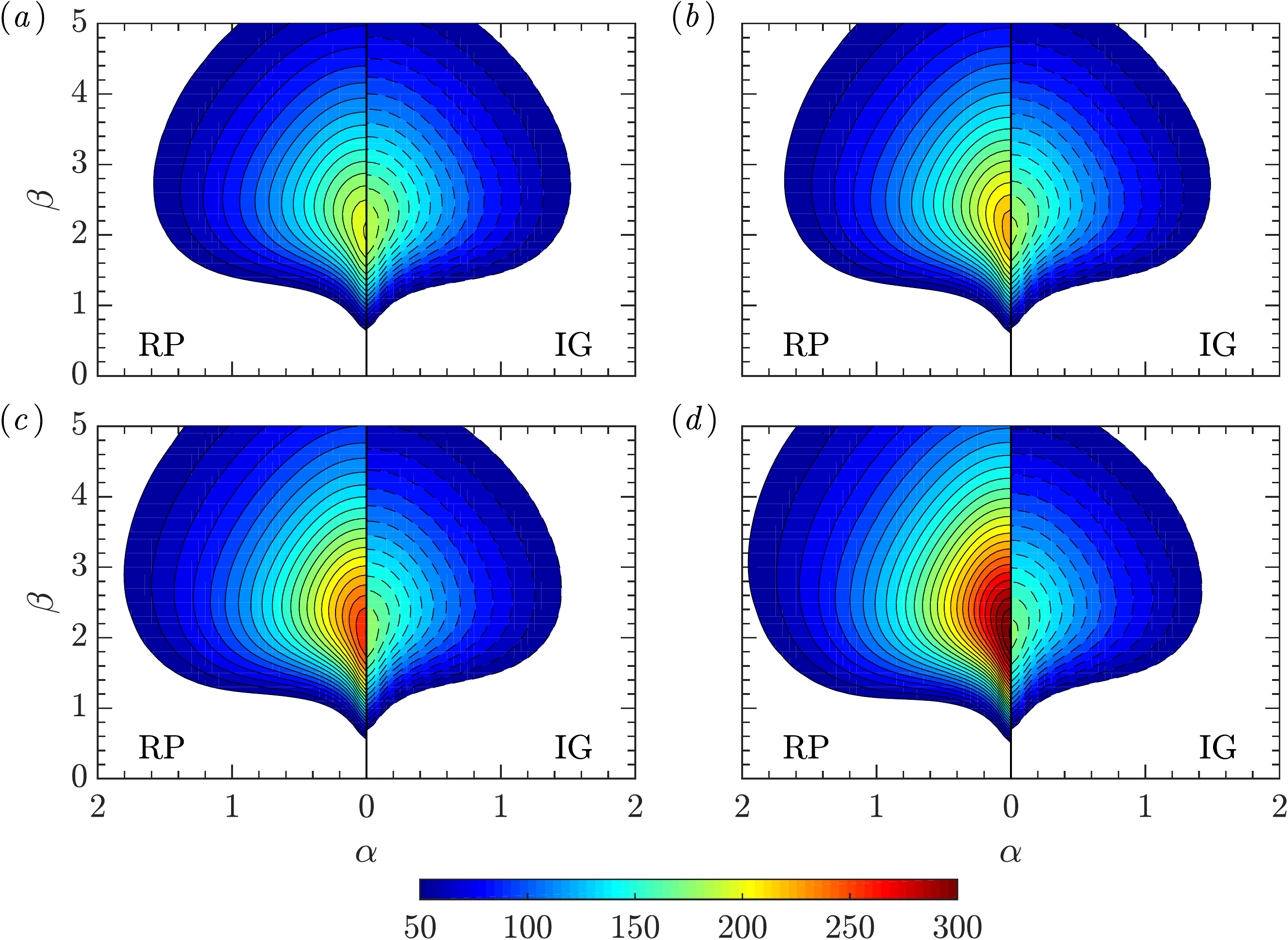}
\end{center}
\caption{Contour plot of $G_{\max}$ at $T_w^*=290$ K. $\Rey=1000$. On the left and right side of each subplot, we show the results for non-ideal (RP) and ideal (IG) gases respectively. (a) $\PrEc=0.01$, (b) $\PrEc=0.03$, (c) $\PrEc=0.05$, (d) $\PrEc=0.07$.}
\label{Fig12}
\end{figure}

\begin{figure}
\begin{center}
\includegraphics[width=0.8\linewidth,clip]{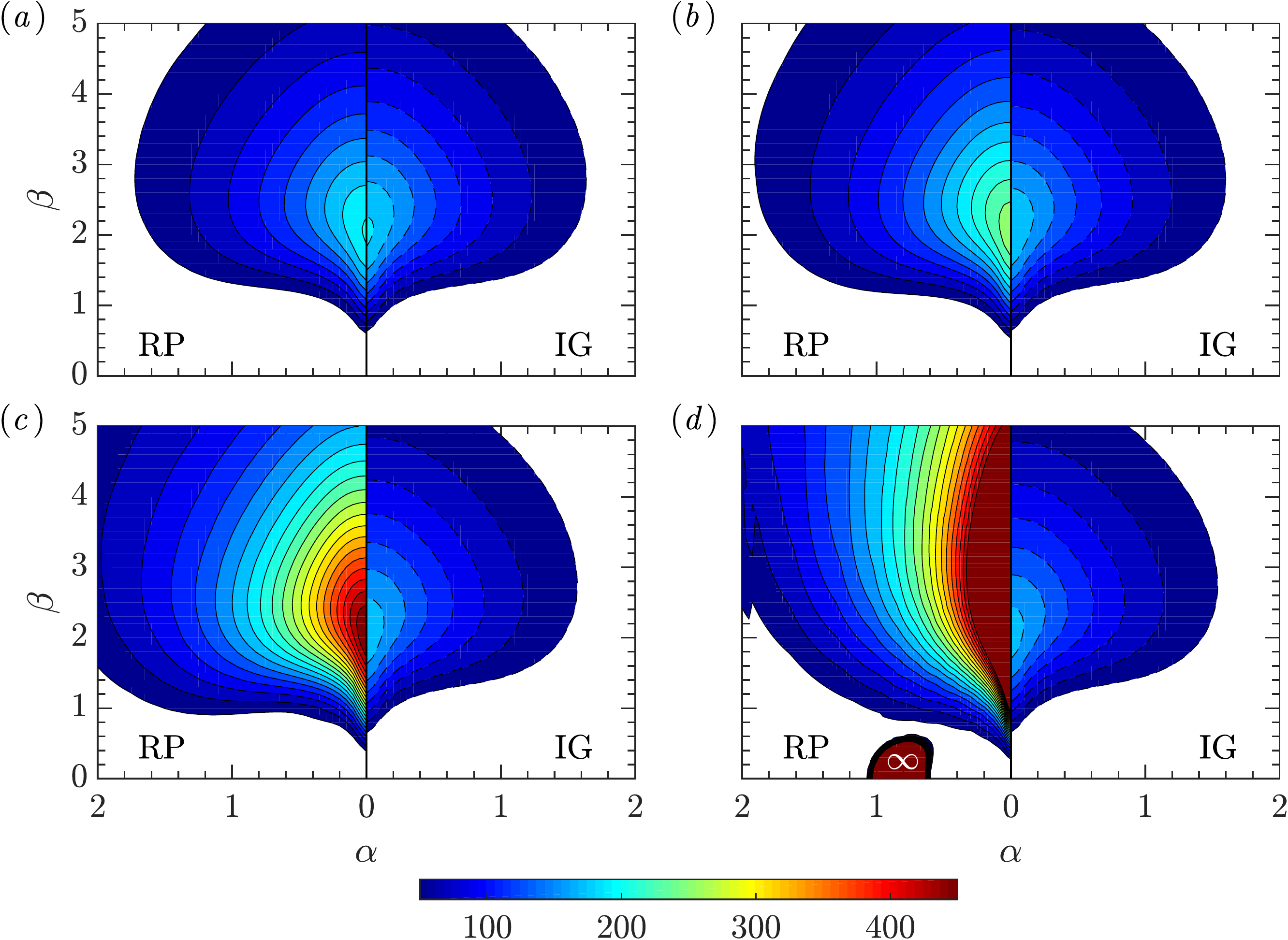}
\end{center}
\caption{Same as figure \ref{Fig12} but for $T_w^*=300$ K.}
\label{Fig13}
\end{figure}

\begin{figure}
\begin{center}
\includegraphics[width=0.8\linewidth,clip]{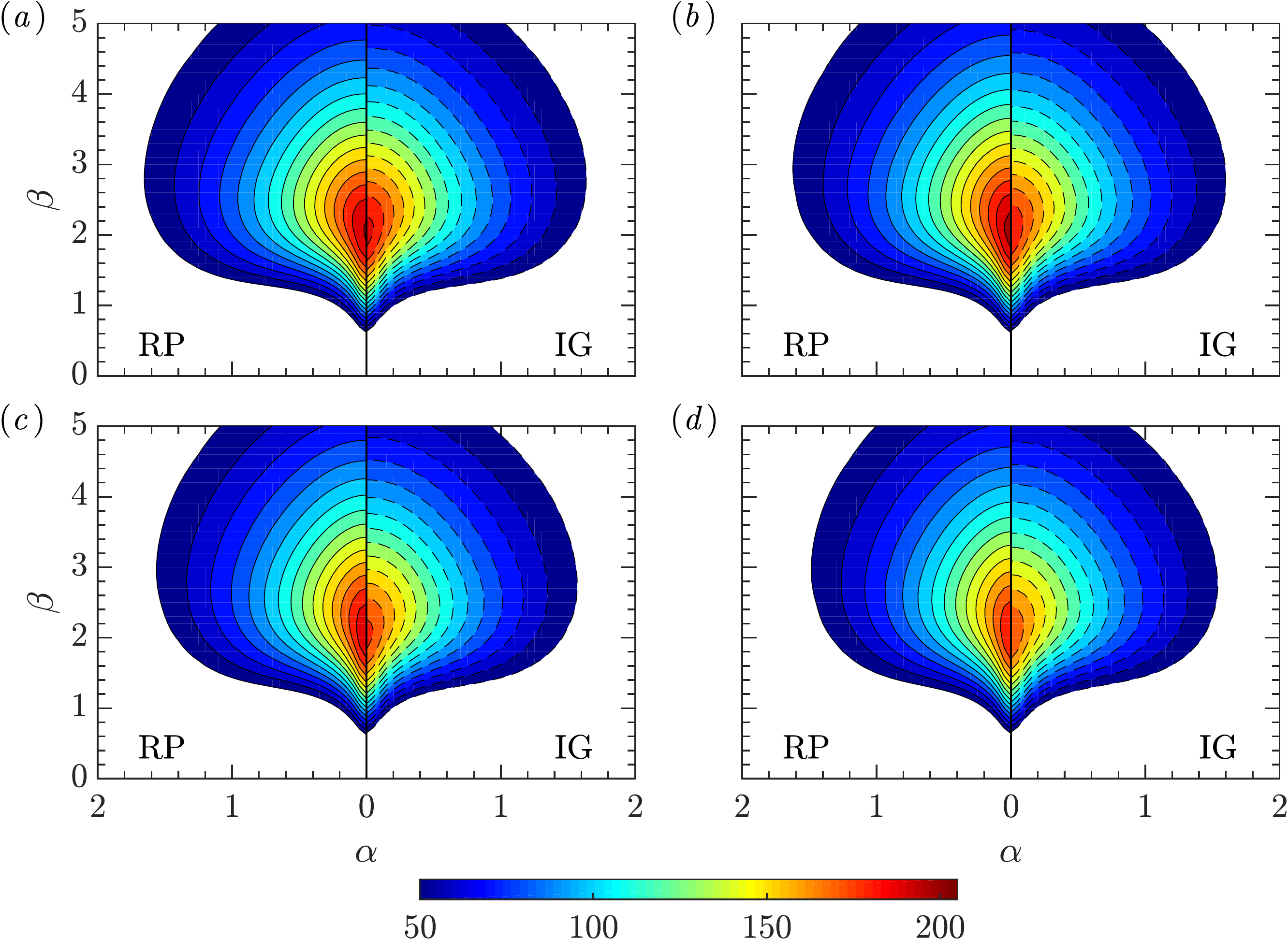}
\end{center}
\caption{Same as figure \ref{Fig12} but for $T_w^*=310$ K.}
\label{Fig14}
\end{figure}

The algebraic growth has been studied for the subcritical, transcritical and supercritical cases at $\Rey=1000$ and $\PrEc=$ 0.01, 0.03, 0.05, 0.07. The optimal energy growth $G_{max}$ for RP model is compared with IG model in figure \ref{Fig12}, \ref{Fig13} and \ref{Fig14}, respectively. The three cases actually start from the same results at the isothermal limit (figure~\ref{Fig11}a). Regardless of the wall temperature and $\PrEc$, the largest transient growth occurs at $\alpha=0$ and $2\leq\beta\leq2.1$ for both ideal and non-ideal gases. In the subcritical and transcritical cases (figure \ref{Fig12} and \ref{Fig13}), when $\PrEc$ is increased, the ideal gas tends to be slightly more stable, while the non-ideal gas becomes more unstable. In fact, due to the Power/Sutherland law (for the transport properties), the results for the ideal gas are weakly dependent on the wall temperature. Notably in figure \ref{Fig13}(d), where $\PrEc=0.07$, an area of $G_{\max}\rightarrow\infty$ stands out. Recall the discussion in \S \ref{Sec4}, the base flow has entered the triangular zone (see figure \ref{Fig5}) and becomes inflectional. Hence, the flow is inviscid unstable  and the critical $\Rey$ is reduced considerably (see figure \ref{Fig7}(c)). As a result, a sub-zone of modal growth (near $\beta=0$) in the $\alpha-\beta$ diagram is observed (where $G_{\max}\rightarrow\infty$). For better display of the results, we have limited the color band to $G_{\max}=450$ in figure~\ref{Fig13}. In the supercritical case (figure \ref{Fig14}), the plots are almost symmetrical, indicating the non-ideal gas effects are rather insignificant. The non-ideal gas is only slightly more unstable than the ideal gas. Table \ref{Table4} summarizes the above maximum transient growth $G_{\max}$. With the increase in $\PrEc$, a similar trend as for the modal growth can be observed. Namely, the ideal gas becomes more stable, while the non-ideal gas tends to be more unstable for the subcritical and transcritical case, and more stable for the supercritical case. \red{On the whole, the non-ideal gas effects increase the algebraic instability in all regimes, most prominently in the transcritical regime}.

\begin{table}
\begin{center}\def~{\hphantom{0}}
  \begin{tabular}{lc|ccc}
& ideal gas (IG) &  & non-ideal gas (RP) & \\[3pt]
& $T_w^*=290/300/310$ K \hspace{10pt} & \hspace{10pt} $T_w^*=290$ K & $T_w^*=300$ K & $T_w^*=310$ K\\[3pt]
       $\PrEc=0.01$\hspace{10pt} & 193.3 & \hspace{10pt}206.5 & 212.2 &201.4\\
       $\PrEc=0.03$\hspace{10pt} & 187.9 & \hspace{10pt}231.6 & 262.7 &204.4\\
       $\PrEc=0.05$\hspace{10pt} & 182.8 & \hspace{10pt}265.6 & 472.3 &199.5\\
       $\PrEc=0.07$\hspace{10pt} & 178.1 & \hspace{10pt}316.7 & $\infty$ & 190.3\\
  \end{tabular}
  \caption{Maximum transient growth $G_{\max}$ of the perturbations at $\Rey=1000$.}\label{Table4}
\end{center}\end{table}

\begin{figure}
\begin{center}
\includegraphics[width=0.8\linewidth,clip]{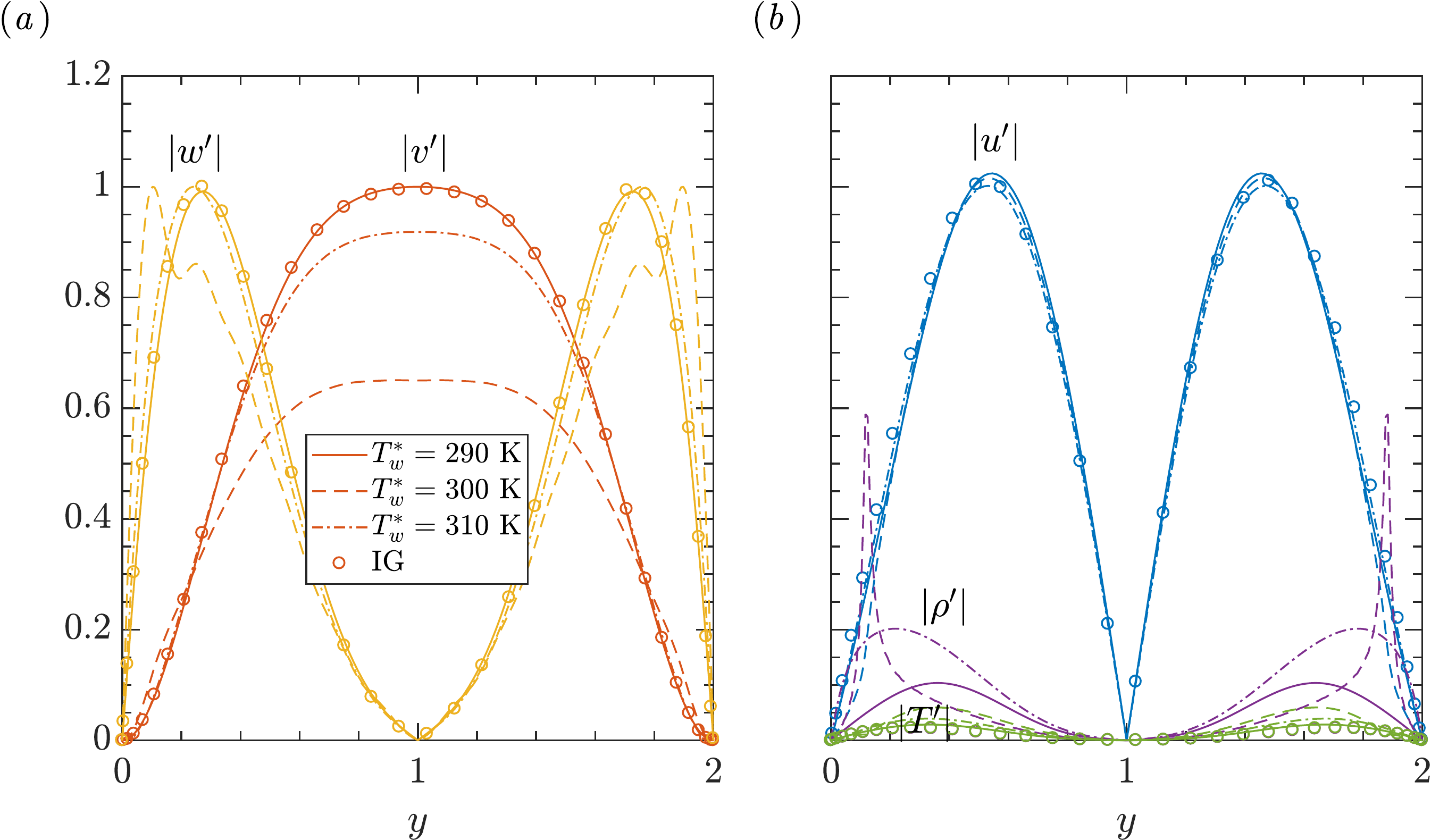}
\end{center}
\caption{Optimal perturbations (a) and the resulting output (b). $\PrEc=0.07$, $\alpha=0$ and $\beta=2$.  Only significant  components are plotted, namely in (a) $|v^\prime|$,  $|w^\prime|$, (b) $|u^\prime|$,  $|\rho^\prime|$ and $|T^\prime|$. }
\label{Fig15}
\end{figure}

The typical optimal perturbation and the resulting output are shown in figure \ref{Fig15} at $\PrEc=0.07$, $\alpha=0$ and $\beta=2$. Similar to an incompressible flow, the streamwise vortices and velocity streaks are recovered as the optimal perturbation and the output, respectively. For compressible flows, thermal streaks ($\rho^\prime$ and $T^\prime$) also become significant. Considering the non-ideal gas effects, the subcritical and supercritical cases share similar optimal perturbations as the ideal gas. In the transcritical case, the profiles are strongly influenced by the inflectional base flow and the strong property variations. On the other hand, the output perturbations are almost the same with regard to the $u^\prime$ component, indicating similar dynamic streaks are being generated. The amplitude of the thermal streak is much larger in the transcritical case close to the wall. 

\begin{figure}
\begin{center}
\includegraphics[width=0.65\linewidth]{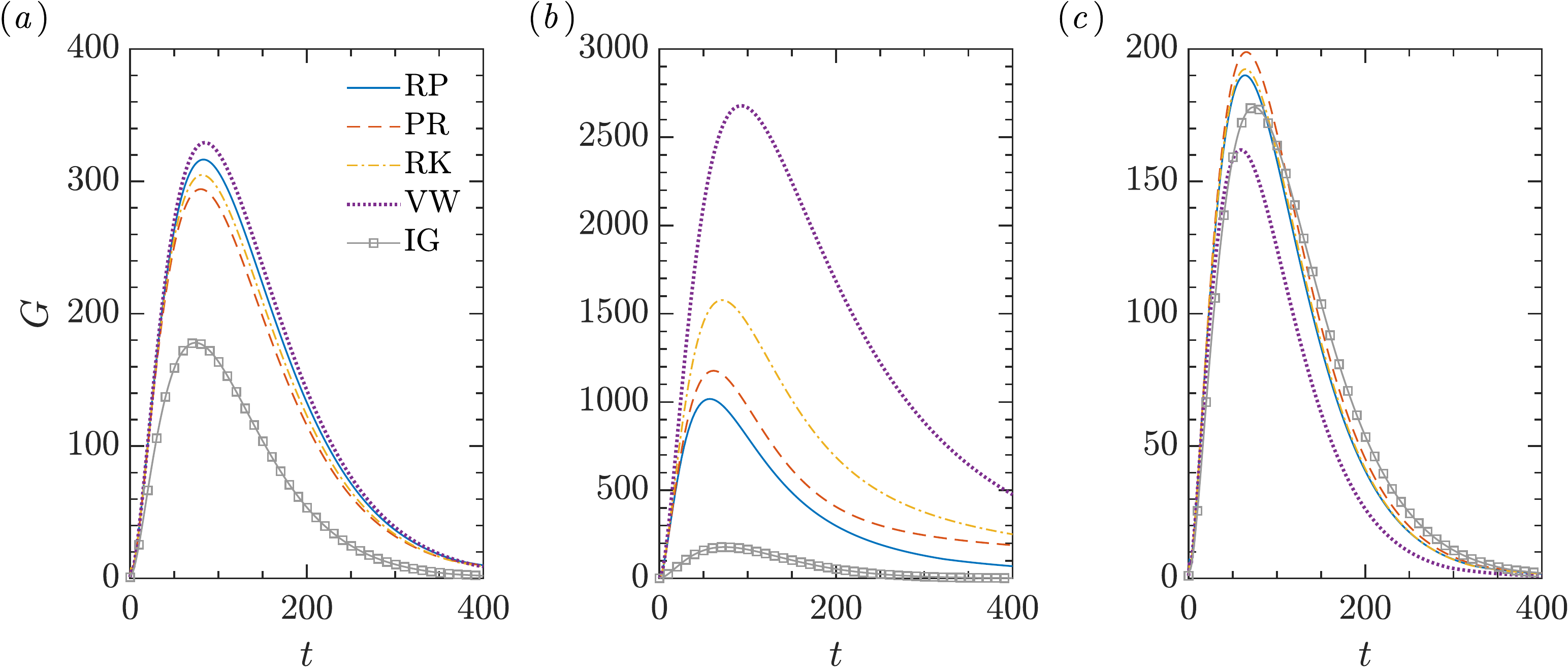}
\end{center}
\caption{The transient amplification curve $G(t)$ at $\PrEc=0.07$, $\alpha=0$ and $\beta=2$. (a) $T_w^*=290$ K, (b)  $T_w^*=300$ K, (c) $T_w^*=310$ K.}
\label{Fig16}
\end{figure}

We have shown in \S \ref{Sec4-2} that cubic EoS cannot guarantee accurate results for the growth rate if compared to results obtained with the accurate REFPROP EoS. This is also true for the algebraic instability as shown in figure~\ref{Fig16}, depicting $G-t$ curves of the three cases with different EoS at $\PrEc=0.07$, $\alpha=0$ and $\beta=2$. For example, the van der Waals EoS over-predicts $G_{max}$ by 270\% for the transcritical case. In the supercritical case, the non-ideal gas effects are less significant, and the results of all considered EoS are close to each other. 

\begin{figure}
\begin{center}
\includegraphics[width=0.32\linewidth]{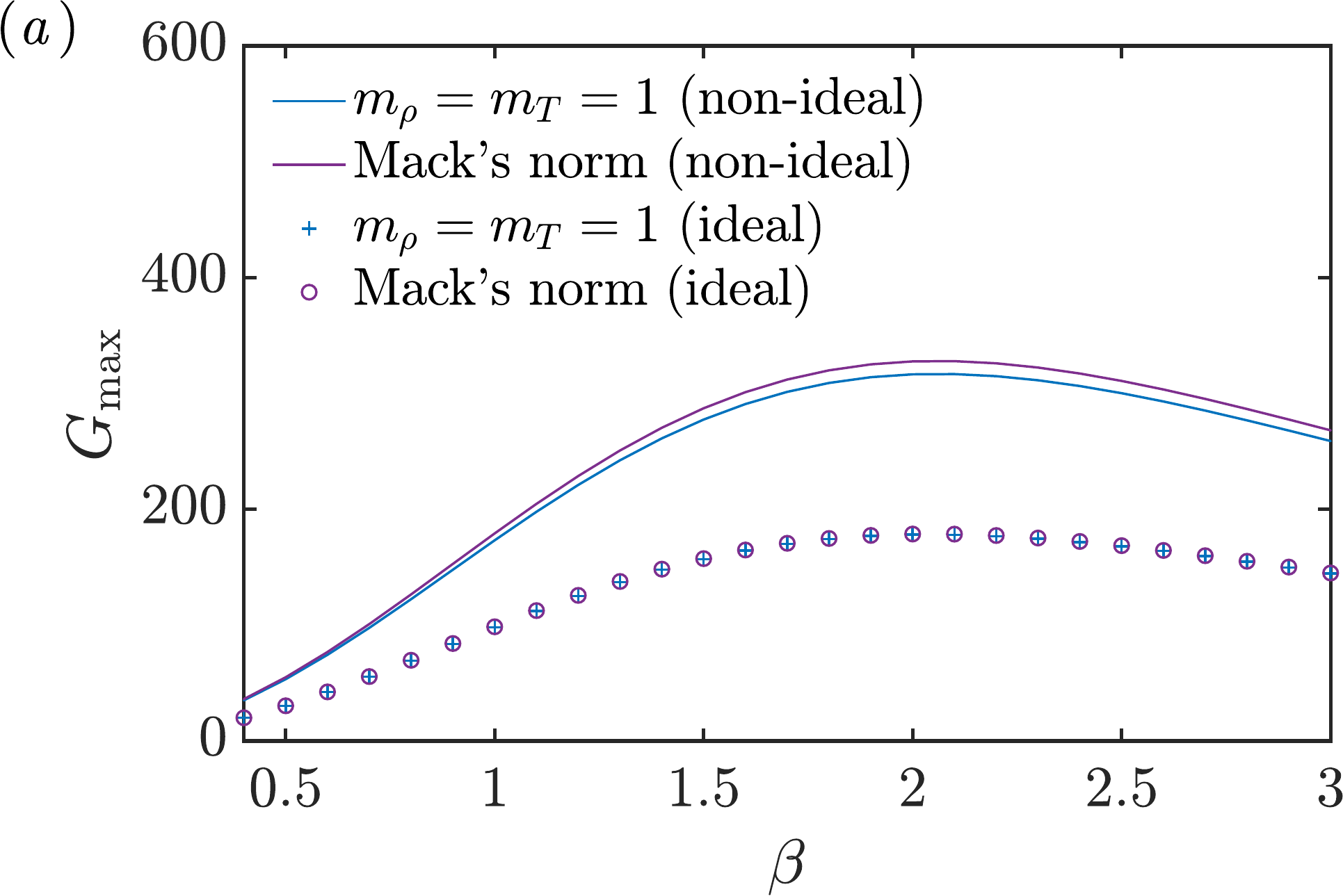}
\includegraphics[width=0.32\linewidth]{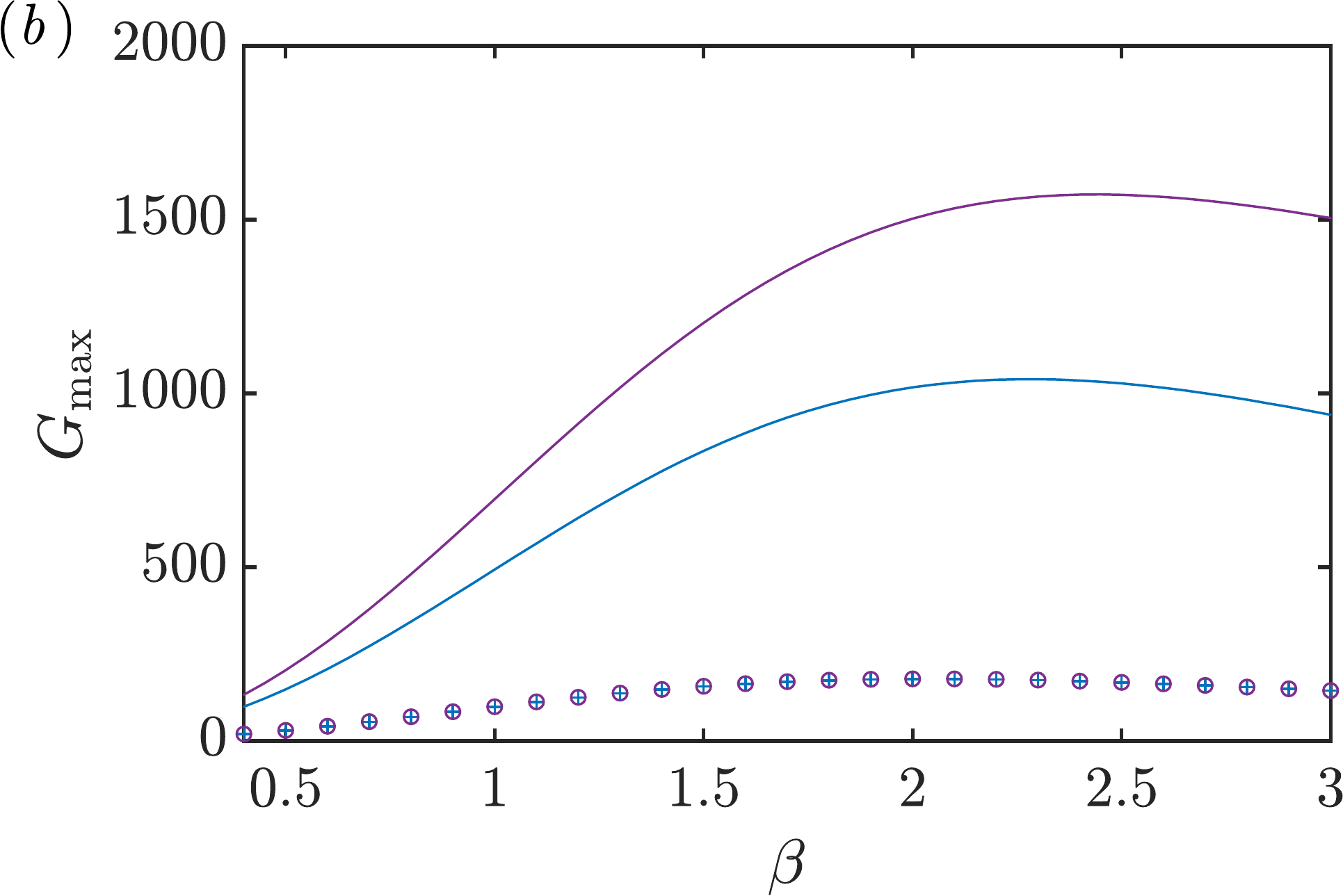}
\includegraphics[width=0.32\linewidth]{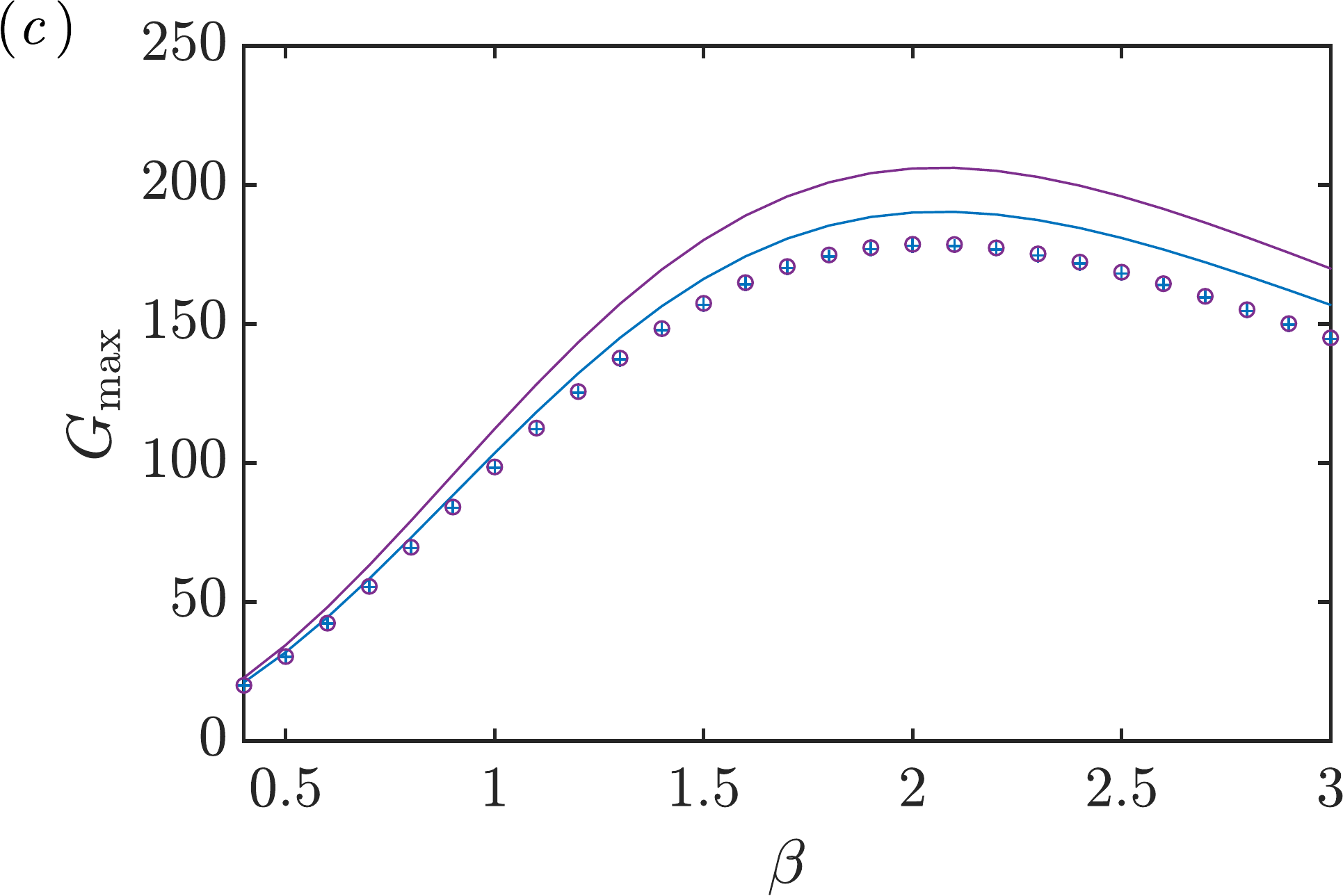}
\end{center}
\caption{Comparison of maximum algebraic growth using Mack's energy norm. Subplots show cases with (a) $T^*_w=290$ K, (b) $T^*_w=300$ K and (c) $T^*_w=310$ K. The other parameters are kept constant: $\alpha=0$, $\Rey=1000$, $\PrEc=0.07$.}
\label{Fig17}
\end{figure}

\red{The main results presented in this section are based on the energy norm: $m_\rho=m_T=1$. When Mack's energy norm is used, figure~\ref{Fig17} provides a comparison for all three regimes with highly non-ideal gas effects ($\PrEc=0.07$, $\alpha=0$). Indeed, the non-ideal gas has a larger algebraic growth in all three cases with Mack's energy norm, while on the other hand, the ideal gas are rather insensitive to different norms. As a result, the conclusion on algebraic growth will not change.}

\section{Conclusion}\label{Sec6}
Linear stability of highly non-ideal plane Poiseuille flows is studied. We have chosen carbon dioxide (CO$_2$) at supercritical pressures ($p^*=$80 bar) as an example of a fluid in a highly non-ideal thermodynamic region. 
The investigation is based on the fully compressible Navier-Stokes equations in which the product of two dimensionless parameters, namely the Prandtl $Pr$ and Eckart $Ec$ numbers, determines the viscous heating and consequently the temperature increase between the two isothermal walls. The investigated range of $\PrEc$ is from the isothermal limit ($\PrEc\rightarrow0$) to typical compressible flows with $\PrEc=0.1$. 
Three cases with wall temperatures in the vicinity of the pseudo-critical point ($T^*_{pc}=307.7$ K) have been investigated in more detail. In particular, the wall temperatures are set such that the temperature profile is subcritical ($T_w^*=290$ K), transcritical ($T_w^*=300$ K) and supercritical ($T_w^*=310$ K). In all cases, the thermodynamic and transport properties are strongly dependent on the thermodynamic state of the fluid (\eg temperature, density) and they influence the stability in a coupled way through the base flow and the linear stability operator. 

In the isothermal limit, the three cases with different wall temperatures have the same base flow as the ideal gas. When $\PrEc$ increases, the base flows of the three cases deviate from the ideal gas solution. In the subcritical regime, as $\PrEc$ increases, the flow becomes more unstable with regard to both the modal and algebraic growth, while for ideal gas the trend is opposite.  When $\PrEc$ is large enough, or $T_w$ is closer to (but lower than) $T_{pc}$, the flow falls in the transcritical regime. Due to the large gradient of the viscosity near $T_{pc}$, the base flow becomes inflectional and inviscid unstable. As a consequence, the stability of the non-ideal gas flow is very different from the ideal gas. The neutral curve is expanded, which results in a very low critical Reynolds number. Moreover, the algebraic growth is also enhanced. It should be expected that the laminar-turbulent transition is more likely to be dominated by modal growth in this regime. When $T_w>T_{pc}$, the fluid is in the thermodynamic supercritical regime. In this case, the results of the modal growth show that the non-ideal gas is substantially more stable than the ideal gas. However, the transient growth shows only a weak dependence on the non-ideal gas effects. Additionally, we show that the linear stability analysis with simple cubic equations of state give qualitatively similar results than using the more accurate multi-parameter equation of state implemented in the REFPROP library. \red{Discussions on the reference scaling indicate that the conclusion is not influenced by the choice of the reference variables.} This investigation constitutes the first systematic study of linear stability with highly non-ideal fluids close to the thermodynamic critical point. Future studies will focus on the validation of the results using direct numerical simulations.

\appendix
\section{The stability equation}\label{appA}
The non-zero elements in the stability equation \eqref{Stability} are given below. For simplicity, the derivative of a thermodynamic quantity with respect to $\rho_0$ at constant $T_0$ (and vice-versa) has been denoted as $\frac{\partial}{\partial \rho_0}$, instead of $\left.\frac{\partial}{\partial\rho_{0}}\right|_{T_{0}}$. The elements are, 

\begin{equation}
\left.\begin{array}{l}
\mathscr{L}_{t}\left(1,1\right)=1\\
\mathscr{L}_{t}\left(2,1\right)=u_{0}\\
\mathscr{L}_{t}\left(2,2\right)=\mathscr{L}_{t}\left(3,3\right)=\mathscr{L}_{t}\left(4,4\right)=\rho_{0}\\
\mathscr{L}_{t}\left(5,1\right)=e_{0}+\rho_{0}\frac{\partial e_{0}}{\partial\rho_{0}}\\
\mathscr{L}_{t}\left(5,5\right)=\rho_{0}\frac{\partial e_{0}}{\partial T_{0}}
\end{array}\right\} 
\end{equation}

\begin{equation}
\left.\begin{array}{l}
\mathscr{L}_{x}\left(1,1\right)=u_{0}\\
\mathscr{L}_{x}\left(1,2\right)=\rho_{0}\\
\mathscr{L}_{x}\left(2,1\right)=u_{0}u_{0}+\frac{\partial p_{0}}{\partial\rho_{0}}\\
\mathscr{L}_{x}\left(2,2\right)=2\rho_{0}u_{0}\\
\mathscr{L}_{x}\left(2,3\right)=-\frac{1}{Re}\frac{\partial\mu_{0}}{\partial y}\\
\mathscr{L}_{x}\left(2,5\right)=\frac{\partial p_{0}}{\partial T_{0}}\\
\mathscr{L}_{x}\left(3,1\right)=-\frac{1}{Re}\frac{\partial\mu_{0}}{\partial\rho_{0}}\frac{\partial u_{0}}{\partial y}\\
\mathscr{L}_{x}\left(3,2\right)=-\frac{1}{Re}\frac{\partial\lambda_{0}}{\partial y}\\
\mathscr{L}_{x}\left(3,3\right)=\mathscr{L}_{x}\left(4,4\right)=\rho_{0}u_{0}\\
\mathscr{L}_{x}\left(3,5\right)=-\frac{1}{Re}\frac{\partial\mu_{0}}{\partial T_{0}}\frac{\partial u_{0}}{\partial y}\\
\mathscr{L}_{x}\left(5,1\right)=e_{0}u_{0}+\rho_{0}u_{0}\frac{\partial e_{0}}{\partial\rho_{0}}\\
\mathscr{L}_{x}\left(5,2\right)=\rho_{0}e_{0}+p_{0}\\
\mathscr{L}_{x}\left(5,3\right)=-\frac{2}{Re}\mu_{0}\frac{\partial u_{0}}{\partial y}\\
\mathscr{L}_{x}\left(5,5\right)=\rho_{0}u_{0}\frac{\partial e_{0}}{\partial T_{0}}
\end{array}\right\} 
\end{equation}

\begin{equation}
\left.\begin{array}{l}
\mathscr{L}_{y}\left(1,3\right)=\rho_{0}\\
\mathscr{L}_{y}\left(2,1\right)=-\frac{1}{Re}\frac{\partial\mu_{0}}{\partial\rho_{0}}\frac{\partial u_{0}}{\partial y}\\
\mathscr{L}_{y}\left(2,2\right)=-\frac{1}{Re}\frac{\partial\mu_{0}}{\partial y}\\
\mathscr{L}_{y}\left(2,3\right)=\rho_{0}u_{0}\\
\mathscr{L}_{y}\left(2,5\right)=-\frac{1}{Re}\frac{\partial\mu_{0}}{\partial T_{0}}\frac{\partial u_{0}}{\partial y}\\
\mathscr{L}_{y}\left(3,1\right)=\frac{\partial p_{0}}{\partial\rho_{0}}\\
\mathscr{L}_{y}\left(3,3\right)=-\frac{2}{Re}\frac{\partial\mu_{0}}{\partial y}-\frac{1}{Re}\frac{\partial\lambda_{0}}{\partial y}\\
\mathscr{L}_{y}\left(3,5\right)=\frac{\partial p_{0}}{\partial T_{0}}\\
\mathscr{L}_{y}\left(4,4\right)=-\frac{1}{Re}\frac{\partial\mu_{0}}{\partial y}\\
\mathscr{L}_{y}\left(5,1\right)=-\frac{1}{RePrEc}\left(\frac{\partial\kappa_{0}}{\partial\rho_{0}}\frac{\partial T}{\partial y}\right)\\
\mathscr{L}_{y}\left(5,2\right)=-\frac{2}{Re}\mu_{0}\frac{\partial u_{0}}{\partial y}\\
\mathscr{L}_{y}\left(5,3\right)=\rho_{0}e_{0}+p_{0}\\
\mathscr{L}_{y}\left(5,5\right)=-\frac{1}{RePrEc}\left(\frac{\partial\kappa_{0}}{\partial y}+\frac{\partial\kappa_{0}}{\partial T_{0}}\frac{\partial T_{0}}{\partial y}\right)
\end{array}\right\} 
\end{equation}

\begin{equation}
\left.\begin{array}{l}
\mathscr{L}_{z}\left(1,4\right)=\rho_{0}\\
\mathscr{L}_{z}\left(2,4\right)=\rho_{0}u_{0}\\
\mathscr{L}_{z}\left(3,4\right)=-\frac{1}{Re}\frac{\partial\lambda_{0}}{\partial y}\\
\mathscr{L}_{z}\left(4,1\right)=\frac{\partial p_{0}}{\partial\rho_{0}}\\
\mathscr{L}_{z}\left(4,3\right)=-\frac{1}{Re}\frac{\partial\mu_{0}}{\partial y}\\
\mathscr{L}_{z}\left(4,5\right)=\frac{\partial p_{0}}{\partial T_{0}}\\
\mathscr{L}_{z}\left(5,4\right)=\rho_{0}e_{0}+p_{0}
\end{array}\right\} 
\end{equation}

\begin{equation}
\left.\begin{array}{l}
\mathscr{L}_{q}\left(1,3\right)=\frac{\partial\rho_{0}}{\partial y}\\
\mathscr{L}_{q}\left(2,1\right)=-\frac{1}{Re}\frac{\partial\mu_{0}}{\partial\rho_{0}}\frac{\partial^{2}u_{0}}{\partial y^{2}}-\frac{1}{Re}\frac{\partial u_{0}}{\partial y}\left(\frac{\partial^{2}\mu_{0}}{\partial\rho_{0}^{2}}\frac{\partial\rho_{0}}{\partial y}+\frac{\partial^{2}\mu_{0}}{\partial\rho_{0}\partial T_{0}}\frac{\partial T_{0}}{\partial y}\right)\\
\mathscr{L}_{q}\left(2,3\right)=\rho_{0}\frac{\partial u_{0}}{\partial y}+u_{0}\frac{\partial\rho_{0}}{\partial y}\\
\mathscr{L}_{q}\left(2,5\right)=-\frac{1}{Re}\frac{\partial\mu_{0}}{\partial T_{0}}\frac{\partial^{2}u_{0}}{\partial y^{2}}-\frac{1}{Re}\left(\frac{\partial^{2}\mu_{0}}{\partial T_{0}^{2}}\frac{\partial T_{0}}{\partial y}+\frac{\partial^{2}\mu_{0}}{\partial T_{0}\partial\rho_{0}}\frac{\partial\rho_{0}}{\partial y}\right)\frac{\partial u_{0}}{\partial y}\\
\mathscr{L}_{q}\left(3,1\right)=\frac{\partial^{2}p_{0}}{\partial\rho_{0}^{2}}\frac{\partial\rho_{0}}{\partial y}+\frac{\partial^{2}p_{0}}{\partial\rho_{0}\partial T_{0}}\frac{\partial T_{0}}{\partial y}\\
\mathscr{L}_{q}\left(3,5\right)=\frac{\partial^{2}p_{0}}{\partial T_{0}^{2}}\frac{\partial T_{0}}{\partial y}+\frac{\partial^{2}p_{0}}{\partial\rho_{0}\partial T_{0}}\frac{\partial\rho_{0}}{\partial y}\\
\mathscr{L}_{q}\left(5,1\right)=-\frac{1}{RePrEc}\left(\frac{\partial^{2}T_{0}}{\partial y^{2}}\frac{\partial\kappa_{0}}{\partial\rho_{0}}+\left(\frac{\partial^{2}\kappa_{0}}{\partial\rho_{0}^{2}}\frac{\partial\rho_{0}}{\partial y}+\frac{\partial^{2}\kappa_{0}}{\partial\rho_{0}\partial T_{0}}\frac{\partial T_{0}}{\partial y}\right)\frac{\partial T_{0}}{\partial y}\right)-\frac{1}{Re}\frac{\partial\mu_{0}}{\partial\rho_{0}}\left(\frac{\partial u_{0}}{\partial y}\right)^{2}\\
\mathscr{L}_{q}\left(5,3\right)=e_{0}\frac{\partial\rho_{0}}{\partial y}+\rho_{0}\frac{\partial e_{0}}{\partial y}\\
\mathscr{L}_{q}\left(5,5\right)=-\frac{1}{RePrEc}\left(\frac{\partial^{2}T_{0}}{\partial y^{2}}\frac{\partial\kappa_{0}}{\partial T_{0}}+\left(\frac{\partial^{2}\kappa_{0}}{\partial T_{0}^{2}}\frac{\partial T_{0}}{\partial y}+\frac{\partial^{2}\kappa_{0}}{\partial\rho_{0}\partial T_{0}}\frac{\partial\rho_{0}}{\partial y}\right)\frac{\partial T_{0}}{\partial y}\right)-\frac{1}{Re}\frac{\partial\mu_{0}}{\partial T_{0}}\left(\frac{\partial u_{0}}{\partial y}\right)^{2}
\end{array}\right\} 
\end{equation}

\begin{equation}
\left.\begin{array}{l}
\mathscr{V}_{xx}\left(2,2\right)=\mathscr{V}_{yy}\left(3,3\right)=\mathscr{V}_{zz}\left(4,4\right)=-\frac{2\mu_{0}+\lambda_{0}}{Re}\\
\mathscr{V}_{xx}\left(3,3\right)=\mathscr{V}_{xx}\left(4,4\right)=-\frac{\mu_{0}}{Re}\\
\mathscr{V}_{yy}\left(2,2\right)=\mathscr{V}_{yy}\left(4,4\right)=-\frac{\mu_{0}}{Re}\\
\mathscr{V}_{zz}\left(2,2\right)=\mathscr{V}_{zz}\left(3,3\right)=-\frac{\mu_{0}}{Re}\\
\mathscr{V}_{xx}\left(5,5\right)=\mathscr{V}_{yy}\left(5,5\right)=\mathscr{V}_{zz}\left(5,5\right)=-\frac{\kappa_{0}}{RePrEc}\\
\mathscr{V}_{xy}\left(2,3\right)=\mathscr{V}_{xy}\left(3,2\right)=-\frac{\mu_{0}+\lambda_{0}}{Re}\\
\mathscr{V}_{xz}\left(2,4\right)=\mathscr{V}_{xz}\left(4,2\right)=-\frac{\mu_{0}+\lambda_{0}}{Re}\\
\mathscr{V}_{yz}\left(3,4\right)=\mathscr{V}_{yz}\left(4,3\right)=-\frac{\mu_{0}+\lambda_{0}}{Re}
\end{array}\right\} 
\end{equation}

\begin{figure}
\begin{center}
\includegraphics[width=0.6\linewidth]{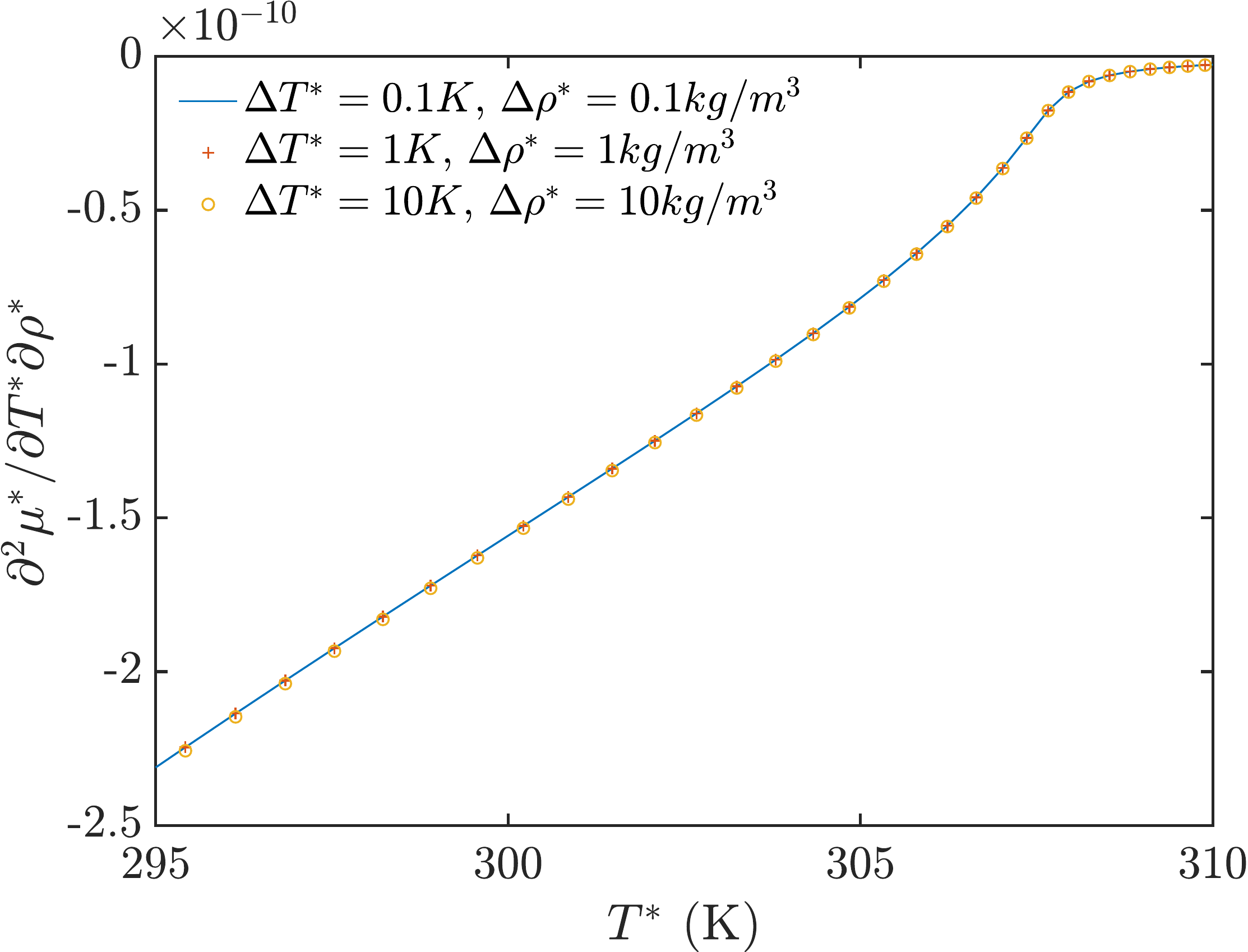}
\end{center}
\caption{Sensitivity of $\dfrac{\pp^2 \mu^*}{\pp T^*\pp\rho^*}$ to $\Delta T^*$ and $\Delta \rho^*$.}
\label{Fig18}
\end{figure}

\red{The second-order finite differences were used to determine the gradients of the properties. For instance, the gradients of viscosity
\begin{equation}
\frac{\partial\mu\left(T_{0},\rho_{0}\right)}{\partial T}=\frac{\mu\left(T_{0}+\Delta T,\rho_{0}\right)-\mu\left(T_{0}-\Delta T,\rho_{0}\right)}{2\Delta T}
\end{equation}
\begin{equation}
\frac{\partial\mu\left(T_{0},\rho_{0}\right)}{\partial\rho}=\frac{\mu\left(T_{0},\rho_{0}+\Delta\rho\right)-\mu\left(T_{0},\rho_{0}-\Delta\rho\right)}{2\Delta\rho}
\end{equation}
\begin{equation}
\frac{\partial^{2}\mu\left(T_{0},\rho_{0}\right)}{\partial T\partial\rho}=\frac{\frac{\partial\mu}{\partial\rho}\left(T_{0}+\Delta T,\rho_{0}\right)-\frac{\partial\mu}{\partial\rho}\left(T_{0}-\Delta T,\rho_{0}\right)}{2\Delta T}
\end{equation}
}
\red{An example of the sensitivity of $\dfrac{\pp^2 \mu^*}{\pp T^*\pp\rho^*}$ to $\Delta T^*$ and $\Delta \rho^*$ is shown in figure~\ref{Fig18}. In fact, the gradients of the thermodynamic \& transport properties became rather robust when $\Delta T^*\leq1$ K and $\Delta \rho^*\leq1$ Kg/m$^3$. In this study, the results are obtained with $\Delta T^*=0.1$ K and $\Delta \rho^*=0.1$ Kg/m$^3$.}

\section{Cubic equation of state}\label{appB}
\red{
The material dependent parameters for CO2 are provided in table~\ref{Table5}. These parameters are necessary inputs for the cubic equation of states detailed below and can be easily replaced for other fluids.} 
\begin{table}
\begin{center}
\begin{tabular}{ccccc}
gas constant  	& heat capacity ratio  	& acentric factor & critical pressure &  critical temperature\vspace{5pt}\\
$R_g^*=188.9$ J/(Kg K)  & $\gamma=1.289$ & $\omega=0.228$ & $p_c^*=73.9$ bar & $T_c^*=304.1$ K\\
\end{tabular}
\caption{The material dependent parameters for CO2.}
\label{Table5}
\end{center}
\end{table}

\subsection{The van der Waals equation of state}
The \citet{VdW_1873} equation of state (EoS) is the simplest cubic equation of state that is capable of accounting phase separation and the critical point \citep[see  the introduction in][]{Zappoli_2015,Engineering_Thermodynamics_7Ed}. The EoS can be written as

\begin{equation}
p=\frac{R_{g}T}{\vartheta-b}-\frac{a}{\vartheta^{2}},
\end{equation}

\noindent
where $R_{g}$ is the specific gas constant, $a$ is a measure of the attraction forces between molecules, and $b$ accounts for the finite volume occupied by the molecules. The constants $a$ and $b$ can be determined at the critical point where 
\begin{equation}
\left.\frac{\partial p}{\partial\vartheta}\right|_{T_c}=\left.\frac{\partial^{2}p}{\partial\vartheta^{2}}\right|_{T_c}=0~~\Rightarrow~~
a=\frac{27}{64}\frac{R_{g}^{2}T_{c}^{2}}{p_{c}},~~
b=\frac{R_{g}T_{c}}{8p_{c}}
\end{equation}
Using the Maxwell relations and the departure function, it is possible to obtain the internal energy  as
\begin{equation}
e=C_vT-a\rho. 
\end{equation}
The required derivatives for stability equations are
\begin{equation}
\left.\frac{\partial p}{\partial T}\right|_{\rho}=\frac{\rho R_{g}}{1-\rho b},~
\left.\frac{\partial p}{\partial\rho}\right|_{T}=\frac{R_{g}T}{\left(1-\rho b\right)^{2}}-2a\rho,
\end{equation}
\begin{equation}
\left.\frac{\partial}{\partial\rho}\right|_{T}\left(\left.\frac{\partial p}{\partial T}\right|_{\rho}\right)=\frac{R_{g}}{\left(1-\rho b\right)^{2}},~
\left.\frac{\partial^{2}p}{\partial T^{2}}\right|_{\rho}=0,~
\left.\frac{\partial^{2}p}{\partial\rho^{2}}\right|_{T}=\frac{2R_{g}Tb}{\left(1-\rho b\right)^{3}}-2a,
\end{equation}
\begin{equation}
\left.\frac{\partial e}{\partial T}\right|_{\rho}=C_v+\left.\frac{\partial C_v}{\partial T}\right|_{\rho}T,~
\left.\frac{\partial e}{\partial\rho}\right|_{T}=-a.
\end{equation}

\subsection{The Redlich-Kwong equation of state}
The Redlich-Kwong \citep{RK_1949} equation of state is given as
\begin{equation}
p=\frac{R_{g}T}{\vartheta-b}-\frac{a\alpha}{\vartheta\left(\vartheta+b\right)}, 
\end{equation}
where $\alpha=\sqrt{T_c/T}$. Similarly, by satisfying the critical condition, the constants $a$ and $b$ are 
\begin{equation}
a=\frac{0.42748R_{g}T_{c}^{2}}{p_{c}},~
b=\frac{0.08664R_{g}T_{c}}{p_{c}}.
\end{equation}
The internal energy is
\begin{equation}
e=C_vT+\frac{3}{2}\frac{a}{b}\alpha\ln\frac{1}{1+b\rho}.
\end{equation}
The derivatives in the stability equations are
\begin{equation}
\left.\frac{\partial p}{\partial T}\right|_{\rho}=\frac{\rho R_{g}}{1-b\rho}+\frac{1}{2}T^{-\frac{3}{2}}T_{c}^{\frac{1}{2}}\frac{a\rho^{2}}{1+b\rho},~
\left.\frac{\partial p}{\partial\rho}\right|_{T}=\frac{R_{g}T}{\left(1-\rho b\right)^{2}}-T^{-\frac{1}{2}}T_{c}^{\frac{1}{2}}\frac{2a\rho+ab\rho^{2}}{\left(1+\rho b\right)^{2}},
\end{equation}
\begin{equation}
\left.\frac{\partial}{\partial\rho}\right|_{T}\left(\left.\frac{\partial p}{\partial T}\right|_{\rho}\right)=\frac{R_{g}}{\left(1-b\rho\right)^{2}}+\frac{1}{2}T^{-\frac{3}{2}}T_{c}^{\frac{1}{2}}\frac{2a\rho+ab\rho^{2}}{\left(1+b\rho\right)^{2}},
\end{equation}
\begin{equation}
\left.\frac{\partial^{2}p}{\partial T^{2}}\right|_{\rho}=-\frac{3}{4}T^{-\frac{5}{2}}T_{c}^{\frac{1}{2}}\frac{a\rho^{2}}{1+b\rho},~
\left.\frac{\partial^{2}p}{\partial\rho^{2}}\right|_{T}=\frac{2bR_{g}T}{\left(1-\rho b\right)^{3}}-T^{-\frac{1}{2}}T_{c}^{\frac{1}{2}}\frac{2a}{\left(1+\rho b\right)^{3}},
\end{equation}
\begin{equation}
\left.\frac{\partial e}{\partial T}\right|_{\rho}=C_v+\left.\frac{\partial C_v}{\partial T}\right|_{\rho}T-\frac{3}{4}\frac{a}{b}T^{-\frac{3}{2}}T_{c}^{\frac{1}{2}}\ln\frac{1}{1+b\rho},~
\left.\frac{\partial e}{\partial\rho}\right|_{T}=-\frac{3}{2}T^{-\frac{1}{2}}T_{c}^{\frac{1}{2}}\frac{a}{1+b\rho}.
\end{equation}

\subsection{The Peng-Robinson equation of state}

The Peng-Robinson \citep{PR_1976} equations of state modifies the original RK and SRK (RK modified by \citet{SRK_1972}) EoS, giving better results regarding the liquid density, vapor pressure and equilibrium ratios. It is one of the most used EoS. It is given as
\begin{equation}
p=\frac{R_{g}T}{\vartheta-b}-\frac{a\alpha}{\vartheta^{2}+2b\vartheta-b^{2}}. 
\end{equation}
The constants $a$, $b$ and parameter $\alpha$ are given by
\begin{equation}
a=\frac{0.457235R_{g}^{2}T_{c}^{2}}{p_{c}},~
b=\frac{0.077796R_{g}T_{c}}{p_{c}},~
\alpha=\left(1+K\left(1-\sqrt{T/T_{c}}\right)\right)^{2}.
\end{equation}
Here $K=0.37464+1.54226\omega-0.26992\omega^{2}$, $\omega$ is the acentric factor of the species. The internal energy 
\begin{equation}
e=C_vT+\frac{a}{2\sqrt{2}b}\left[\left(1+K\right)^{2}-K(1+K)\sqrt{T/T_{c}}\right]\ln\frac{1+b\left(1-\sqrt{2}\right)\rho}{1+b\left(1+\sqrt{2}\right)\rho}.
\end{equation}
The derivatives used in the linear stability equations are give by
\begin{equation}
\left.\frac{\partial p}{\partial T}\right|_{\rho}=\frac{\rho R_{g}}{1-\rho b}+K\sqrt{\frac{\alpha}{TT_{c}}}\frac{a\rho^{2}}{1+2b\rho-b^{2}\rho^{2}},
\end{equation}
\begin{equation}
\left.\frac{\partial p}{\partial\rho}\right|_{T}=\frac{R_{g}T}{\left(1-\rho b\right)^{2}}-\alpha\frac{2a\rho+2ab\rho^{2}}{\left(1+2b\rho-b^{2}\rho^{2}\right)^{2}},
\end{equation}
\begin{equation}
\left.\frac{\partial}{\partial\rho}\right|_{T}\left(\left.\frac{\partial p}{\partial T}\right|_{\rho}\right)=\frac{R_{g}}{\left(1-\rho b\right)^{2}}+K\sqrt{\frac{\alpha}{TT_{c}}}\frac{2a\rho+2ab\rho^{2}}{\left(1+2b\rho-b^{2}\rho^{2}\right)^{2}},
\end{equation}
\begin{equation}
\left.\frac{\partial^{2}p}{\partial T^{2}}\right|_{\rho}=-\frac{K\left(1+K\right)}{2\sqrt{T^{3}T_{c}}}\frac{a\rho^{2}}{1+2b\rho-b^{2}\rho^{2}},
\end{equation}
\begin{equation}
\left.\frac{\partial^{2}p}{\partial\rho^{2}}\right|_{T}=\frac{2R_{g}bT}{\left(1-\rho b\right)^{3}}-\alpha\frac{2a\left(2b^{3}\rho^{3}+3b^{2}\rho^{2}+1\right)}{\left(1+2b\rho-b^{2}\rho^{2}\right)^{3}},
\end{equation}
\begin{equation}
\left.\frac{\partial e}{\partial T}\right|_{\rho}=C_v+\left.\frac{\partial C_v}{\partial T}\right|_{\rho}T+\frac{a}{4\sqrt{2}b}\left[-K(1+K)\sqrt{1/TT_{c}}\right]\ln\frac{1+b\left(1-\sqrt{2}\right)\rho}{1+b\left(1+\sqrt{2}\right)\rho},
\end{equation}
\begin{equation}
\left.\frac{\partial e}{\partial\rho}\right|_{T}=-\frac{a}{1+2b\rho-b^{2}\rho^{2}}\left[\left(1+K\right)^{2}-K(1+K)\sqrt{T/T_{c}}\right].
\end{equation}

\section{Influence of the bulk viscosity}\label{appC}
The dynamics of a fluid are described by the Navier-Stokes equation, which in its simplest form contain a linear relation between deformation of a fluid element and the resulting stress, with the shear viscosity $\mu$ the coefficient of proportionality. Phenomenologically, another coefficient is possible, the second viscosity $\lambda$, which was introduced by \citet{Stokes1845}. Stokes anticipated that the second viscosity might play a role in compressible fluids. However, for the cases he considered, the fluids can be assumed incompressible with negligible dilatational effects, such that the bulk viscosity within the second viscosity can be ignored. This is known as the Stokes approximation. Consequently, setting the bulk viscosity $\mu_b$ to zero, has been broadly adopted in numerical simulations of compressible flows \citep[see a succinct review by][]{Graves1999}. 

\begin{figure}
\begin{center}
\includegraphics[width=0.45\linewidth,clip]{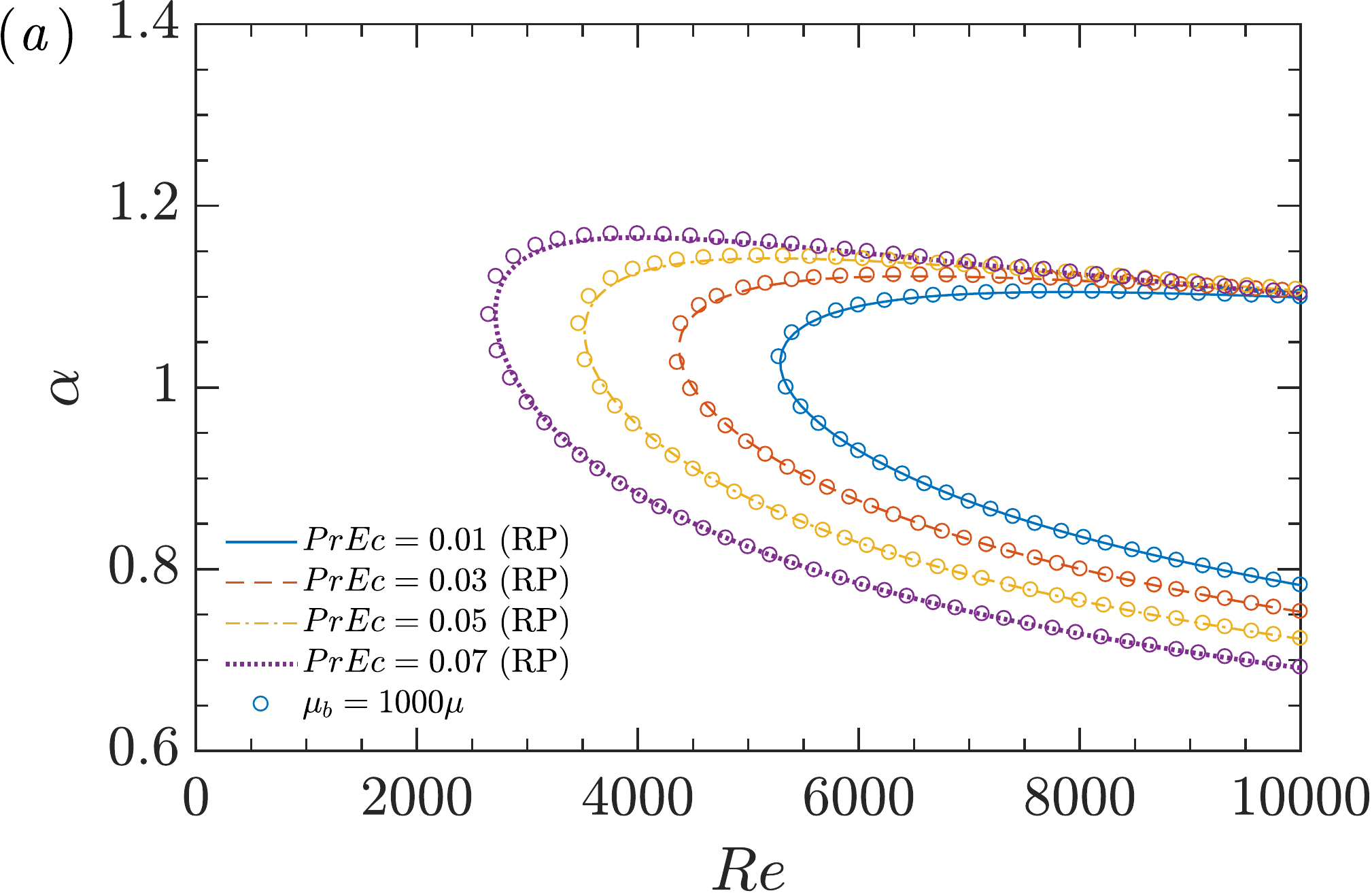}
\includegraphics[width=0.465\linewidth,clip]{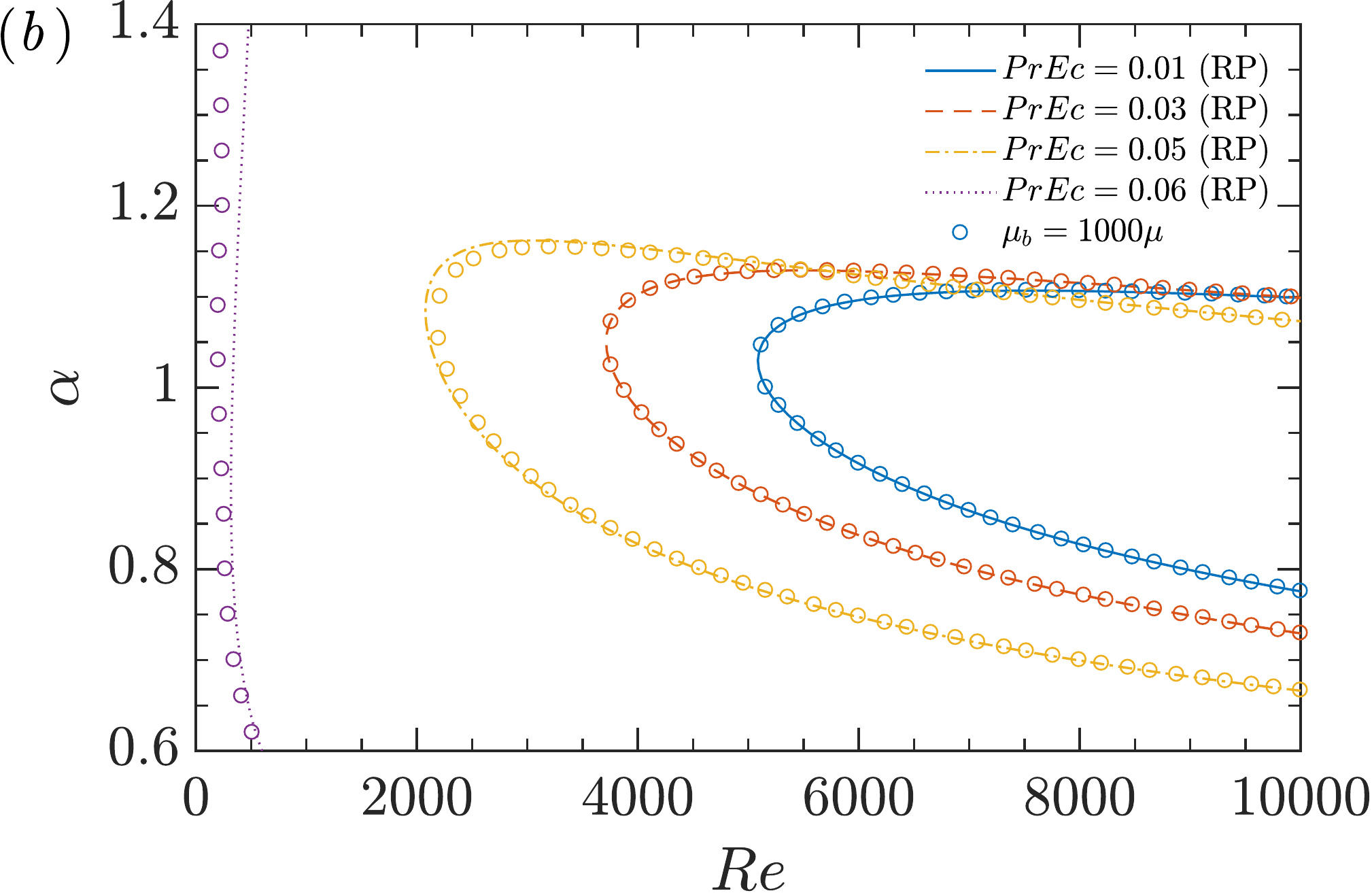}\\\vspace{10pt}
\includegraphics[width=0.465\linewidth,clip]{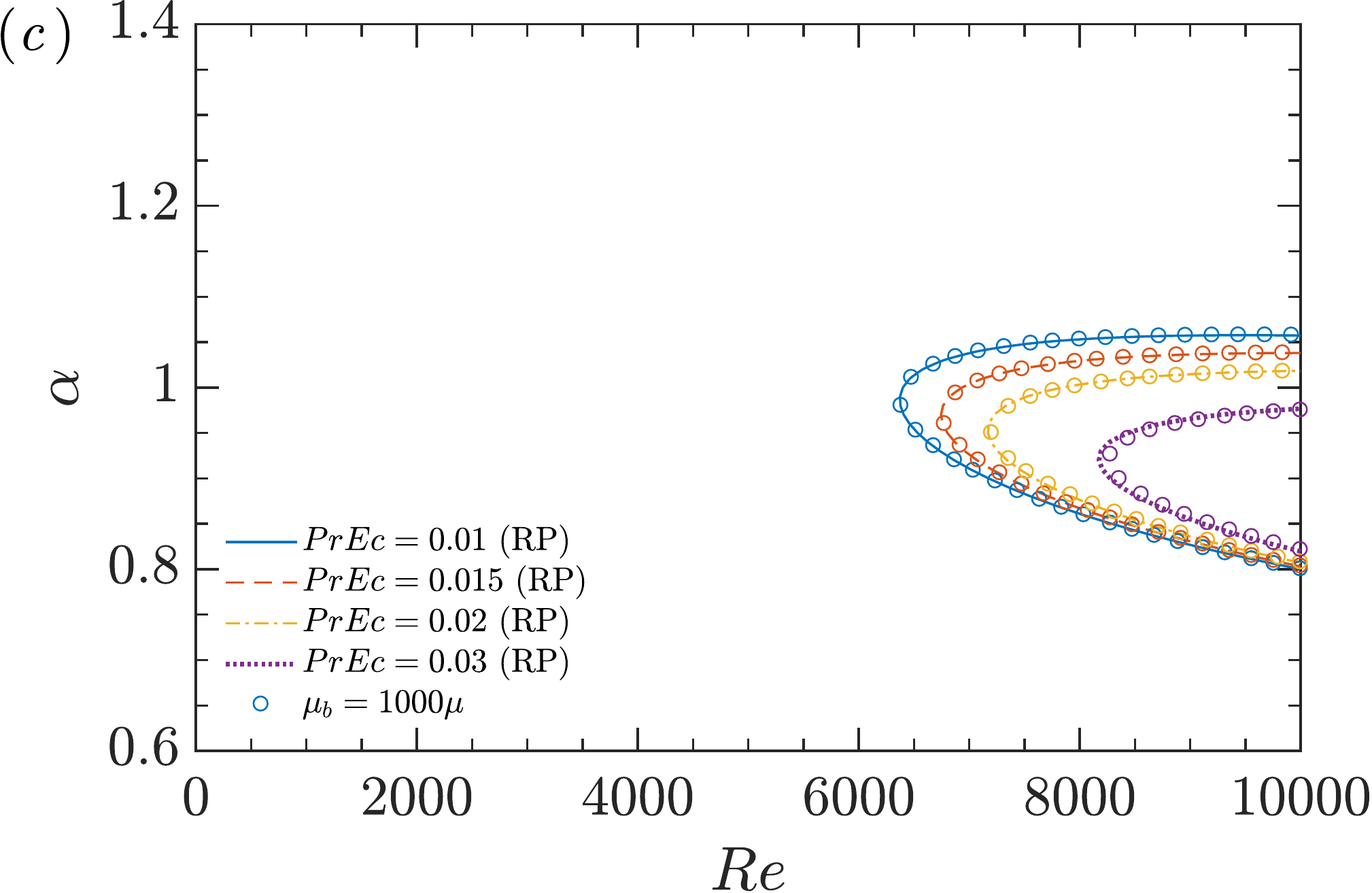}
\end{center}
\caption{Influence of bulk viscosity on neutral curves in (a) subcritical (b) transcritical and (c) supercritical case.}
\label{Fig19}
\end{figure}

\begin{figure}
\begin{center}
\includegraphics[width=0.45\linewidth,clip]{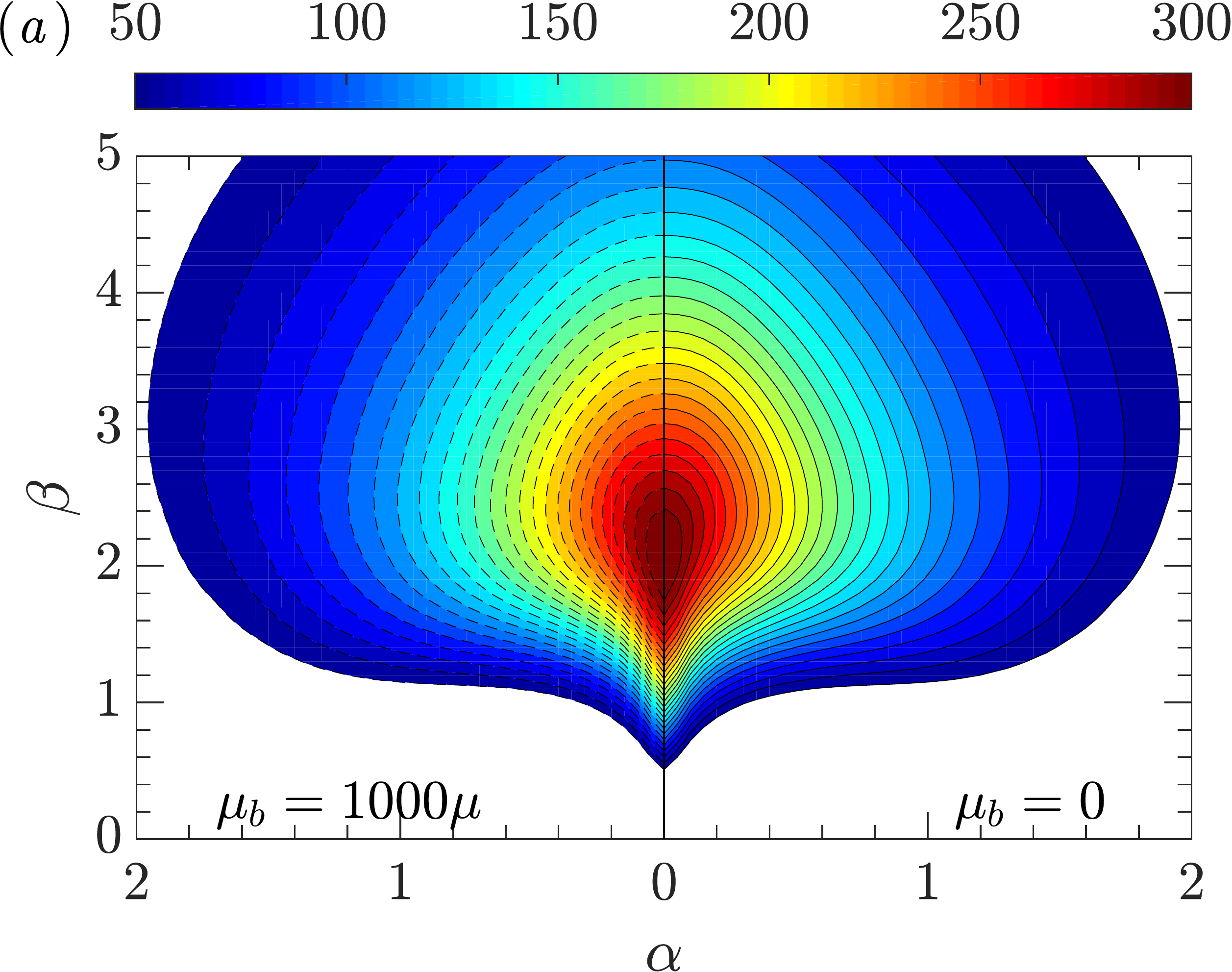}
\includegraphics[width=0.45\linewidth,clip]{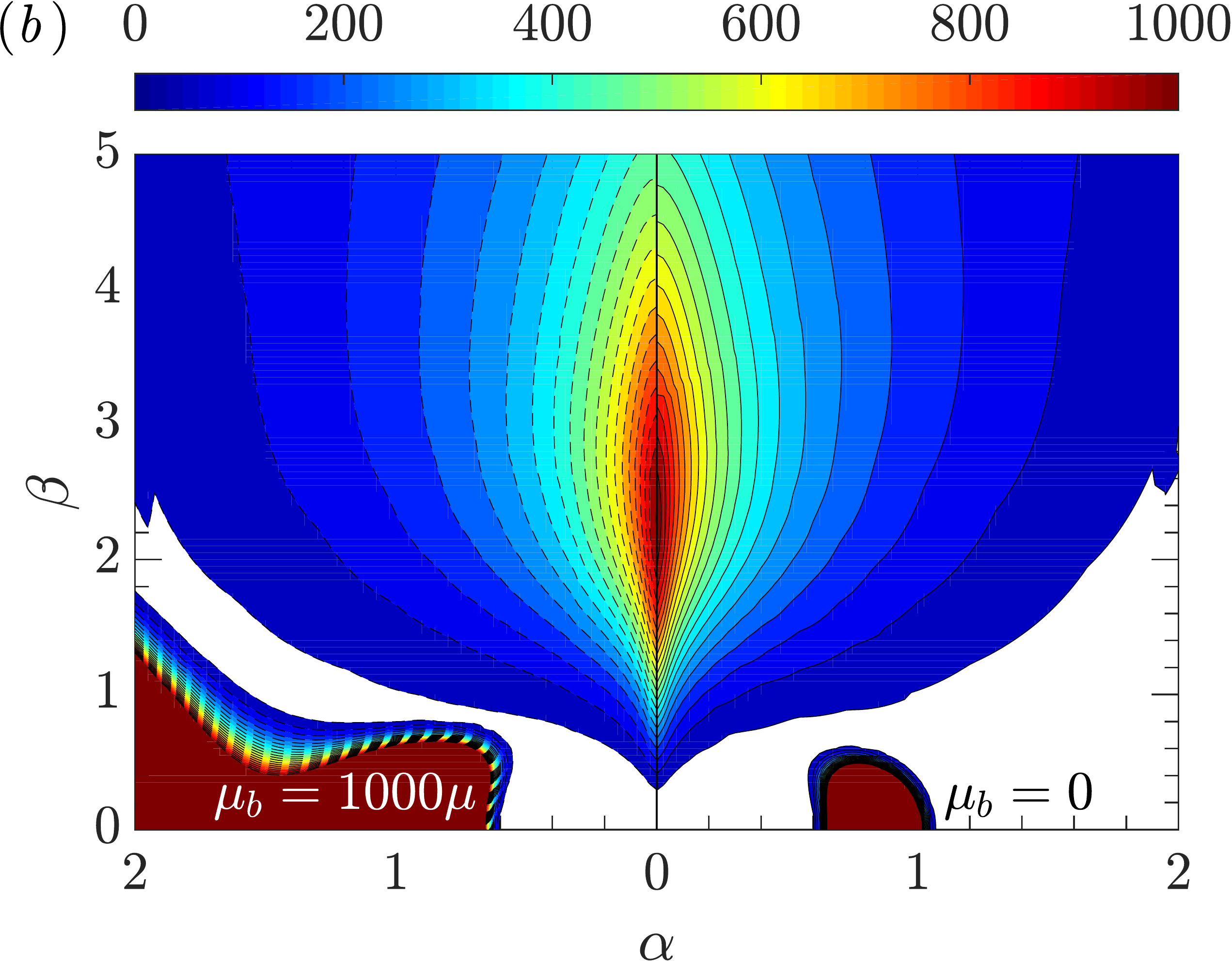}\\\vspace{10pt}
\includegraphics[width=0.45\linewidth,clip]{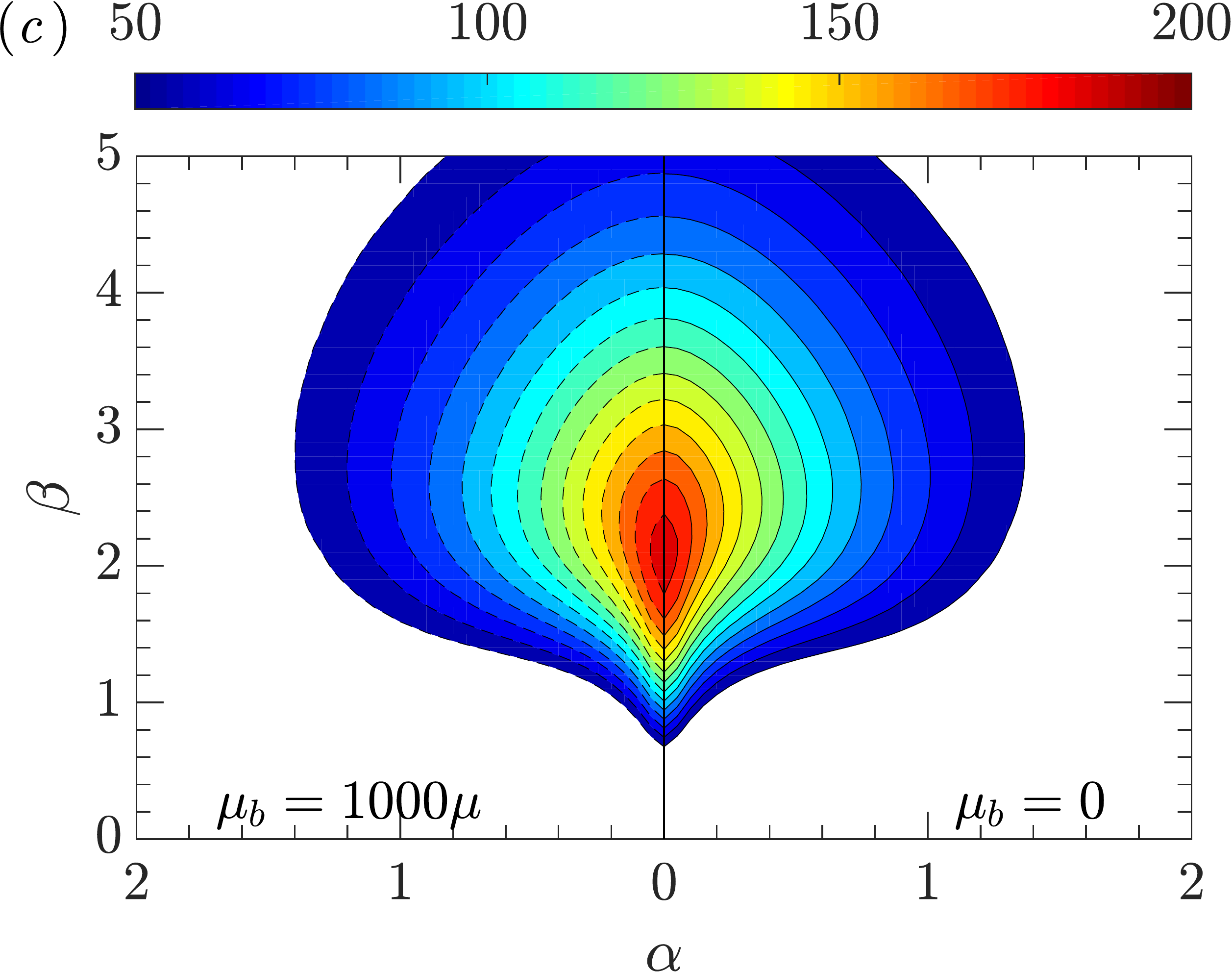}
\end{center}
\caption{Influence of bulk viscosity on the algebraic growth. $\PrEc=0.07$, $\Rey=1000$. (a) subcritical (b) transcritical and (c) supercritical case.}
\label{Fig20}
\end{figure}

\citet{Cramer2012}'s numerical estimates indicate that $\mu_b/\mu$ of some common gases can reach $O(10^3)$. To investigate the influence of $\mu_b$ on the results of the linear stability, we performed simulations with $\mu_b=1000\mu$. The results are shown in figure~\ref{Fig19} and \ref{Fig20}, which show the comparison of the linear stability results for $\mu_b=1000\mu$ and $\mu_b=0$, using the RP model (the other parameters are kept the same). Figure~\ref{Fig19} shows that the neutral curves are barely affected. A discernible difference only exists in the transcritical case ($T_w^*=300$ K, \PrEc=0.06), where the neutral curve with $\mu_b=1000\mu$ becomes slightly more expanded. On the other hand, the algebraic instability does not vary with bulk viscosity. Only the modal growth region ($G_{max}=\infty$) in figure~\ref{Fig20}(b) becomes larger with $\mu_b=1000\mu$ and is consistent with the results shown in figure~\ref{Fig19}(b). 

The above comparisons support the Stokes' hypothesis used in this study. In fact, $\mu_b$ is frequency dependent, this means that depending on the perturbation one prescribes in the stability analysis, the bulk viscosity will have different values. Therefore for a more rigorous investigation we would need reliable frequency resolved data for the bulk viscosity, either from theories, experiments \citep{Karim1952}, or molecular dynamics simulations \citep{Hoover1980} .
\section{\red{Influence of the reference scaling}}\label{appD}
\red{
Previous studies have shown that the scaling of the governing equations may have a large influence on the results. For example, if the viscosity at the cold wall is used as a reference value, \citet{sahu2010a} concluded that increasing the temperature difference between both walls destabilizes the flow, while \citet{SAMEEN2007} concluded the opposite behaviour if the viscosity at the hot wall is used. On the other hand, \citet{Rinaldi2018} investigated the edge state solutions of viscosity-stratified flows where they showed that a different reference value for viscosity does not qualitatively change the results. In this appendix, we show how the definition of the non-dimensional quantities will influence the results presented in the paper. }

\begin{figure}
\begin{center}
\includegraphics[width=0.45\linewidth]{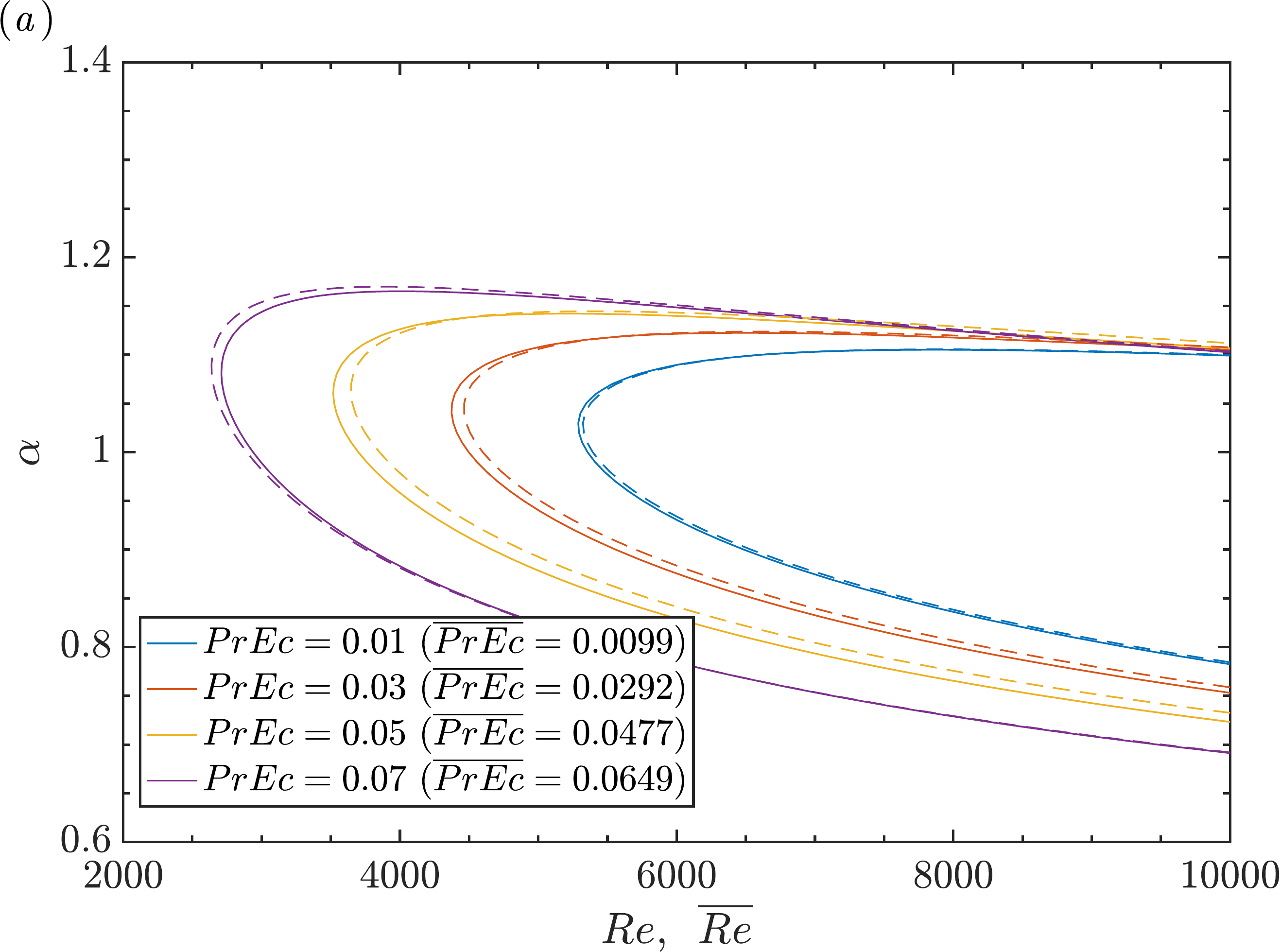}
\includegraphics[width=0.45\linewidth]{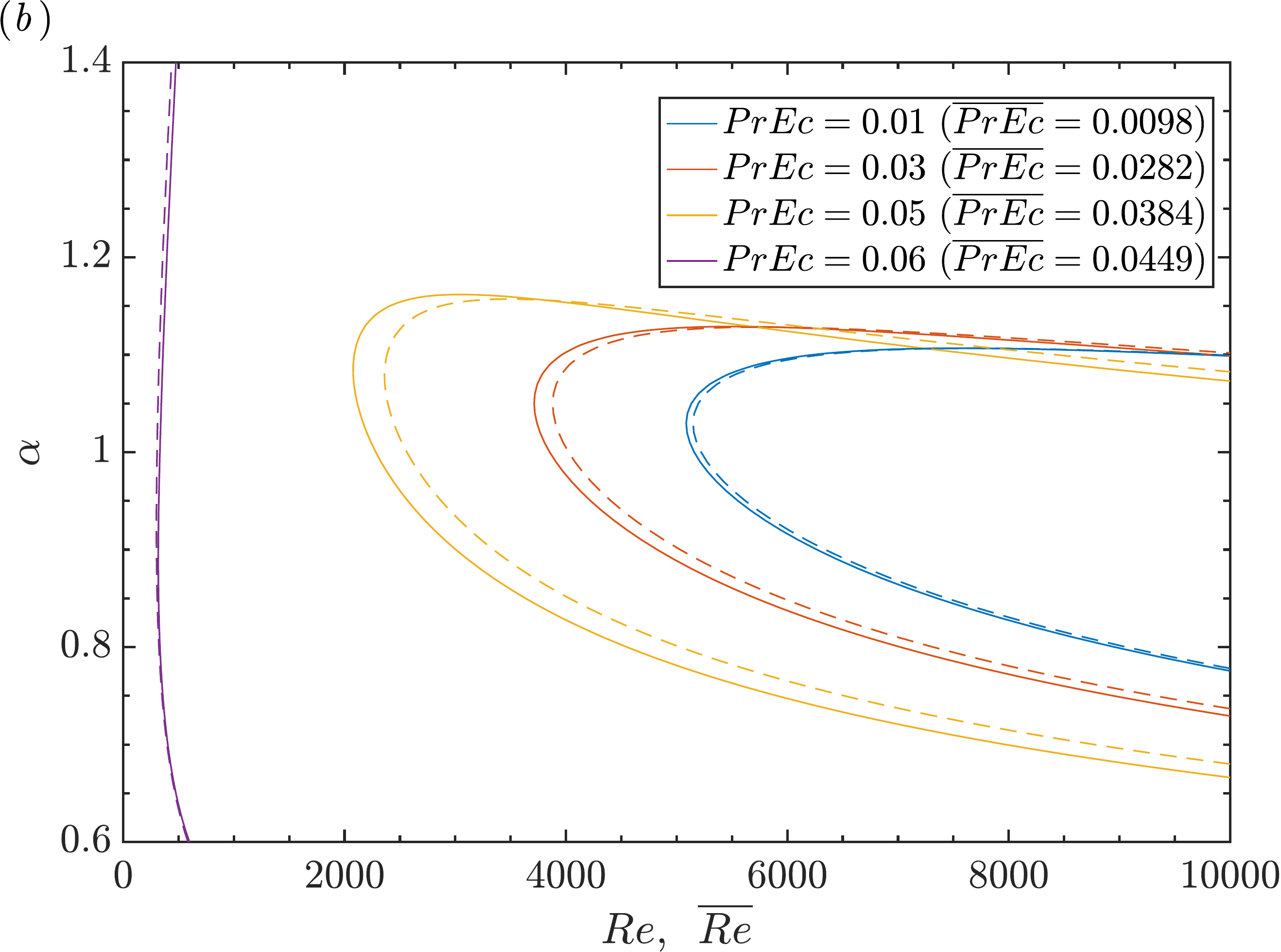}\\\vspace{10pt}
\includegraphics[width=0.45\linewidth]{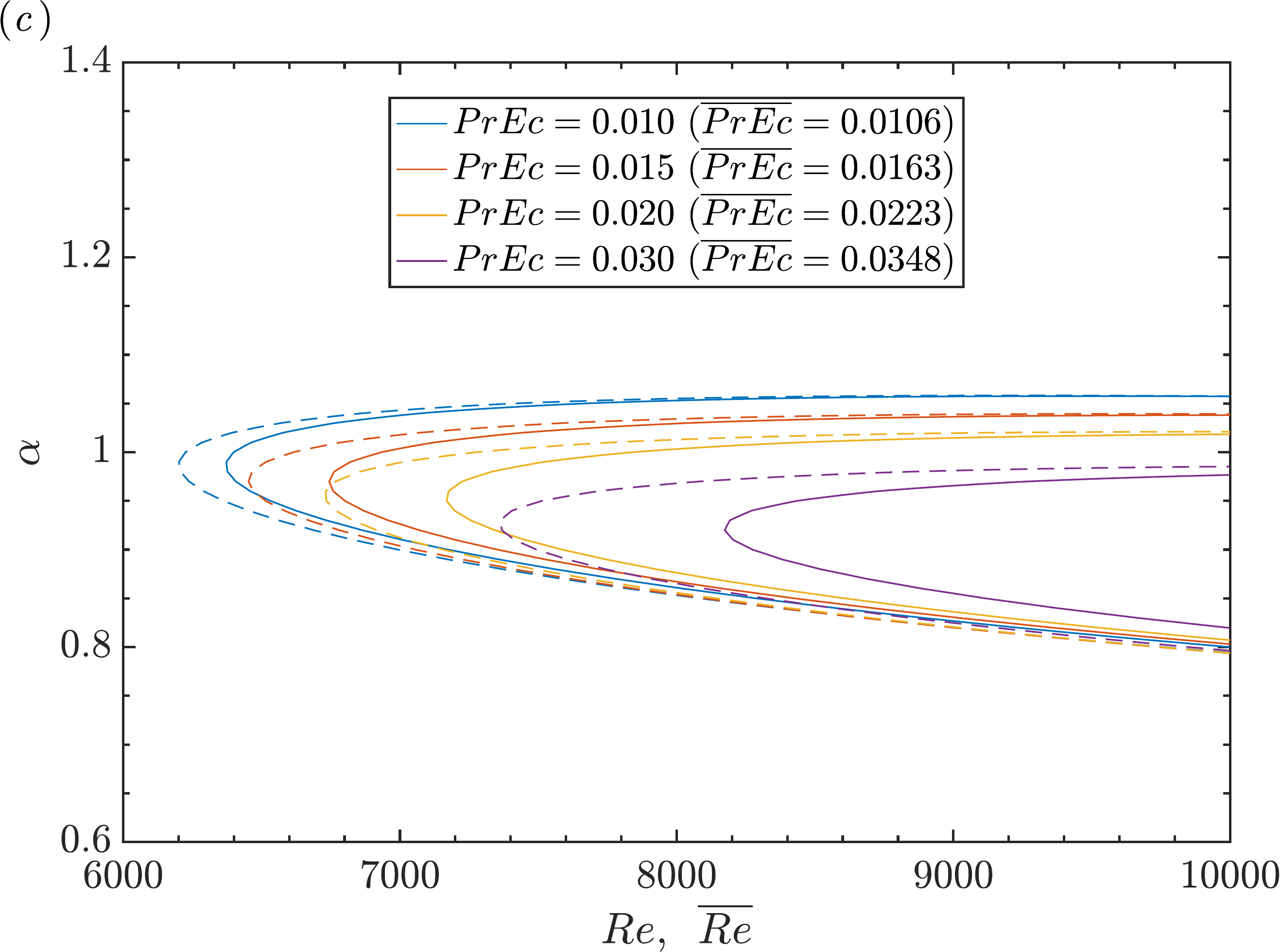}
\end{center}
\caption{Influence of the reference scaling on the neutral curve for the (a) subcritical (b) transcritical and (c) supercritical cases. Solid lines show the results with wall scaling (in the $\alpha-\Rey$ diagram), while dashed lines indicate the average scaling (in the $\alpha-\overline{\Rey}$ diagram). These changes will}
\label{Fig21}
\end{figure}

\red{We introduce the averaged values of the thermodynamic \& transport properties:
\begin{gather}
T^*_{av}=\dfrac{1}{h^*}\int T^*\dd y,~
\rho^*_{av}=\dfrac{1}{h^*}\int \rho^*\dd y,\\
\mu^*_{av}=\dfrac{1}{h^*}\int \mu^*\dd y,~
\kappa^*_{av}=\dfrac{1}{h^*}\int \kappa^*\dd y.
\end{gather}
When the governing equations are scaled by the above averaged values, one obtains the averaged Reynolds number, $\overline{\Rey}$, and the product of the averaged Prandtl and Eckert number, $\overline{\PrEc}$:
\begin{equation}
\overline{\Rey}=\frac{\rho_{av}^{*}u_{r}^{*}h^{*}}{\mu_{av}^{*}},~
\overline{\PrEc}=\frac{\mu_{av}^{*}u_{r}^{*2}}{\kappa_{av}^{*}T_{av}^{*}}.
\end{equation}
We name it the \emph{average scaling}, to distinguish from the \emph{wall scaling} presented in \S\ref{Sec2-1} of the paper.
Note that the reference velocity is not independent, and is given by
\begin{equation}
u_r^*=\sqrt{\PrEc\frac{\kappa^*_wT_w^*}{\mu_w^*}}=\sqrt{\overline{\PrEc}\frac{\kappa^*_{av}T_{av}^*}{\mu_{av}^*}}.
\end{equation}
Using both scalings resulted in qualitatively similar conclusion as shown in figure~\ref{Fig21} for the modal instability. That is, the flow becomes more unstable in the subcritical regime, inviscid unstable in the transcritical regime, and more stable in the supercritical regime.
}

\begin{figure}
\begin{center}
\includegraphics[width=0.45\linewidth]{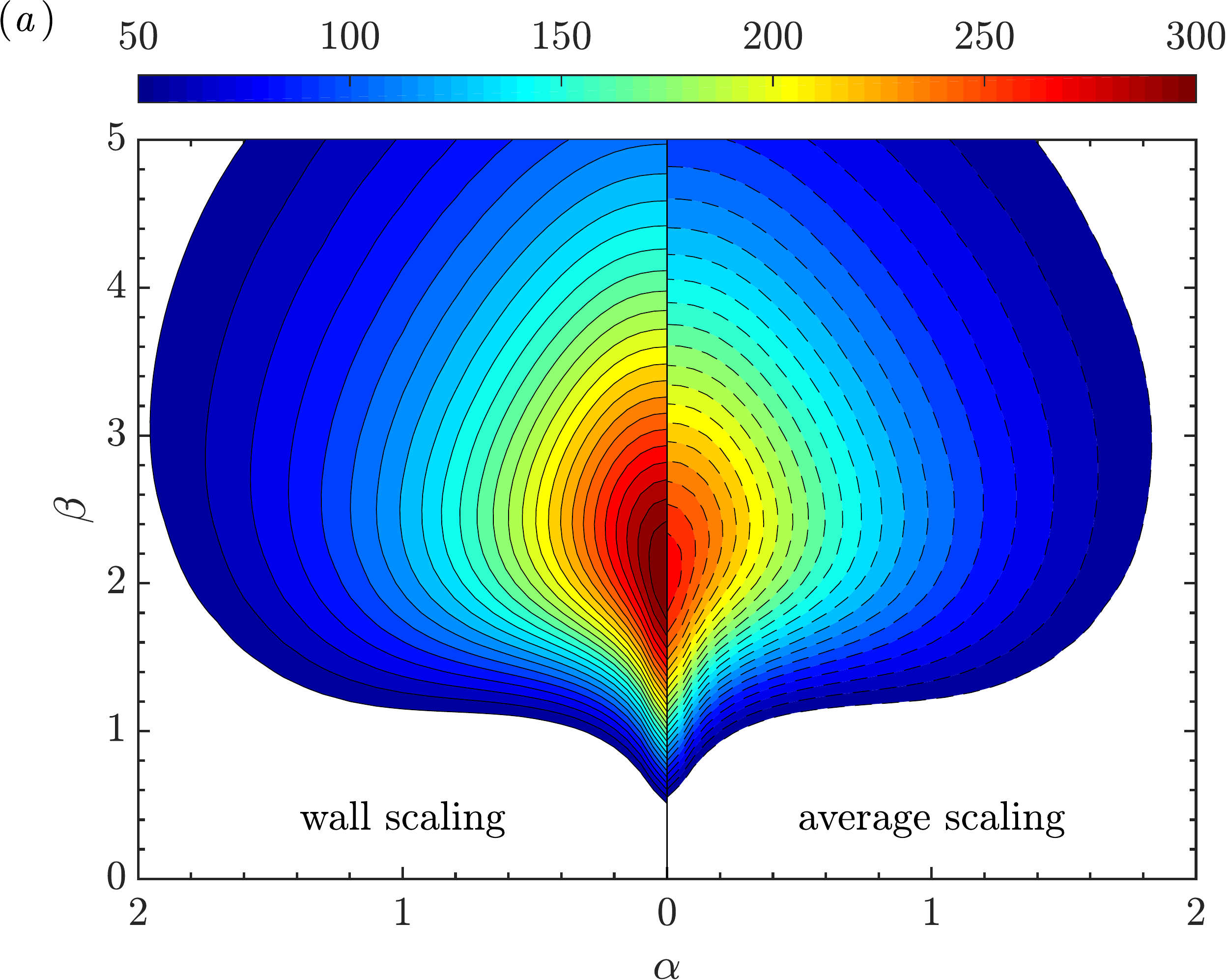}
\includegraphics[width=0.45\linewidth]{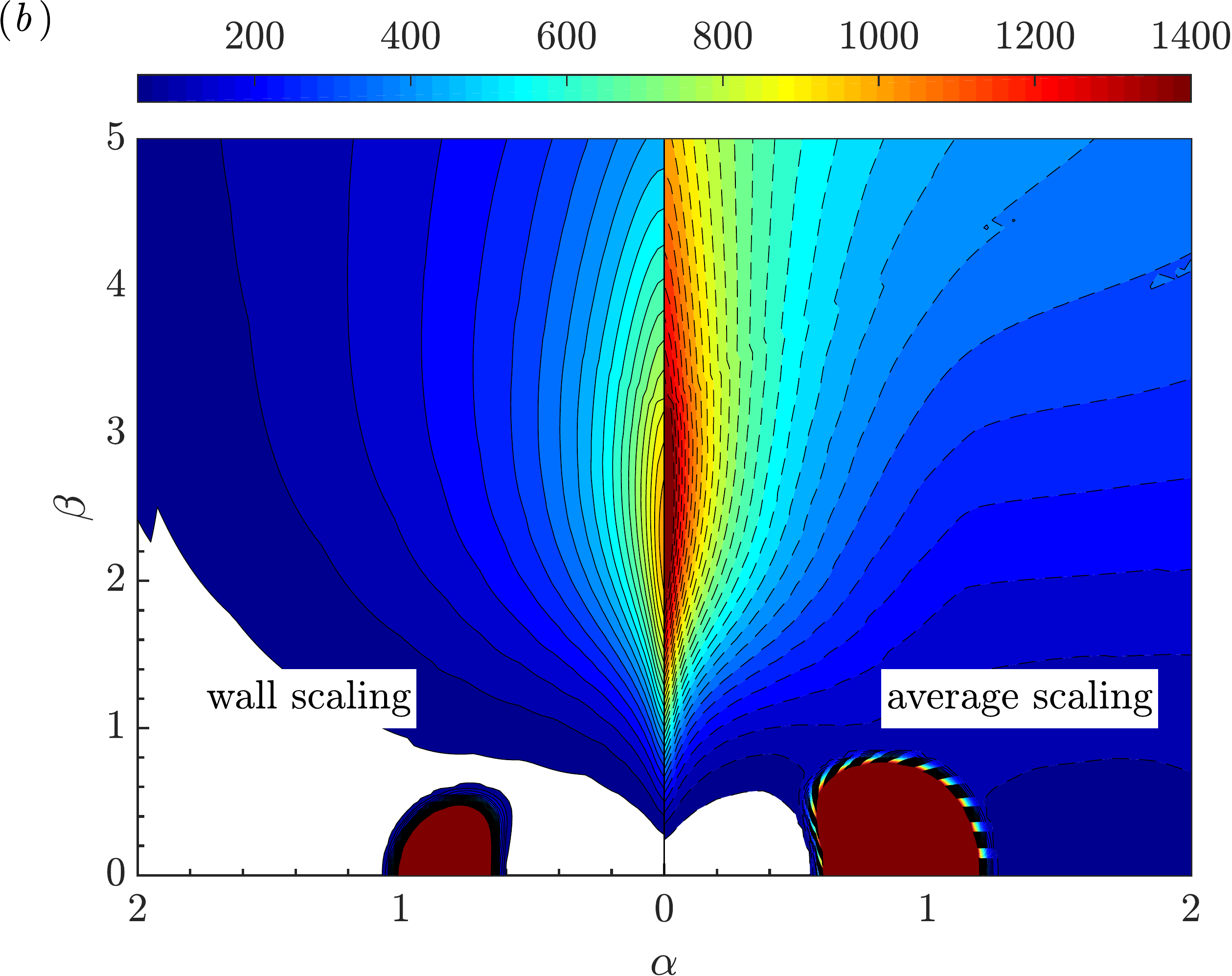}\\\vspace{10pt}
\includegraphics[width=0.45\linewidth]{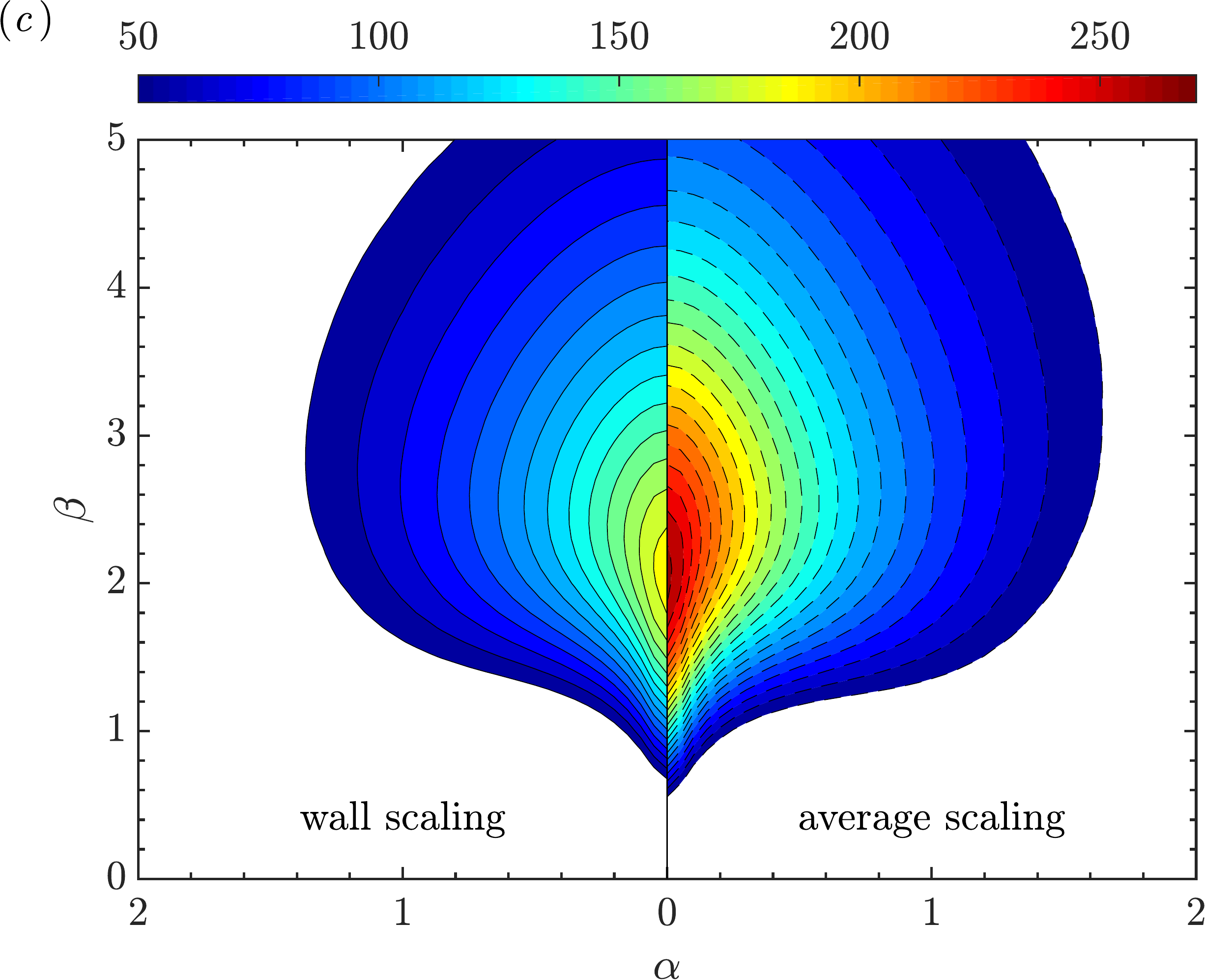}
\end{center}
\caption{Influence of the reference scaling on the maximum algebraic growth $G_{\max}$ for the (a) subcritical (b) transcritical and (c) supercritical cases. The left and right half show the results with wall scaling and average scaling respectively. The non-ideal gas model is RP, $\PrEc=0.07$, $\Rey=1000$.}
\label{Fig22}
\end{figure}

\red{Regarding the algebraic instability using the average scaling, as can be seen from figure~\ref{Fig22}, the maximum growth shows a minor reduction in the subcritical regime. Increases in $G_{\max}$ are noticed for the trans- and supercritical regimes. Comparisons with the ideal gas have been summarized in table~\ref{Table6}. The ideal gas are not sensitive to the wall temperature under both scalings. With average scaling, the conclusion for the algebraic instability will not change.}

\begin{table}
\begin{center}
\begin{tabular}{lcc}
       &            \multicolumn{2}{c}{Scaling} \\
 	   &              wall-based 	  & average-based\vspace{5pt}\\
$T^*_w=290$ K (IG)  & 178.1 	      & 189.9\\
$T^*_w=290$ K (RP) 	& 316.7 		  & 271.3\vspace{5pt}\\
$T^*_w=300$ K (IG)  & 178.1 		  & 189.9\\
$T^*_w=300$ K (RP) 	& 1040.8	      & 1582.7\vspace{5pt}\\
$T^*_w=310$ K (IG)  & 178.1			  & 189.9\\
$T^*_w=310$ K (RP) 	& 190.3			  & 268.2
\end{tabular}
\caption{\red{Maximum algebraic growth $G_{\max}$ over $\alpha$ and $\beta$. $\PrEc=0.07$, $\Rey=1000$ (wall scaling) or $\overline{\Rey}=1000$ (average scaling). The values for the $T^*_w=300$ case are for $\alpha=0$ where modal instability is absent.}}
\label{Table6}
\end{center}
\end{table}

\bibliographystyle{jfm}
\bibliography{jfm}

\end{document}